\documentclass[fleqn,usenatbib]{mnras}
\usepackage{newtxtext,newtxmath}
\usepackage[T1]{fontenc}
\usepackage{graphicx}	% Including figure files
\usepackage{amsmath}	% Advanced maths commands
\usepackage{svg}        % to load ORCID logo
\usepackage{xspace}     % adds a space after a macro
\usepackage{booktabs}   % extra commands for tables
\usepackage{subdepth}   % equalise the height of subscripts
\usepackage{subcaption} % to use with multiple panel figures
\newcommand{\orcid}[1]{\href{https://orcid.org/#1}{\includesvg[width=7pt]{images/orcid}}}
\newcommand{\kms}{\,km\,s$^{-1}$\xspace} % km/s
\newcommand{\cms}{\,cm\,s$^{-2}$\xspace} % cm/s^-2
\newcommand{\Msun}{$\,M_\odot$\xspace}  % solar mass
\newcommand{\Rsun}{$\,R_\odot$\xspace}  % solar radius
\newcommand{\Lsun}{$\,L_\odot$\xspace}  % solar luminosity
\newcommand{\vsini}{$\varv \sin i$\xspace} % vsini
\newcommand{\vmac}{$\varv_{\rm mac}$\xspace} % v_mac
\newcommand{\vmic}{$\varv_{\rm mic}$\xspace} % v_mic
\newcommand{\chisq}{$\chi^2$\xspace} % chi^2
\newcommand{\Teff}{$T_{\rm eff}$\xspace} % 
\newcommand{\logg}{$\log g$\xspace} % 
\newcommand{\porb}{$P_{\rm orb}$\xspace} %
\newcommand\spline[3]{\text{#1\,\textsc{\lowercase{#2}}\,$\lambda$#3}}

\NewCommandCopy{\mc}{\multicolumn}

\newlength{\VSpaceBeforeTabBib}
\newlength{\VSpaceBeforeTabFoot}
\newcommand\tablefoot[1]{\VSpaceBeforeTabBib=1ex%
  \par\vspace{\VSpaceBeforeTabFoot}
  \noindent
  \begin{minipage}{\linewidth}
    {\small\bfseries Notes.}~%
    \small
    \ignorespaces
    #1%
  \end{minipage}%
}

\defcitealias{villasenor+21}{Paper~I}

\title[VFTS~291: A stripped star from a recent mass transfer phase]{The B-type Binaries Characterisation Programme II.\\ VFTS~291: A stripped star from a recent mass transfer phase}

\author[J.~I. Villaseñor et al.]{
J.~I. Villaseñor\textsuperscript{\orcid{0000-0002-7984-1675}},$^{1}$\thanks{E-mail: jaime.villasenor@kuleuven.be} 
        D.~J. Lennon,$^{2,3}$
        A. Picco,$^{1}$
        T. Shenar\textsuperscript{\orcid{0000-0003-0642-8107}},$^{4}$
        P. Marchant\textsuperscript{\orcid{0000-0002-0338-8181}},$^{1}$
        N. Langer,$^{5,6}$
        P.~L. Dufton,$^{7}$
        \newauthor
        F. Nardini,$^{1}$
        C.~J. Evans\textsuperscript{\orcid{0009-0005-7288-6407}},$^{8}$
        J. Bodensteiner,$^{9}$
        S.~E. de Mink\textsuperscript{\orcid{0000-0001-9336-2825}},$^{10,4}$
        Y. Götberg,$^{12}$
        I. Soszyński,$^{13}$
        \newauthor
        W.~D. Taylor,$^{14}$
        and H. Sana\textsuperscript{\orcid{0000-0001-6656-4130}}$^{1}$\vspace{2.5mm}
\\
    $^{1}$Institute of Astronomy, KU Leuven, Celestijnenlaan 200D, 3001 Leuven, Belgium\\
    $^{2}$Instituto de Astrof\'isica de Canarias,E-38\,200 La Laguna, Tenerife, Spain\\
    $^{3}$Dpto. Astrof\'isica, Universidad de La Laguna, E-38\,205 La Laguna, Tenerife, Spain\\
    $^{4}$Anton Pannekoek Institute for Astronomy, University of Amsterdam, Science Park 904, 1098 XH Amsterdam, The Netherlands\\
    $^{5}$Argelander-Institut f\"ur Astronomie, Universit\"at Bonn, Auf dem H\"ugel 71, 53121 Bonn, Germany\\
    $^{6}$Max-Planck-Institut f\"ur Radioastronomie, Auf dem H\"ugel 69, 53121 Bonn, Germany\\
    $^{7}$Astrophysics Research Centre, School of Mathematics \& Physics,  Queen’s University, Belfast, BT7 1NN, UK\\
    $^{8}$European Space Agency (ESA), ESA Office, Space Telescope Science Institute, 3700 San Martin Drive, Baltimore, MD 21218, USA\\
    $^{9}$European Organisation for Astronomical Research in the Southern Hemisphere (ESO), Karl-Schwarzschild-Str. 2, 85748 Garching b. München, Germany\\
    $^{10}$Max-Planck-Institut für Astrophysik, Karl-Schwarzschild-Stra{\ss}e 1, 85740 Garching bei München, Germany\\
    $^{11}$The Observatories of the Carnegie Institution for Science, 813 Santa Barbara Street, Pasadena, CA 91101, USA\\
    $^{12}$Astronomical Observatory, University of Warsaw, Al. Ujazdowskie 4, 00-478 Warszawa, Poland\\
    $^{13}$UK Astronomy Technology Centre, Royal Observatory, Blackford Hill, Edinburgh, EH9 3HJ, UK\\
}

\date{Accepted XXX. Received YYY; in original form ZZZ}

\pubyear{2023}

\begin{document}
\label{firstpage}
\pagerange{\pageref{firstpage}--\pageref{lastpage}}
\maketitle

\begin{abstract}
Recent studies of massive binaries with putative black-hole companions have uncovered a phase of binary evolution that has not been observed before, featuring a bloated stripped star that very recently ceased transferring mass to a main sequence companion. In this study, we focus on the candidate system VFTS~291, a binary with an orbital period of 108\,d and a high semi-amplitude velocity ($K_1=93.7\pm0.2$\kms). Through our analysis of the disentangled spectra of the two components, together with dynamical and evolutionary arguments, we identify a narrow-lined star of $\sim$1.5-2.5\Msun dominating the spectrum, and an early B-type main-sequence companion of $13.2\pm1.5$\Msun. The low mass of the narrow-lined star, and the high mass ratio, suggest that VFTS~291 is a post mass-transfer system, with the narrow-lined star being bloated and stripped of its hydrogen-rich envelope, sharing many similarities with other recently discovered stripped stars. Our finding is supported by our detailed binary evolution models, which indicate that the system can be well explained by an initial configuration consisting of an 8.1\Msun primary with an 8\Msun companion in a 7\,d orbital period. While some open questions remain, particularly concerning the surface helium enrichment of the stripped star and the rotational velocity of the companion, we expect that high-resolution spectroscopy may help reconcile our estimates with theory. Our study highlights the importance of multi-epoch spectroscopic surveys to identify and characterise binary interaction products, and provides important insights into the evolution of massive binary stars.
\end{abstract}


%%%%%%%%%%%%%%%%% BODY OF PAPER %%%%%%%%%%%%%%%%%%

\section{Introduction}

The majority of massive stars ($M\gtrsim8$\Msun) are members of multiple systems \citep[e.g.][]{kobulnicky+fryer07, mason+09, kiminki+kobulnicky12, sana+12, sana+13, sana+14, kobulnicky+14, dunstall+15, moe+distefano17, banyard+22}, with a fraction that could go as high as 100\% on the zero-age main sequence \citep[ZAMS,][see also \citealt{offner+22}]{bordier+22}. Due to the large preference for short periods \citep{sana+12, kobulnicky+14, almeida+17, barba+17, villasenor+21, banyard+22}, it is common for massive binaries to interact during their lives, with about a third of them exchanging mass while still on the MS \citep{sana+12}. Binary interaction products constitute a large fraction of current populations of massive stars, with the occurrence of mergers predicted to be 10-30\% \citep{sana+12, demink+14}. Stripped stars, that is stars that have lost their hydrogen-rich envelopes due to mass transfer to a companion \citep{paczynski67, podsiadlowski+92, goetberg+17, goetberg+18}, should also be a common outcome of binary evolution \citep{sana+12, eldridge+13}, however, the small number of identified objects has been a long standing problem for stellar evolution \citep{wellstein+01}.

In the last years, and motivated by the first detections of gravitational waves by the LIGO Scientific and Virgo Collaborations \citep[see][and references therein]{abbott+21}, the search for progenitors of binary black holes (BHs) and neutron stars (NSs) has intensified. Only a handful of high-mass X-ray binaries \citep[HMXBs,][]{vandenheuvel19} with BHs are known, so it is possible that many OB+BH systems are in a `X-ray-quiet'' phase, where no accretion disk has been formed, and therefore are very challenging to observe \citep{sen+21}. Several studies have claimed the detection of dormant BHs as companions to B-type stars \citep{casares+14, liu+19, rivinius+20, saracino+22} but have been challenged with scenarios involving low-mass stripped stars with large RV shifts produced by more massive B/Be companions, but no BHs \citep{abdul-masih+20, irrgang+20, shenar+20, bodensteiner+20b, el-badry+burdge22, frost+22, rivinius+22}. However, the recent announcement of two stellar-BH detections around massive stars has been made. \citet{mahy+22} presented the galactic binary HD~130298 as a strong  OB+BH candidate, consisting of a O6.5~III(n)(f) primary with an unseen companion of at least 7.7\Msun. From the Tarantula Massive Binary Monitoring \citep[TMBM,][]{almeida+17}, \citet{shenar+22b} analysed the companions to 51 single-lined spectroscopic binary (SB1) systems, confirming VFTS~243 as an OB+BH system (and two additional less certain candidates), with a 25\Msun O7~V:(n)((f)) star plus a 10\Msun BH \citep{shenar+22a}.

The aforementioned recent findings of stripped stars and OB+BH systems, further highlight the relevance of spectroscopic surveys and monitoring campaigns of OB-type stars in identifying products of binary interactions at various evolutionary stages. Moreover, from detailed binary evolution models, \citet{langer+20a} predicted that there should be of the order of 120 OB+BH binaries currently in the LMC, of which 60 are expected to be B+BH systems. Surveys such as the FLAMES Survey of Massive Stars \citep[FSMS,][]{evans+05, evans+06} and the VLT-FLAMES Tarantula Survey \citep[VFTS,][]{evans+11} might have observed some of these systems serendipitously. Given their expected high semi-amplitude velocities \citep[$K_1 \gtrsim 30$\kms][]{langer+20a}, these should have been detected as SB1 systems and followed-up by the B-type Binaries Characterisation (BBC) programme \citep{villasenor+21} with multi-epoch spectroscopy.

\citet[][henceforth Paper~I]{villasenor+21} presented the orbital properties of 88 B-type binaries from the 30 Doradus (30 Dor) region, obtained from 29 epochs of optical spectroscopy with VLT-FLAMES (the observational strategy and data reduction can be found in \citetalias{villasenor+21}). Using the RV semi-amplitudes of the primaries ($K_1$) and the orbital periods of the system ($K$-$P$ diagram), they showed that several B-type binaries were located in the higher probability regions of having BH companions from the predictions for OB+BH systems \citep{langer+20a}. One of those binaries immediately stands out: VFTS~291 has a significantly higher semi-amplitude velocity ($K_1$) than the rest of the binaries with similar orbital periods (their fig. 20), a region also occupied by the newly discovered stripped stars LB-1 and HR 6819 (with values from \citet{shenar+20} and \citet{bodensteiner+20b}, respectively).

VFTS~291 was previously classified as a SB1 with a B5 II-Ib primary \citep{evans+15}. A complementary atmospheric analysis by \citet{mcevoy+15} found an effective temperature of $T_{\rm eff}=13\,500\,{\rm K}$, a surface gravity of $\log g = 2.35\, {\rm cm}\,{\rm s}^{-2}$, a spectroscopic mass of $M_{\rm sp}=6$ \Msun, an evolutionary mass of $M_{\rm ev}=14$ \Msun, and a luminosity of $\log (L_B/\text{\Lsun}) = 4.3$. \citet{mcevoy+15} suggested that the objects in their sample that are well beyond the terminal age MS could possibly be single core-He-burning (CHeB) stars, with the exception of VFTS~291, given its binary status \citep{dunstall+15}. The high mass function found by \citet{villasenor+21} suggested that VFTS~291 could be an ideal candidate for hosting a BH companion, as a primary mass of 6\Msun would require a minimum secondary mass of 13.7\Msun. However, no clear signs of such a bright companion were detected in the spectrum of VFTS~291.

In this paper we present a detailed analysis of the available observational data for VFTS~291 that lead us to conclude that it is actually composed of a bloated stripped star with an OB companion. In Sec.~\ref{sec:meth} we describe the spectral disentangling procedure and further analysis of the individual spectra. Section~\ref{sec:phot} presents the analysis of the available photometric data, which includes the fit of the spectral energy distribution and light curve analysis. The evolution of the system up to its current state is discussed in Sect.~\ref{sec:evol} by means of detailed binary evolutionary models. In Sect.~\ref{sec:disc}, we discuss the implications and uncertainties of our findings. We finally present our conclusions in Sec.~\ref{sec:concl}.

%%%%%%%%%%%%%%%%%%%%%%%%%%%%%%%%%%%%%%%%%%%%%%
% Sect. 2
\section{Spectral analysis} \label{sec:meth}
%%%%%%%%%%%%%%%%%%%%%%%%%%%%%%%%%%%%%%%%%%%%%%

In this section we present our spectral disentangling procedure that we applied on the 29 BBC epochs, and the subsequent spectral fitting to the disentangled spectra.

\subsection{Previous spectral classifications and peculiarities}

\begin{figure}
\centering
    \includegraphics[width=\columnwidth]{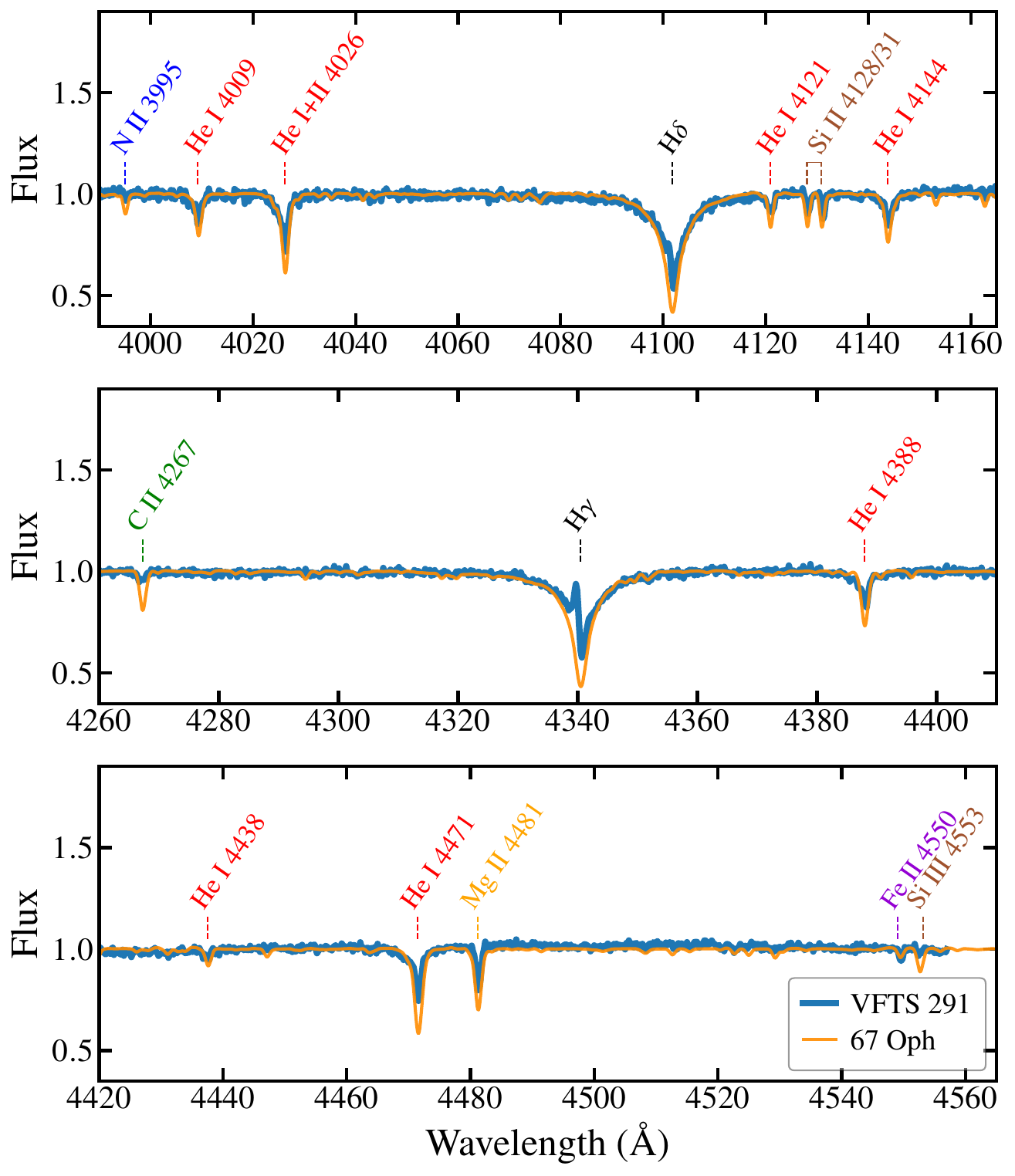}
    \caption{Comparison of VFTS~291 to the B5~Ib star 67 Oph. The spectrum of 67 Oph has been downgraded to FLAMES resolution. Data for 67 Oph is available at ESO archives, programme ID194.C-0833 (P.I. Nick Cox).}\label{fig:67oph}
\end{figure}

Spectral classifications were given to the VFTS sample of B-type stars by \citet{evans+15}, where VFTS~291 was classified as a B5~II-Ib star. We compare one of the original BBC spectra to 67~Oph in Fig.~\ref{fig:67oph}. The star 67~Oph is a well known B5 Ib star \citep[][also classified as B5 II by \citealt{lennon+92}]{walborn+bohlin96, reed03}, with an effective temperature (\Teff) of 15\,000 K, surface gravity (\logg) of 2.75\cms and projected rotational velocity (\vsini) of 44\kms.  It is possible to see the similarities in Fig.~\ref{fig:67oph}, specially in the wings of the Balmer lines---indicative of surface gravity---and in the ratio \spline{He}{i}{4471}/\spline{Mg}{ii}{4481}, used in spectral-type determination. There are some differences, particularly in the metal lines, that might be attributable to the lower metallicity of the LMC, or to a possible slightly higher temperature  of 67~Oph, as suggested by the weaker \spline{Si}{iii}{4553} line presented by VFTS~291.

The spectrum of VFTS~291 has been also studied before by \citet[][star 8 in their identification chart]{walborn+blades97} in their spectral classification study of 30 Dor OB stars. It was classified as B5 :p (peculiar), and surprisingly, the authors noted no \spline{Si}{ii}{4128--30} visible in the spectrum and \spline{He}{i}{4471}$\,<\,$\spline{Mg}{ii}{4481}. While the former difference might be due to the lower S/N ($\sim30$) of their spectra, the ratio \spline{He}{i}{4471}/\spline{Mg}{ii}{4481}$\,<1$ is opposite to what we see in the BBC spectra. Variability in the ratio \spline{He}{i}{4471}/\spline{Mg}{ii}{4481} has been observed in Be stars \citep{vogt+90} on timescales of a few days up to years, possibly related to non-radial pulsations, but VFTS~291 does not present the H$\alpha$ emission characteristic of Be stars.

\begin{figure}
\centering
    \includegraphics[width=\columnwidth]{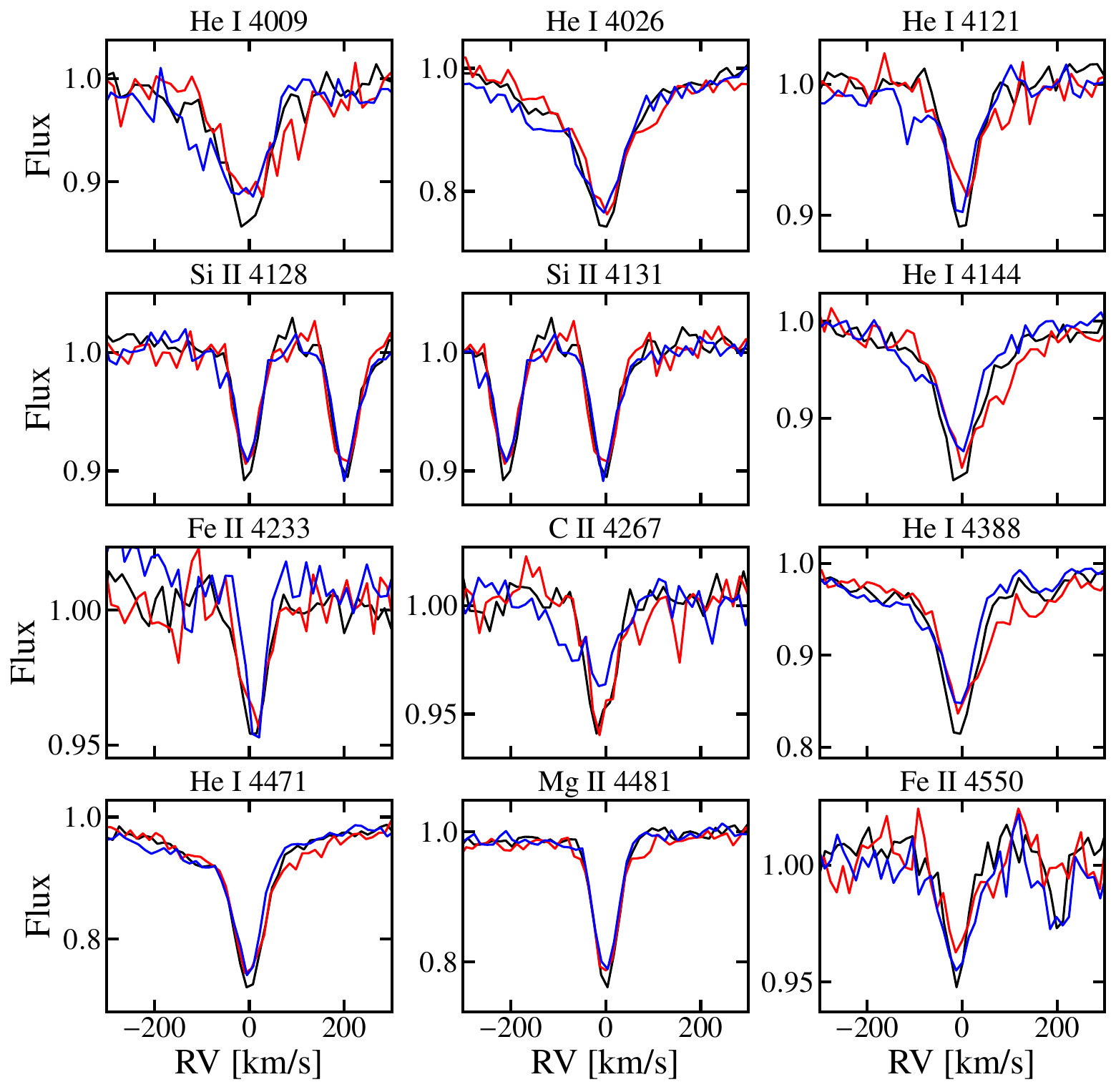}
    \caption{Co-added spectra for epochs close to the maximum RV semi-amplitude (blue), close to the systemic velocity (black), and close to the minimum RV semi-amplitude (red). Three spectra were co-added in each case. The co-added spectra have been shifted by the mean RV of the individual epochs, so that a potential companion could be identified in the red and blue spectra. Line-profile variability is observed predominantly in the wings of the \ion{He}{i} lines, but not in all metal lines (see \ion{Si}{ii} and \ion{Mg}{ii}).}\label{fig:SLshifts}
\end{figure}

It was noted in \citetalias{villasenor+21} (individual notes) that VFTS~291 presented line-profile variability in its \ion{He}{i} lines and that this could be a sign of a weak secondary hiding in the spectrum. To further consider this possibility, we have grouped VFTS~291 spectra into those with RVs close to the highest value, close to the systemic velocity, and close to the lowest RV value, by co-adding them and discarding epochs with intermediate values. The result is shown in Fig.~\ref{fig:SLshifts}, where we can see blue and red shifts, especially in the wings of the \ion{He}{i} lines, associated to our highest and lowest RV measurements, respectively. These apparent shifts are present to a different degree through the different spectral lines, but are less evident in the \ion{Si}{ii} and \ion{Mg}{ii} lines. To really understand the source of this variability it is necessary to disentangle the possible contribution from a companion. 

\subsection{Spectral disentangling}\label{ssec:disent}

Spectral disentangling is a technique used to separate the spectra of the components of a spectroscopic binary. Here we have followed the shift-and-add technique \citep{marchenko+98, gonzalez+levato06} as implemented in \citet[][see also \citealt{shenar+22a}]{shenar+20}. In general terms, a grid of RV semi-amplitudes ($K_1$ and $K_2$) can be used to iteratively compute representative spectra  for both components. With the orbital parameters and RV semi-amplitudes known, the RVs of the components at each epoch can be computed. The computed spectra at any given iteration are shifted and then subtracted from the observations, where a flat spectrum is assumed for the secondary component for the first iteration (i.e., the first approximation for the spectrum of the primary is the co-added spectrum).  At each step the resulting spectra are co-added to increase the S/N of the computed spectra, forming the spectra of the next iteration.  An additional static component is introduced to mimic the nebular emission \citep{abdul-masih+19, shenar+22a}. However, since the secondary component is close to static (see below), the nebular contamination cannot be fully removed. Typically, convergence is attained within a few tens of iterations (generally depending on the line profile and RV amplitudes, see \citealt{shenar+22b}).

By minimising the chi-square statistic ($\chi^2$) between computed and observed spectra, the RV semi-amplitudes $K_1, K_2$ can be constrained. The disentangled spectra are obtained for the set of RV semi-amplitudes that minimise the $\chi^2$. To avoid spurious "emission wings" in the line profiles of the disentangled spectrum of the secondary and speed up convergence, we enforce the disentangled spectrum to lie below the continuum \citep{marchenko+98, Shenar+2019, Quintero+2020}. For this reason, the disentangled spectrum can contain "flat" continuum regions. The light ratio of the two stars cannot be derived from the method without additional assumptions, and only affects the overall scaling of the intensities of the spectral lines.

\begin{figure*}
\centering
    \includegraphics[width=\textwidth]{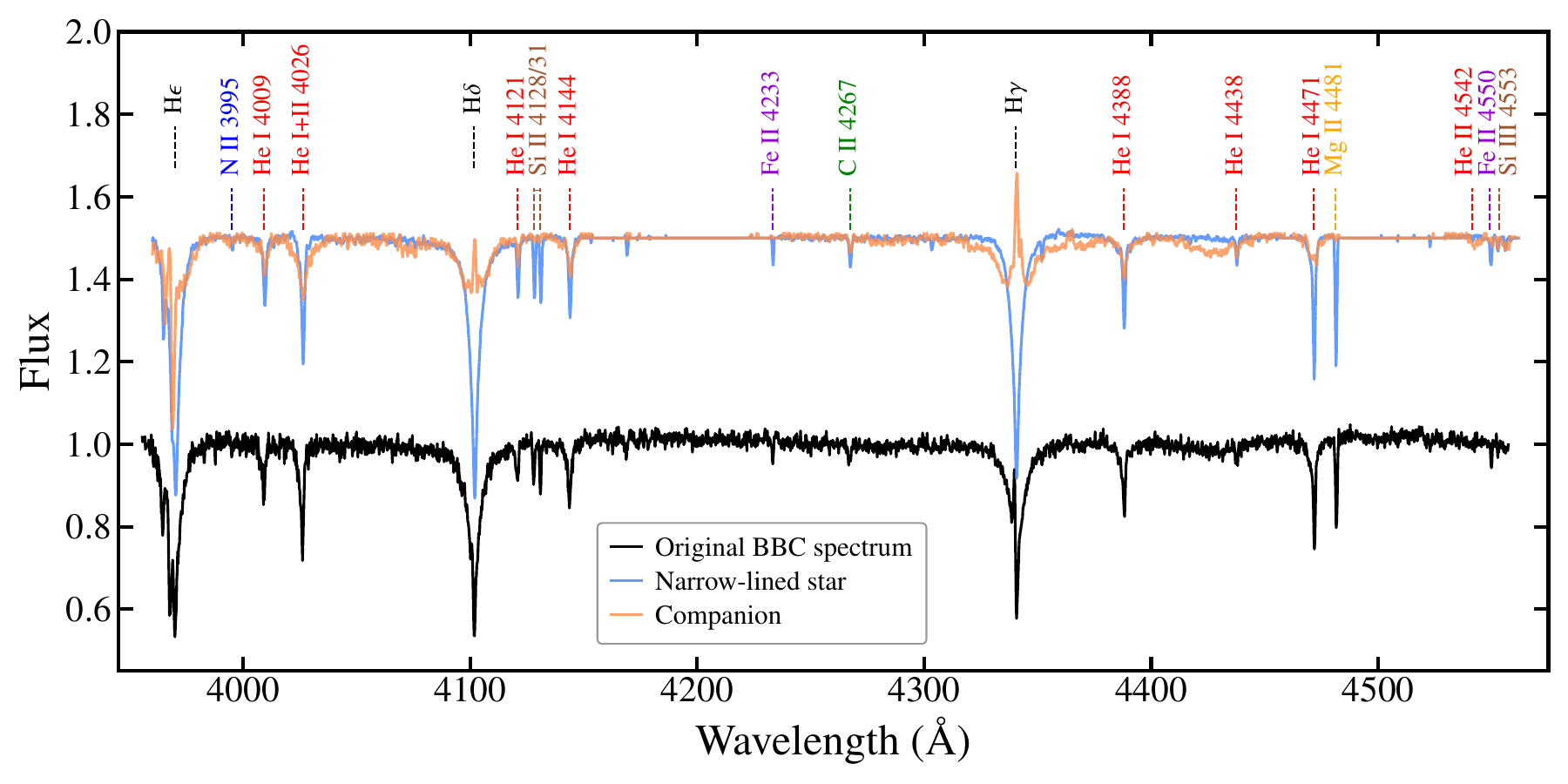}
    \caption{Disentangled spectra re-scaled to a 0.38 light ratio (or 38\% contribution from the secondary/companion star), shifted by 0.5 flux units. One of the BBC spectra is shown in black for comparison. Flat regions, e.g. around 4200\AA, are due to the enforced continuum limit (see Sect.~\ref{ssec:disent}). For the derivation of the light ratio, see Sect.~\ref{ssec:spfit} .}\label{fig:disentang}
\end{figure*}

The disentangled spectra are shown in Fig.~\ref{fig:disentang}. It confirms the presence of a second luminous star and that the strong metal lines (\ion{Si}{ii} and \ion{Mg}{ii}) belong to the dominant star in the spectrum, which we refer to as the narrow-lined star. The spectral type previously assigned by \citet[B5 II-Ib]{evans+15} seems consistent with the spectrum of the narrow-lined star. In the case of the companion, there are few spectral lines that could be used to determine the spectral type. We note a weak feature at 4542\AA\xspace that could be a \ion{He}{ii} line, which would be a signature of an early-B star. Given the intermediate resolution of the data, the profile of the Balmer lines are strongly affected by the blending between the two components. There is also considerable nebular emission affecting the Balmer lines as can be seen in the disentangled spectrum of the companion (Fig.~\ref{fig:disentang}). Nebular emission can lead to spurious features in the disentangled spectrum, and so the profile of the Balmer lines should be taken with caution. It is also worth noting that the Balmer lines present a significantly broad profile whereas some of the \ion{He}{i} lines are significantly narrower, e.g. \spline{He}{i}{4121} and \spline{C}{ii}{4267}, but also the cores of \spline{He}{i}{$\lambda$4144, 4388}.

\begin{figure}
    \centering
    \includegraphics[width=\columnwidth]{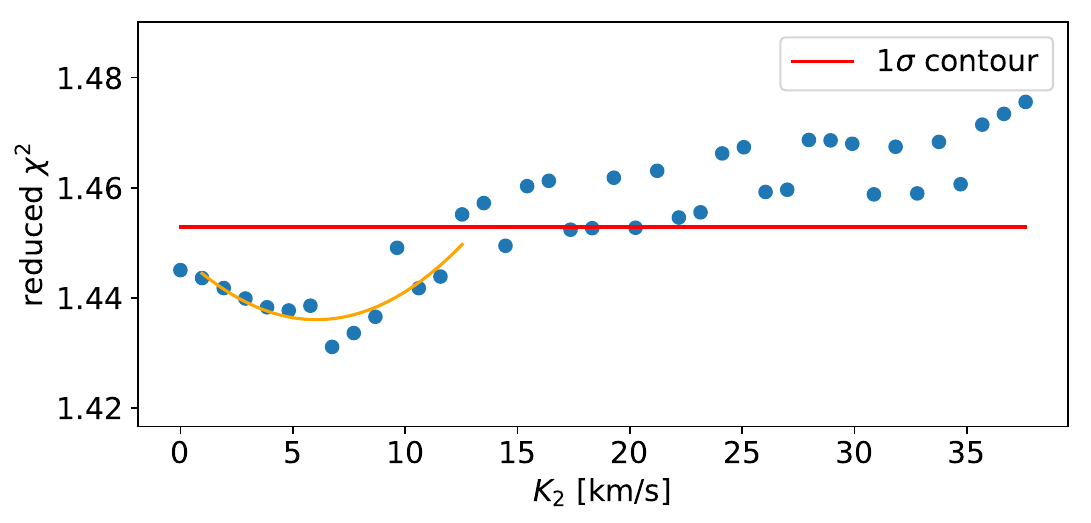}
    \caption{Reduced $\chi^2$ of the disentangling as a function of the RV semi-amplitudes of the companion ($K_2$) for the \spline{He}{i}{4388} line. The red horizontal line indicates 1-$\sigma$ uncertainties.}\label{fig:chi2-K2}
\end{figure}

Figure~\ref{fig:chi2-K2} shows the results of our $\chi^2$ minimisation for the RV semi-amplitude of the companion from the \spline{He}{i}{4388} line. The minimum at 6\kms suggests a low semi-amplitude velocity for the companion, although other lines show that $K_2$ is not well constrained (see Appendix~\ref{ap:dist_chi2}). If the companion star is moving with such a low velocity, it must be a much more massive object than the narrow-lined star, but based only on this evidence we cannot rule out that the initially unseen star is a stationary star, e.g. a third star in a wider orbit or a spurious line-of-sight alignment. The latter possibilities would then require a further massive object to account for the movement of the narrow-lined star. In the case of the primary, we obtained values ranging from 92 to 99\kms, with a weighted mean of  $K_1 = 95 \pm 4$\kms obtained from the He\,{\sc i} lines (avoiding  \spline{He}{i}{4471} due to the strong nebular contamination). If we take $K_2\leq15$\kms as a conservative guess---based on the results from the disentangling---, that leads to a minimum mass ratio of $q_{\rm min}=K_1/K_2\approx6$.

\begin{table}
\caption{VFTS 291 orbital solution computed exclusively from metal line RVs. Eccentricity and argument of the periastron have been kept fixed due to the low eccentricity of the system. Also the semi-amplitude velocities obtained from the disentangling for both components are shown. In the case of the companion ($K_2$), we can only set a rough upper limit based on the fit to different \ion{He}{i} lines.}
\label{tab:orbsol}
\center
\begin{tabular}{lr}
\toprule
\toprule
Parameter			&Value\\
\midrule
$P_{\rm orb}$ (d)		&108.062 $\pm$ 0.038\\
$T_p$ (HJD)		&2457325.42 $\pm$ 0.09\\
$e$			&0\\
$\omega$ (deg)		&90\\
$\gamma$ (km/s)	&270.85 $\pm$ 0.19\\
$K_1$ (km/s)		&93.651 $\pm$ 0.22\\
$a_1\sin i$ (\Rsun)	&200.029 $\pm$ 0.529\\
$f(m_1,m_2)$ ($M_\odot$)	&9.197 $\pm$ 0.067\\
\midrule
\mc{2}{c}{Disentangling}\\
\midrule
$K_1$ (km/s)		&95 $\pm$ 4\\
$K_2$ (km/s)		&$\lesssim$15\\
\bottomrule
\end{tabular}
\tablefoot{$T_p$ is the time of periastron passage, and the argument of periastron has been defined as $\omega=90^\circ$ to match $T_p$ with inferior conjunction.}
\end{table}

\begin{figure}
\centering
    \includegraphics[width=\columnwidth]{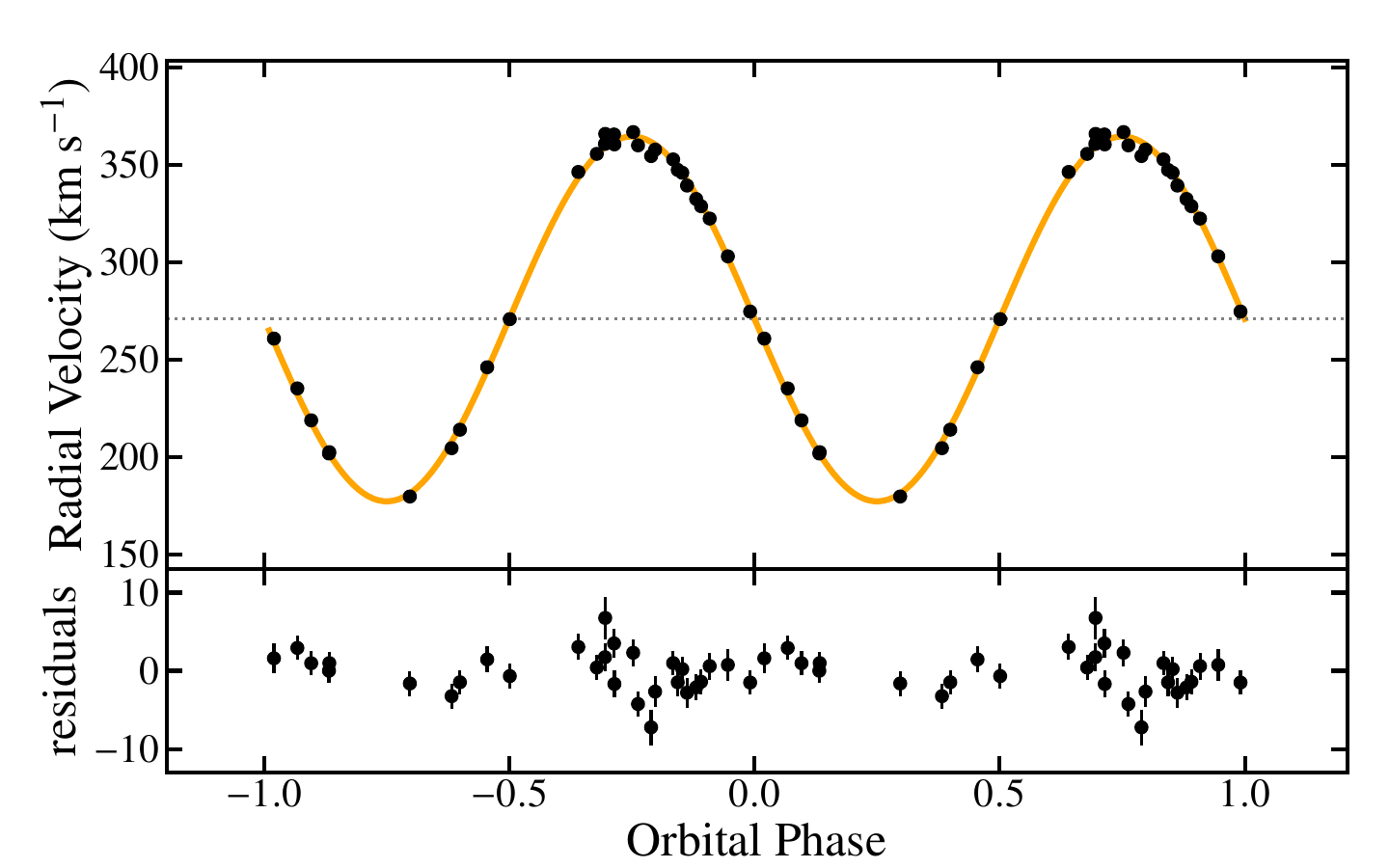}
    \caption{Radial velocity (RV) curve for VFTS~291 computed solely from the metal lines \spline{Si}{ii}{$\lambda$4128, 4131} and \spline{Mg}{ii}{4481}, orange curve shows our orbital solution. The systemic velocity ($\gamma$) is marked with a dotted line. Residuals from the RVs and the orbital solution are shown in the bottom panel.}\label{fig:rvcurve}
\end{figure}

The $K_1$ value obtained from the disentangling procedure ($K_1=95\pm4$\kms) is significantly higher than the one from the orbital solution published in \citetalias{villasenor+21} ($K_1=84.2\pm0.5$\kms) and we investigate this further below. The new disentangled spectra in Fig.~\ref{fig:disentang} have confirmed that the metal lines (\ion{Si}{ii} and \ion{Mg}{ii}) almost exclusively belong to the narrow-lined star, so we have remeasured the RVs and updated the orbital solution based only on the \ion{Si}{ii} and \ion{Mg}{ii} lines (as opposed to \citetalias{villasenor+21} where we used the Balmer and \ion{He}{i} lines).  To do so, we have used the same tools as in \citetalias{villasenor+21} to measure RVs\footnote{These python tools are now part of the Massive bINaries Analysis TOols (MINATO) package, available at \url{https://github.com/jvillasr/MINATO}}, based on Gaussian fitting of the individuals spectral lines. The final RV at each epoch is computed by taking the weighted mean of the RVs obtained for each line. The new solution is available in Table~\ref{tab:orbsol} and the RV curve is shown in Fig.~\ref{fig:rvcurve}. We found a new RV semi-amplitude value of $K_1=93.7\pm0.2$, in excellent agreement with results from the disentangling, and increased by almost 10\kms in comparison to our previous orbital solution.  This is to be expected considering that in this new solution the contribution from the RVs of the companion are minimal. We note that the solution obtained via orbital fitting is more accurate than that obtained with disentangling due to the additional freedom in the spectral shape of the primary in the disentangling procedure. The systemic velocity increased from 262 to 271\kms (which could have implications for cluster membership, see Sect.~\ref{ssec:membership}), the orbital period, however, remains in close agreement with our previous determination. 

\subsection{Spectral fitting}\label{ssec:spfit}

To determine the physical properties of the two stars, we have followed a spectral fitting procedure that minimises the \chisq for both spectra simultaneously. We have selected all the main spectral features present in the disentangled spectra, and performed local fits (i.e., in the region of each spectral line) simultaneously to all spectral lines. We constructed a grid of values for $T_{{\rm eff}, A}$, $\log g_A$, $\varv\sin i_{A}$, $T_{{\rm eff}, B}$, $\log g_B$ and $\varv\sin i_{B}$ where $A$ corresponds to the narrow-lined star and $B$ to the companion. Due to the low temperature found by \citet[][13.5\,kK]{mcevoy+15}, and the observed \spline{He}{i}{4471}/\spline{Mg}{ii}{4481} ratio in the disentangled spectrum, we have included low-temperature synthetic spectra calculated from the grid of LTE ATLAS9 \citep[local thermodynamics equilibrium,][]{kurucz05} models published by \citet[][private communication]{howarth11}. The synthetic spectra have been convolved with an instrumental profile to match FLAMES resolution. Temperature and surface gravity of the narrow-lined star have been covered in the grid by values going from 11000 to 15000\,K in steps of 1000\,K for \Teff, and from 1.8 to 2.8\,dex in steps of 0.2\,dex for \logg. We have extended the temperature range for star $A$ to 18000\,K using TLUSTY models from the BSTAR2006 grid \citep{lanz+hubeny07}, although these have a different step in \logg (0.25\,dex). For component $B$ (or in general for temperatures above 15\,kK), we have used the TLUSTY BSTAR2006 grid for \Teff$\le30\,000$\,K and the OSTAR2002 grid \citep{lanz+hubeny03} for higher temperatures, with values going from 16000 to 37500\,K in steps of 2000\,K if \Teff$\le30\,000$\,K and 2500\,K otherwise, and from 2.0 to 4.5\,dex for \logg in steps of 0.25\,dex. 

For the ATLAS9 models, the LMC abundances can be found in \citet[][see their Table A1]{howarth11}. The TLUSTY models have been computed with a solar helium abundance (${\rm He/H}=0.1$, by number), while the other abundances are scaled from the solar value \citep{lanz+hubeny07}.

All B-type models (ATLAS9 and TLUSTY) were computed with a microturbulent velocity ($\varv_t$) of 2\kms, whereas the hotter O-type models assumed $\varv_t=10$\kms.

Rotational velocities for both stars were also included in the grid, by convolution of the model spectra by a rotational-broadening function. For the narrow-lined star we have covered a range between 0 and 70\kms, spaced by 10\kms, whereas a broader range was used for the companion, going from 0 to 600\kms in steps of 50\kms.

The light ratio of the disentangled spectrum is a completely free parameter, and we have included it in the grid as the contribution from the companion star, with values from 15 to 50\% in steps of 5\%. Given the possibility of an enhanced fraction of He in the surface of the narrow-lined star in the case of mass transfer towards the companion star, we have also used the He/H ratio of the narrow-lined star as part of our grid. The LMC \textsc{ATLAS9} models were computed with a helium abundance by number of 10.88 \citep{howarth11}, which translates into a ratio by number of ${\rm He/H}=0.076$ for a hydrogen abundance of 12.0 dex. Therefore, we have artificially modified the He/H fraction in the synthetic spectra by re-scaling the regions of the spectrum containing these lines. For the re-scaling, we have used the relations:

\begin{equation}
    F_1=(F_0 - 1) \times (f_1/f_0) + 1
\end{equation}
\begin{equation}
    F_1=(F_0 - 1) \times \frac{1-f_1}{1-f_0} + 1
\end{equation}
for helium and hydrogen respectively, where $F_0$ is the original flux, $f_0=0.076$ is the He/H ratio of the \textsc{ATLAS9} models (or 0.1 for TLUSTY models), and $f_1$ is the He/H ratio provided by our grid. We used a range of values between 0.04 and 0.15, in steps of 0.01, with values $<f_0$ essentially as an evaluation of our procedure since these are not expected in an envelope-stripped star. It is worth noting that the scaling of the \ion{He}{i} and Balmer lines only provides a test of helium enrichment. For a more detailed analysis, the computation of new synthetic spectra is needed.

\begin{figure}
\centering
    \includegraphics[width=\columnwidth]{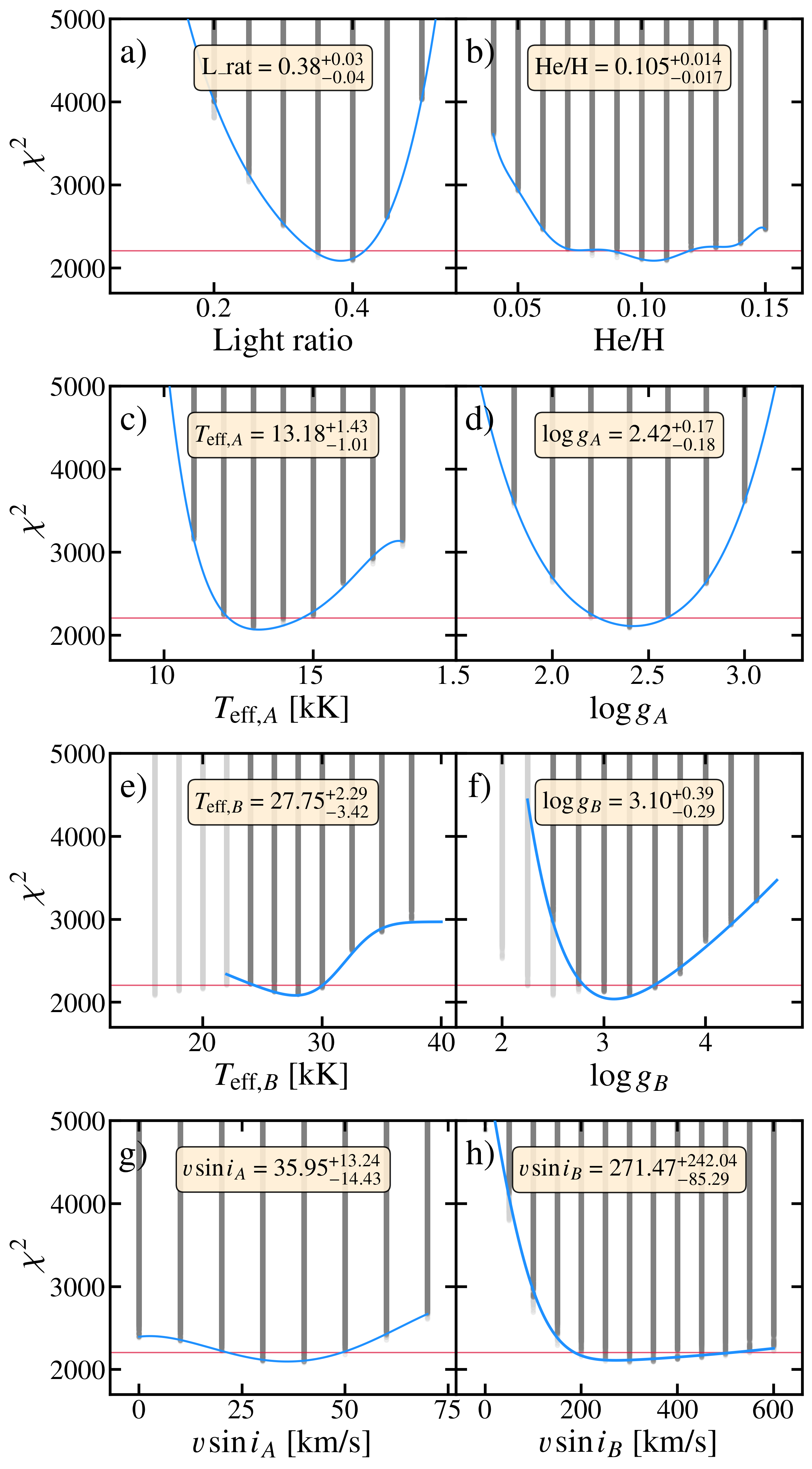}
    \caption{Results from our \chisq analysis. The eight panels, labelled from a) to h) show our \chisq determinations (grey dots) for each value of each parameter in our grid. The minimum \chisq values for each parameter have been fitted with a spline interpolation, whose minimum gives us the best-fit value of the parameters in our grid, and the intercept of the curve with the 95\% confidence level (horizontal red line) provides the error in our determined values. Light grey dots correspond to \chisq values from $T_{{\rm eff}, B}\le24000$\,K, see text for details. Best-fit values and corresponding errors are shown in the yellow boxes. \chisq values have been normalised to the degrees of freedom. We note that the He/H ratio has only been applied to the narrow-lined star.}\label{fig:chi2_er}
\end{figure}

Our grid is then composed of 8 parameters and $>5\times10^7$ models. The results from our \chisq analysis for these models are shown in Fig.~\ref{fig:chi2_er}, through panels a) to h), each for one of the parameters in the grid. The grey dots are the \chisq values for each point in the grid, the blue curves are spline fits to the minimum \chisq values, and the horizontal red lines mark the 95\% confidence level, which is used to determine the error for the best values at the intercept with the blue curve. The best-fitting value for each parameter is given by the minimum of the spline curve. In the case of component $B$, we have found two minima in its temperature and surface gravity values. We believe this is related to the lack of features in the spectrum that constrain temperature, therefore, our fit relies mainly on the \ion{He}{i} lines, which have their peak in strength at spectral type B2 \citep[20\,kK, e.g.][]{didelon82} and can be equally fitted by models at both sides of the peak temperature value. The lack of \ion{Si}{ii} and \ion{Mg}{ii} in the spectrum of component $B$ are a strong constraint, however, these are narrow (and potentially faint) lines that can be strongly diluted by stellar rotation and will not contribute much to the \chisq value in comparison to \ion{He}{i} and Balmer lines. The low \chisq values at $T_{{\rm eff}, B}=16000$\,K could therefore be favoured by the right combination of parameters as long as the Balmer lines are well fitted. Based on the above-discussed absent features in the spectrum of component $B$, we have discarded the low temperatures from our fit, and all \chisq values coming from $T_{{\rm eff}, B}<24000$\,K are treated as spurious results and are showed with light grey dots in Fig.~\ref{fig:chi2_er}.

\begin{figure*}
\centering
    \includegraphics[width=0.9\textwidth]{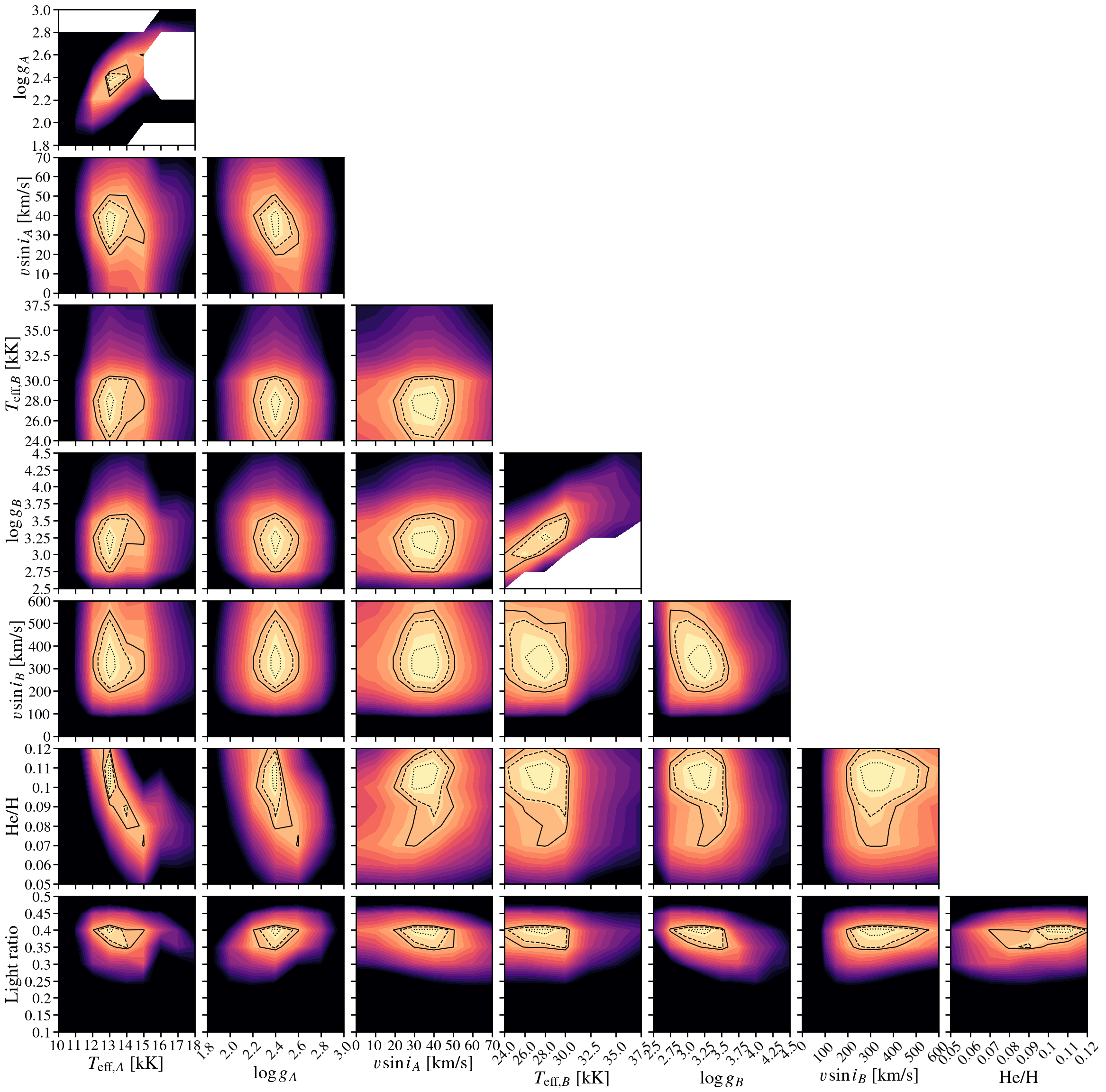}
    \caption{2D \chisq contour map resulting from our spectral fitting procedure. Black contours show 1-$\sigma$ (dotted), 2-$\sigma$ (dashed), and 3-$\sigma$ (solid) levels.}\label{fig:chi2_map}
\end{figure*}

\begin{figure*}
\centering
    \includegraphics[width=0.9\textwidth]{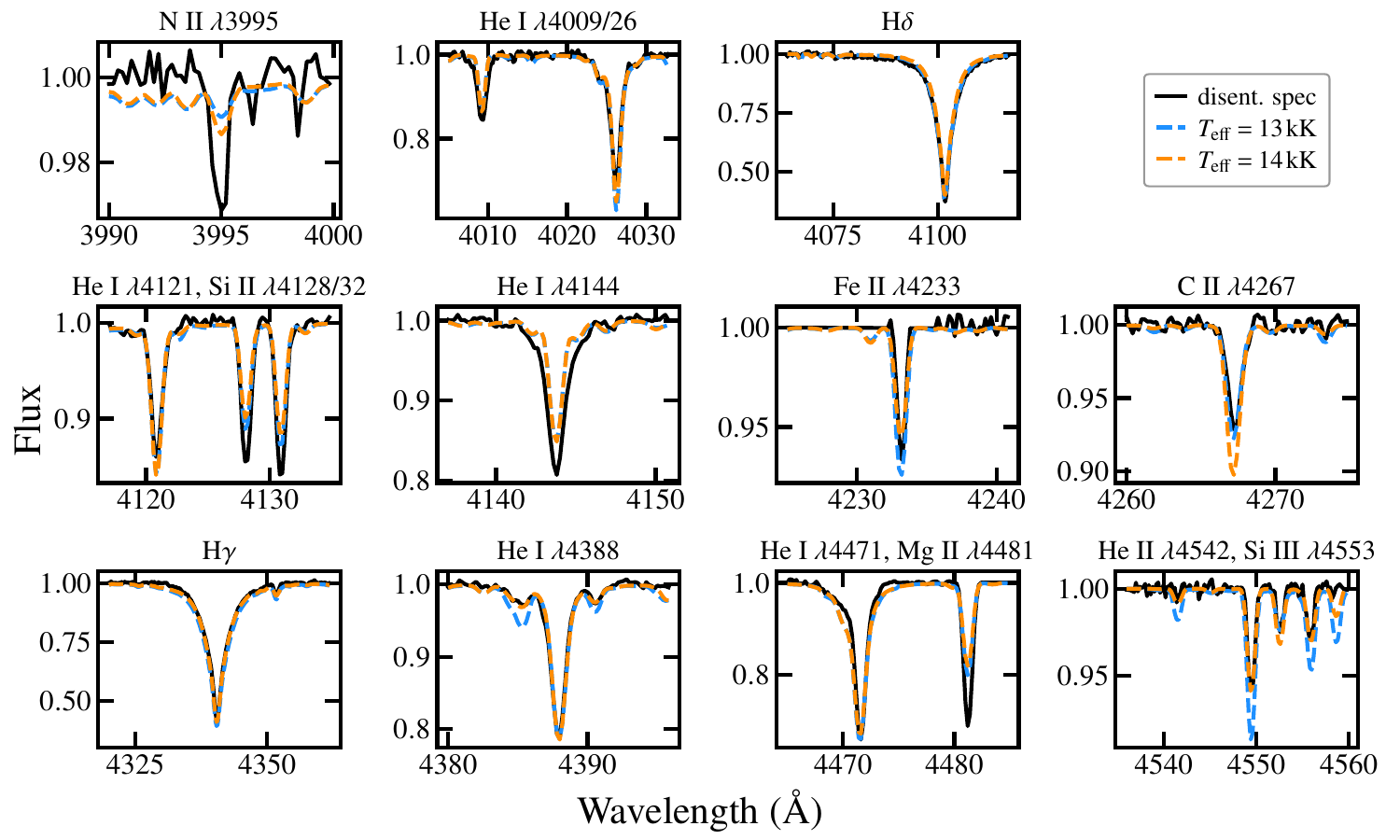}
    \caption{Disentangled spectrum of the narrow-lined star (black) with the two closest models to our best-fit values. A model with \Teff$=13000$\,K, \logg$=2.4$\,dex, \vsini$=40$\kms and He/H$=0.11$ is shown in blue, and \Teff$=13000$\,K, \logg$=2.4$\cms, \vsini$=40$\kms, and He/H$=0.09$ is shown in orange. The disentangled spectrum has been re-scaled to a light ratio of 0.38. All spectral lines shown in the figure have been included in the fit, using the shown wavelength ranges.} 
    \label{fig:spfitA}
\end{figure*}

Panel a) shows the \chisq statistic for the light ratio of the system, for which we found a best value of $0.38^{+0.03}_{-0.04}$, meaning that the companion star contributes 38\% to the total the flux of the system. We have obtained an helium-to-hydrogen ratio of ${\rm He/H}=0.105^{+0.014}_{-0.017}$, as shown in panel b), this is an increment of only 3\% from the initial ratio considered in the models, which might be indicative of an episode of mass transfer, but too low to confirm it.

In panels c) and d) we present the temperature and surface gravity for the narrow-lined star, and our best-fit values are $T_{{\rm eff}, A}=13.2^{+1.4}_{-1.0}$\,kK and $\log g_A=2.4^{+0.2}_{-0.2}$. These are in good agreement (within errors) with the values from \citet{mcevoy+15}, supporting a scenario where the contribution from the companion to the composite spectrum is relatively low and/or rapid stellar rotation is diluting its spectral lines. 

Temperature and surface gravity of component $B$ are displayed in panels e) and f); in contrast to component $A$, these are less well constrained, firstly due to the second minimum in the \chisq values discussed previously, but also the bounds of the values are a factor 2--3 larger than for star $A$. In the case of the temperature, we found a best-fit value of $T_{{\rm eff}, B}=27.8^{+2.3}_{-3.4}$\,kK, whereas for the surface gravity we found $\log g_B=3.1^{+0.4}_{-0.3}$. The temperature values correspond to an early B-type star, which is consistent with the lack/weakness of \ion{Mg}{ii} and \ion{Si}{ii} lines, but also with the absence of \ion{He}{ii}, which are characteristic of O-type stars. However, the surface gravity is lower than expected; it corresponds to an evolved star for which we could expect to see \ion{Si}{iv} lines. The most likely reason for this discrepancy is the effect of the low velocity of the companion and the nebular emission on the resulting spectrum from the disentangling process, as discussed in Sect~\ref{ssec:disent}. 

The last two panels, g) and h), correspond to the rotational velocities of components $A$ and $B$ respectively. For the narrow-lined star, as expected, we found a low projected rotational velocity of $\varv\sin i_{A}=36^{+13}_{-14}$\kms. Despite the large errors, the range is consistent with the narrow profile of the spectral lines, and given Medusa LR02 spectral resolution ($\Delta\lambda=0.71$\AA\xspace from February 2015) which corresponds to 49.8\kms at central wavelength (4272\,\AA), we can only determine an upper limit for \vsini. The case of the companion is more complex; we obtained $\varv\sin i_{B}=271^{+242}_{-85}$\kms, i.e. a lower limit just below 200\kms and a large upper limit in excess of 500\kms. However, from visual inspection (see discussion below), it is hard to reconcile these large rotational velocities with the \ion{He}{i} lines and the \spline{C}{ii}{4267} line. Similarly as with the discrepancy in \logg, the \chisq value is dominated by the Balmer lines, which, particularly for the companion, have a more uncertain profile due to the discussed effects on the disentangling, and are better fitted by large rotational velocities in most cases. 

The 2D \chisq contour maps are shown in Fig.~\ref{fig:chi2_map}. We did not find strong correlations between parameters, except in the case of \Teff and \logg as expected, since \ion{He}{i} lines are sensitive to both parameters. No degeneracies are observed either, although it is clear that \vsini of the companion and the He/H ratio are the less constrained parameters, whereas \Teff and \logg of the narrow-lined star and the light ratio are the better constrained ones. There are some models missing in the \chisq map of \Teff and \logg for both components. In the case of star $A$, it is due to the change in the \logg step from the ATLAS9 to the TLUSTY models. Since the TLUSTY models (16-18\,kK) have higher \chisq values, they have been omitted from the contour map, i.e. \logg values of 2.25, 2.5 and 2.75 are not shown in that temperature range. For star $B$, the missing models are not available in the TLUSTY grid due to the approximate location of the Eddington limit. There is also a steep drop in \chisq values for the temperature of the companion between the B and O TLUSTY grid, and a change in step from 2 to 2.5\,kK. 

Figure \ref{fig:spfitA} shows two of our best-fitting models for the narrow-lined star. Spectral lines and wavelength ranges are the ones we have included in the fitting process. Both models have \logg$=2.4$\,dex and \vsini$=40$\kms. In blue we show a 13\,kK model, it can be seen that it is a slightly better fit to some of the metal lines (e.g., \ion{Si}{ii}, \ion{C}{ii}, \ion{Mg}{ii}), but the 14\,kK model (in orange) improves on the \spline{Fe}{ii} {4550/56} lines and the \spline{Mg}{ii} {4385/91} lines next to the wings of \spline{He}{i}{4388}. In general, the spectrum is well fitted but there is a clear underprediction of the \ion{Si}{ii} lines and especially in the \spline{Mg}{ii}{4481} line (also the case of \spline{He}{i}{4144} but this could be related to the disentangling process since it is the exception between the He lines), raising the question of whether these discrepancies could be related to the observations made by \citet{walborn+blades97} concerning these lines, which would suggest short-term variability. We also see nitrogen enrichment and carbon depletion, the latter in the case of the 14\,kK model, which are indicative of the mixing processes from the CNO cycle occurring in deeper layers of the star, a suggestion of mass transfer and envelope stripping. 

\begin{figure*}
\centering
    \includegraphics[width=0.9\textwidth]{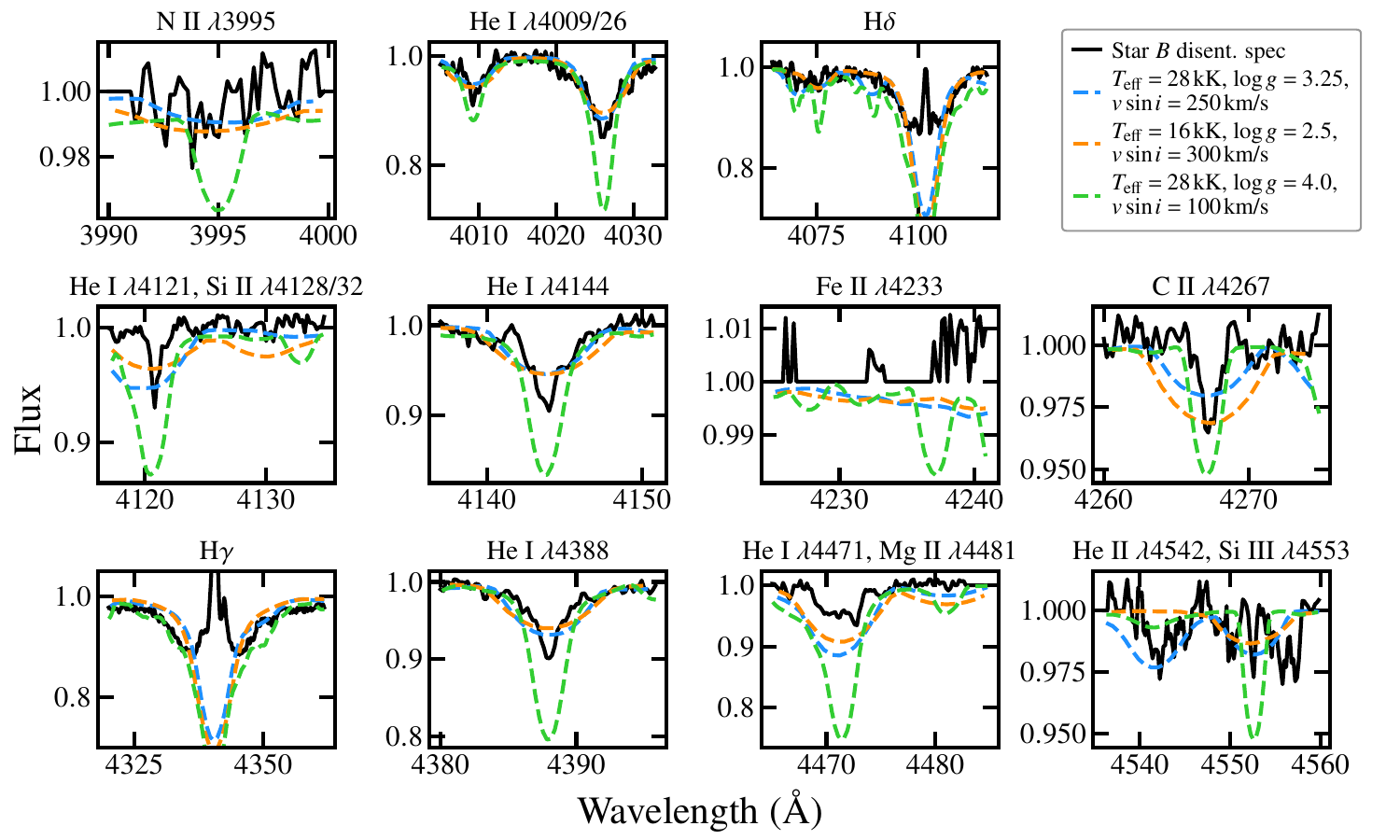}
    \caption{Best-fit atmosphere models to the spectrum of the companion star (black). In colours we show three different models that exemplify the intricacies of fitting this spectrum. The blue model correspond to our best-fit model; in orange we show the low-temperature model with the lowest \chisq value; and in green we show a model that would be closer to the expected model from visual inspection of the disentangled spectrum (see main text for details). The disentangled spectrum has been re-scaled to a light ratio of 0.38. In contrast to the fit for the narrow-lined star, here we have not included the regions between 4225-4241\,\AA\xspace and 4465-4485\,\AA\xspace due to the dubious outcome of the disentangling in these regions.}
    \label{fig:spfitB}
\end{figure*}

Similarly, we show a set of models to compare to the disentangled spectrum of the companion (in black) in Fig~\ref{fig:spfitB}. The first model (in blue), is the closest to our determined values shown in Fig.~\ref{fig:chi2_er}. It is clear that this model presents \ion{He}{i} lines that are too broad compared to the disentangled spectrum, the same can be seen for the \ion{C}{ii} line, which presents an even narrower profile. 

In orange, we show the low-temperature model with the lowest \chisq value. This is one of the models that were not considered further, but it is worth comparing with our high-temperature model. The model has a temperature of \Teff$=16$\,kK, \logg$=2.5$\,dex, and \vsini$=300$\kms. Its \chisq value is comparable to that of the blue model, however there are a few points that need to be noticed and that argue for our temperature range selection. There is a weak feature at the wavelength of the \spline{He}{ii}{4542} line (lower rightmost panel), it is at the level of the noise, but if it is real would be consistent with a spectral type around B0.5 and B0.7 which corresponds to a temperature of roughly 26-28 kK \citep[e.g.][]{garland+17} for MS stars. However, at higher temperatures we expect stronger \ion{He}{ii} lines but also \ion{Si}{iv} lines, which are absent in the spectrum. The latter are even stronger for lower surface gravities which are associated to lower temperatures in our results. The absence/weakness of \spline{Mg}{ii}{4481} and \spline{Si}{ii}{4128/31} also supports this reasoning. 

Finally, we included a model with \Teff$=28$\,kK, \logg$=4.0$\,dex, and \vsini$=100$\kms, shown in green. This model is closer to what we would expect from the visual inspection of the spectrum, i.e. for our constrained temperature (${\sim}25$--30\,kK), the lack/weakness of \ion{Si}{iv} lines strongly suggest a main sequence companion, and narrow \ion{C}{ii} line in addition to the narrow cores of the \ion{He}{i} lines are difficult to reconcile with the large rotational velocities we obtained. However, the \ion{He}{i} lines are too strong in this model for this light ratio. In fact, imposing a much lower flux contribution from the companion (0.15--0.2) improves the fit, but at the expense of a much worse fit of the narrow-lined star. 

The spectral analysis of the companion in VFTS~291 makes clear that the profiles of Balmer and \ion{He}{i} lines are not consistent with only one model. This result could reflect on the difficulties of the disentangling to recover the signal of the companion star due to the nebular contamination and the close-to-static orbital motion of the companion, or, it could be due to the presence of a tertiary star contributing to the flux. Specifically, the main uncertainty in the disentangling outcome is in the Balmer lines profile, with implications for the surface gravity, rotational velocities, and He/H ratio. While these uncertainties affect mainly the weaker companion spectrum, it is possibly that they could expand to the narrow-lined star, which could explain the small fraction of helium enrichment and even the discrepancies in the metal lines (e.g., \ion{Si}{ii} and \ion{Mg}{ii}, see Fig.~\ref{fig:spfitA}). High-resolution spectra would help in this regard by better isolating the nebular emission and by allowing a more accurate constraint on the movement and spectral-line profiles of the companion, thus greatly improving the results from the disentangling.

\subsection{Rotational velocity}

\citet{mcevoy+15} determined rotational velocities and other physical parameters for the VFTS sample of B supergiants. They estimated for VFTS~291 a rotational velocity of $\varv\sin i \leq 20$\kms. We have repeated the analysis on the disentangled spectra to independently test our \vsini values. For this analysis we have used the IDL package \texttt{IACOB-broad} \citep{simon-diaz+herrero14} which uses Fourier transform (FT) in addition to a goodness-of-fit (GoF) procedure to determine stellar rotational velocities from individual spectral lines. For the narrow-lined star, we have used \ion{Si}{ii}~$\lambda4128/31$, \ion{C}{ii}~$\lambda4267$ and \ion{Mg}{ii}~$\lambda4481$. The results obtained with \texttt{IACOB-broad} are shown in Table~\ref{tab:vsini}; for a fixed macroturbulent velocity ($\varv_{\rm mac}=0$), the \vsini results are in good agreement in both FT and GoF methods, with most values between 33 and 41\kms, also in good agreement with our findings from the spectral fitting, the main exception is \spline{C}{ii}{4267} for which the GoF \vsini is considerably higher (59.3\kms). When fixing \vsini to the value found from FT, \texttt{IACOB-broad} determines $\varv_{\rm mac}=12-21$\kms for the \ion{Si}{ii} and \ion{Mg}{ii} lines, and $\varv_{\rm mac}=50.0^{+20.8}_{-19.0}$\kms for \ion{C}{ii}. Finally, when both parameters are considered as free parameters, we found \vsini$=14-22$\kms and \vmac$=25-40$\kms, excluding the results for the weaker \ion{C}{ii} line which presents considerably larger errors. These results suggest that the main source of broadening is macroturbulence, but it should be kept in mind that at the moderate FLAMES resolving power it is very difficult to separate the contribution from rotational and macroturbulent broadening. Furthermore, low values of the microturbulent velocity (\vmic), of the order of $\sim15-25$\kms, can produce a minimum in the Fourier transform analysis in the range 30-45\kms \citep{simon-diaz+herrero14,holgado+22}, so our \vsini values obtained from the spectral fitting might actually be caused by higher values of \vmic than those of our used models and should be taken as upper limits. In the same way, the values obtained for \vmac from the GoF procedure should also be taken as upper limits due to the unaccounted contribution from \vmic.

Determining the rotational velocity of the secondary is even more complex. The only non-diffuse \ion{He}{i} line available that is strong enough to be measured is \ion{He}{i}~$\lambda$4121, but it is very narrow and led to rotational velocities inconsistent with the rest of the spectrum. We have therefore relied on \ion{C}{ii}~$\lambda$4267, the results are at the bottom of Table~\ref{tab:vsini}. By considering a null \vmac, we found values of 74 and 84\kms for \vsini from the FT and GoF analysis respectively. The \vsini and \vmac from GoF present significantly large errors but their central values are also below 100\kms, in accordance with what was shown in Fig.~\ref{fig:spfitB}. However, it is still hard to reconcile this velocity with the wings of the Balmer and \ion{He}{i} lines.

\begin{table}
\caption{Projected rotational velocities (\kms) for both components of VFTS~291 obtained with \texttt{IACOB-broad}. }
\label{tab:vsini}
\center
\resizebox{\columnwidth}{!}{%
\begin{tabular}{lccccc}
\toprule
\toprule
Sp. line & \vsini(GoF) & \vsini(FT) & \vmac(GoF) & \vsini(GoF) & \vmac(GoF) \\
         & \vmac$=0$ & \vmac$=0$ & \vsini$=\,$\vsini(FT) & & \\
\midrule
\mc{6}{c}{Narrow-lined star} \\
\midrule
\spline{Si}{ii}{4128} &  37.9  &  33.7  &  $20.9^{+6.5}_{-6.6}$  &  $22.4^{+16.5}_{-9.9}$  &  $32.8^{+10.0}_{-19.6}$  \\
\spline{Si}{ii}{4131} &	 38.4  &  37.0  &  $20.6^{+8.0}_{-7.6}$  &  $16.0^{+19.2}_{-3.5}$  &  $39.6^{+5.6}_{-18.6}$  \\
\spline{C}{ii}{4267}  &  59.3  &  40.7  &  $50.0^{+20.8}_{-19.0}$ & $12.5^{+48.4}_{-0.0}$  &  $65.7^{+18.3}_{-47.4}$ \\
\spline{Mg}{ii}{4481} &  27.3  &  36.5  &  $12.5^{+0.6}_{-0.0}$  &  $14.2^{+9.3}_{-1.7}$   &  $25.4^{+3.1}_{-7.5}$   \\
\midrule
\mc{6}{c}{Companion} \\
\midrule
\spline{C}{ii}{4267}  &  83.9  &  73.6  &  $42.4^{+54.0}_{-30.0}$ & $40.0^{+67.5}_{-27.6}$ &  $85.5^{+47.3}_{-73.0}$ \\
\bottomrule
\end{tabular}}%
\end{table}

\subsection{Physical parameters}\label{ssec:physparams}

\subsubsection{Extinction}

Using our determination of the light ratio, we can obtain the light contribution from each source, and determine luminosities and spectroscopic masses. However, in addition to the large errors found for the physical parameters---particularly for the companion---there is another strong source of uncertainty: extinction, that can heavily affect the flux and masses of stars. A few studies have investigated the extinction laws in the LMC \citep[e.g.][]{gordon+03,gao+13}, including more focused studies towards 30 Dor \citep{maiz+14, DeMarchi+16,brands+23,fahrion+demarchi23}. The latter works have independently found ratios of total-to-selective extinction of $R_V\approx4.5$, significantly higher than the commonly used LMC average value $R_V=3.41\pm0.06$ \citep{gordon+03}, which can lead to underestimations in the mass of upper-MS stars by a factor of 1.5 on average \citep{DeMarchi+16}. 

VFTS~291 lies in the periphery of the small cluster Hodge~301 (H301), which is about 3\arcmin northwest of the much more massive and younger cluster R136, at the centre of NGC~2070. In a dedicated work on H301, based on $UVI$ data from the Hubble Space Telescope (HST), \citet{grebel+chu00} estimated a mean reddening of $\langle E_{B-V}\rangle=0.28\pm0.05\,$mag and mean extinction of $\langle A_V\rangle=0.96\pm0.13\,$mag for $R_V=3.35$. \citet{cignoni+16} presented an in-depth analysis of the stellar content of H301, framed in the HST Hubble Tarantula Treasury Project (HTTP) survey \citep{sabbi+13}, finding an average reddening of $E(B-V)\approx0.22\text{--}0.24\,$mag, in good agreement with the value from \citeauthor{grebel+chu00}, with a dispersion of 0.004\,mag due to differential reddening. On the other hand, \citet{britavskiy+19} analysed candidates red supergiants (RSGs) observed by the VFTS, and found effectively the same extinction for all H301 members ($A_V=1.14\pm0.25$ for three members and $A_V=1.17\pm0.24$ for one star, the only one at the centre of the cluster) using $R_V=4.5$ \citep[from][]{DeMarchi+16}. Given the small angular size of H301 (20\arcsec radius equivalent to $\sim4.9$\,pc) we do not expect a significant variation in the extinction throughout the periphery of H301, which is consistent with findings from \citet{DeMarchi+16} and \citet{cignoni+16}, so we have adopted the extinction found by \citet{britavskiy+19} for the H301 members. 

\subsubsection{Error propagation}

Given the asymmetric errors obtained in Sect.~\ref{ssec:spfit}, we have estimated the errors of the physical parameters of VFTS~291 by Monte Carlo simulations. We have used the asymmetric errors to construct a skewed normal distribution from which we generated a random sample with $10^5$ values for each of the parameters. For the apparent magnitude ($m_V$), extinction ($A_V$), and distance ($d$), we have generated random normal samples. In the case of the luminosity ($L$), radius ($R$), and mass ($M$), the resulting distributions are lognormal distributions, and therefore are normal in logarithmic space (see subsection below). The only exception is the mass of the companion which shows appreciable skewness in its distribution and did not pass the Anderson–Darling test for normality. In this case the median is given as central tendency indicator and the 16th and 84th percentiles as errors, otherwise, quoted errors correspond to 1-$\sigma$ in logarithmic space in the following.

\subsubsection{Spectroscopic mass}

The VFTS found a $V$ magnitude for VFTS~291 of $V=14.85$ from the Wide Field Imager (WFI) at the 2.2\,m Max-Planck-Gesellschaft (MPG)/ESO telescope at La Silla Observatory. We have considered an uncertainty of 0.1 mag, corresponding to the dispersion of the calibrations performed by \citet{evans+11}. After correcting the magnitude ($m_V$) for the light ratio, we used the precise distance measurement to the LMC ($d=49.59\pm0.54$\,Kpc) from \citet{pietrzynski+19}, and the extinction from \citet[][$A_V=1.14\pm0.25$]{britavskiy+19}, to compute the absolute magnitude of the two stars from $M_V = m_V - 5\log_{10}(d) + 5 - A_V$, obtaining $M_{V, A}=-4.24 \pm 0.28$ and $M_{V, B}=-3.69 \pm 0.29$. We applied the bolometric correction from \citet[][see also \citealt{flowers96}]{torres10} to our absolute magnitude values to find bolometric magnitudes of $M_{{\rm bol}, A}=-5.17 \pm 0.36$ and $M_{{\rm bol}, B}=-6.32 \pm 0.41$. With this, we can directly find the luminosities from the well known relation between luminosity ratios and bolometric magnitude differences, from which it follows that:

\begin{equation}
    L= \frac{L_0}{\text{\Lsun}} 10^{-M_{\rm bol}/2.512}
\end{equation}
where $L_0=3.0128\times10^{28}$\,W and \Lsun$=3.828\times10^{26}$\,W. The luminosities found for the two components are $\log (L_A/\text{\Lsun}) = 3.96 \pm 0.14$ and $\log (L_B/\text{\Lsun}) = 4.41 \pm 0.16$. From the luminosities and temperatures, we used $R= L^{1/2} (4\pi \sigma T_{\rm eff}^4)^{-1/2} $ to compute radii of $\log (R_A/\text{\Rsun}) = 1.25 \pm 0.07$ and $\log (R_B/\text{\Rsun}) = 0.86 \pm 0.07$, which correspond to $17.93^{+2.90}_{-2.49}$\Rsun and $7.27^{+1.27}_{-1.08}$\Rsun respectively. Finally, using these radii and our \logg determinations from the spectral fitting, we found spectroscopic masses of $\log(M_{{\rm sp}, A}/\text{\Msun}) = 0.48\pm0.22$ and $\log(M_{{\rm sp}, B}/\text{\Msun}) = 0.40^{+0.38}_{-0.34}$, equivalent to $3.05^{+1.98}_{-1.20}$\Msun and $2.51^{+3.51}_{-1.36}$\Msun respectively. 

The uncertainties in the spectroscopic mass are quite large, 45\% for component $A$ and close to 90\% for component $B$, this is mainly due to the uncertainties in our determined values for \logg from the spectral-fitting process. Despite the error bars, there are a few interesting things to comment about. In Sect.~\ref{ssec:disent}, we inferred a minimum mass ratio $q_{\rm min}\approx6$ from the well constrained RV semi-amplitude of the narrow-lined star (component $A$), $K_1$, and from a conservative upper limit for the poorly constrained $K_2$ of component $B$. This mass ratio is at least a factor 2 higher than what we obtain from our spectroscopic masses when considering the lowest $M_{{\rm sp}, A}$ and the largest $M_{{\rm sp}, B}$. The major source of uncertainty here comes from our value of \logg for the companion; our spectroscopic data are not sufficient to tightly constrain the mass of the companion, and it could be even outside of our determined range. If the narrow-lined star is 3\Msun, a companion of about 18\Msun would be required to explain our dynamically estimated mass ratio, and given the radius of 7.3\Rsun, this would imply a \logg close to 4\,dex. However, this is at the mass were we would expect to see stronger \ion{He}{ii} lines. If the mass of the narrow-lined star is at the lower end (1.85\Msun), we would expect a companion mass close to 11\Msun, for which a surface gravity of \logg$\sim3.75$\,dex is needed. Given the spectral classification of VFTS~291, and if this were a single star, one would expect a mass closer to $\sim15$\Msun \citep[e.g.][]{markova+puls}. However, the mass of the narrow-lined star is fairly low, definitely lower from what one would expect from a regular mid-B supergiant. One possible explanation is that VFTS~291 could have gone through an episode of mass transfer, where the initially more massive star transferred a substantial fraction of its mass to its companion, as has already been proposed for systems like HR~6819 and LB-1. We will explore this scenario through detailed binary evolutionary models in Sect.~\ref{sec:evol}. 

\subsubsection{Evolutionary mass}

A robust check on our spectroscopic masses are the evolutionary masses. Given that \Teff and $L$ of the companion are relatively well constrained, we can use these values with their respective uncertainties to compute the evolutionary mass of the companion. Here we assume that the companion has been rejuvenated by the accreted mass and that it behaves as a single MS star \citep{braun+langer95}. This assumption should hold for most stars, and even at the most extreme case---where the mass ratio is very close to 1 and the central helium mass fraction at the moment of accretion is $Y_{\rm acc}=1$---the error on the mass in comparison to the single star mass-luminosity relation would be of 0.13\,dex at most \citep[see equations 4 and 7 of][]{braun+langer95}. We can therefore use the Bayesian code \textsc{BONNSAI}\footnote{The \textsc{BONNSAI} web-service is available at \url{http://www.astro.uni-bonn.de/stars/bonnsai}.} \citep{schneider+14} to compute the evolutionary mass of the companion, resulting in $M_{{\rm ev}, B}=13.2\pm1.5$\Msun. This value is in much better agreement with the spectroscopic mass of the narrow-lined star and the minimum mass ratio, and would correspond to a MS early B-type star (B0.5-B1~V), in full agreement with our determined temperature \citep[$T_{{\rm eff}, B}=27.8^{+2.3}_{-3.4}$\,kK, e.g.][]{garland+17}.

Using the evolutionary mass of the companion, we can put further constrains on the dynamical mass of the narrow-lined star using the mass function equation \citep{hilditch01}:

\begin{equation}\label{eq:massfunc}
    f(m)=\frac{M_2^3 \sin^3i}{(M_1+M_2)^2} = (1.0361\times 10^{-7})(1-e^2)^{3/2}K_1^3 P_{\rm orb}\,M_{\odot}
\end{equation}
where subscripts 1 and 2 refer to the narrow-lined star and the companion respectively, and $e=0$. Solving numerically, and for $\sin i=1$, this give us an upper limit on the dynamical mass of the narrow-lined star of $M_{{\rm dyn},1} < 2.7 \pm 1.2$.

Another possibility, is to use our adopted minimum mass ratio. If we assume an uncertainty of 2\kms in the already conservative upper limit for $K_2$ from the disentangling, i.e. $K_2\leq15\pm2$\kms, then the minimum mass ratio becomes $q_{\rm min}=6\pm1$, leading to $M_1<2.2\pm0.4$\Msun. This upper limit is in agreement with the one derived from the mass function equation, and shows that the narrow-lined star is a low mass star of around 2\Msun, too luminous for a star of such mass at the Hertzsprung gap that has evolved without interactions, and too massive for a hot post asymptotic giant branch (PAGB) star of similar luminosity \citep{ikonnikova+20}, providing further arguments in favour of a bloated stripped star as a product of binary interactions.

%%%%%%%%%%%%%%%%%%%%%%%%%%%%%%%%%%%%%%%%%%%%%%
% Sect. 3
\section{Photometry}\label{sec:phot}
%%%%%%%%%%%%%%%%%%%%%%%%%%%%%%%%%%%%%%%%%%%%%%

\subsection{SED fitting}\label{sec:sed}

\begin{table}
\caption{Photometric magnitudes in different bands.}
\label{tab:photometry}
\center
\begin{tabular}{clrcr}
\toprule
\toprule
 Filter & Region & Value & Mission/Project & Reference  \\
\midrule
 B1     & FUV     & 12.89  & UIT  & \citeauthor{parker+98}  \\
 B5     & FUV     & 13.02  & UIT  & \citeauthor{parker+98}  \\
 uvw2   & NUV     & 14.16  & Swift-UVOT & \citeauthor{roming+05} \\
 uvm2   & NUV     & 14.17  & Swift-UVOT & \citeauthor{roming+05} \\
 uvw1   & NUV     & 14.13  & Swift-UVOT & \citeauthor{roming+05} \\
 F275W  & NUV     & 14.13 & HTTP  & \citeauthor{sabbi+16}  \\
 F336W  & NUV     & 14.19 & HTTP  & \citeauthor{sabbi+16}  \\
 B      & Optical & 14.97 & VFTS  & \citeauthor{evans+11}  \\
 V      & Optical & 14.85 & VFTS  & \citeauthor{evans+11}  \\
 V      & Optical & 14.92 & OGLE  & \citeauthor{udalski+15}  \\
 F555W  & Optical & 14.90 & HTTP  & \citeauthor{sabbi+16}  \\
 F775W  & Optical & 14.61 & HTTP  & \citeauthor{sabbi+16}  \\
 I      & NIR     & 14.60 & OGLE  & \citeauthor{udalski+15}  \\
 F110W  & NIR     & 14.39 & HTTP  & \citeauthor{sabbi+16}  \\
 F160W  & NIR     & 14.24 & HTTP  & \citeauthor{sabbi+16}  \\
 J      & NIR     & 14.33 & 2MASS  & \citeauthor{cutri+03}  \\
 H      & NIR     & 14.29 & 2MASS  & \citeauthor{cutri+03}  \\
 K$_s$      & NIR     & 14.14 & 2MASS  & \citeauthor{cutri+03}  \\
\bottomrule
\end{tabular}
\end{table}

We have found several photometric measurements in the literature: far-ultraviolet (FUV) photometry from the Ultraviolet Imaging Telescope \citep[UIT,][]{parker+98}, near-ultraviolet (NUV) photometry from the Swift Ultra-Violet/Optical Telescope \citep[UVOT,][]{roming+05}, NUV, optical, and near infra-red (NIR) photometry from the Hubble Tarantula Treasury Project \citep[HTTP,][]{sabbi+16}, $B$- and $V$-band photometry from the VFTS \citep{evans+11}, $V$- and $I$-band photometry from the Optical Gravitational Lensing Experiment \citep[OGLE,][]{udalski+08, udalski+15}, and $J$-, $H$- and $K_s$-band photometry from the 2MASS all-sky catalog \citep{cutri+03}. All the available measurements are presented in Table~\ref{tab:photometry}. We also found three measurements from XMM-Newton Optical/UV Monitor Telescope (XMM-OM) in the UVW1 filter, that were almost 1\, mag fainter than the other NUV measurements so we have not included it among our values. In the case of the UVOT photometry, we found 8 measurements for the uvw2 and uvm2 filters, and 9 for the uvw1 filter, mean values from these measurements are presented in Table~\ref{tab:photometry}, and we have used the standard deviation as the uncertainty, so it is higher for these three bands than for the rest of the available photometry. Due to the present variability in the measurements, we analysed another close source in the field which turned out to be fairly constant (${\overline\sigma}=0.04$ vs ${\overline\sigma}=0.15$), suggesting than the UV variability could be due to binarity. We have checked the UVOT field of view and VFTS~291 is well isolated and therefore not affected by crowding.

\begin{figure}
\centering
    \includegraphics[width=\columnwidth]{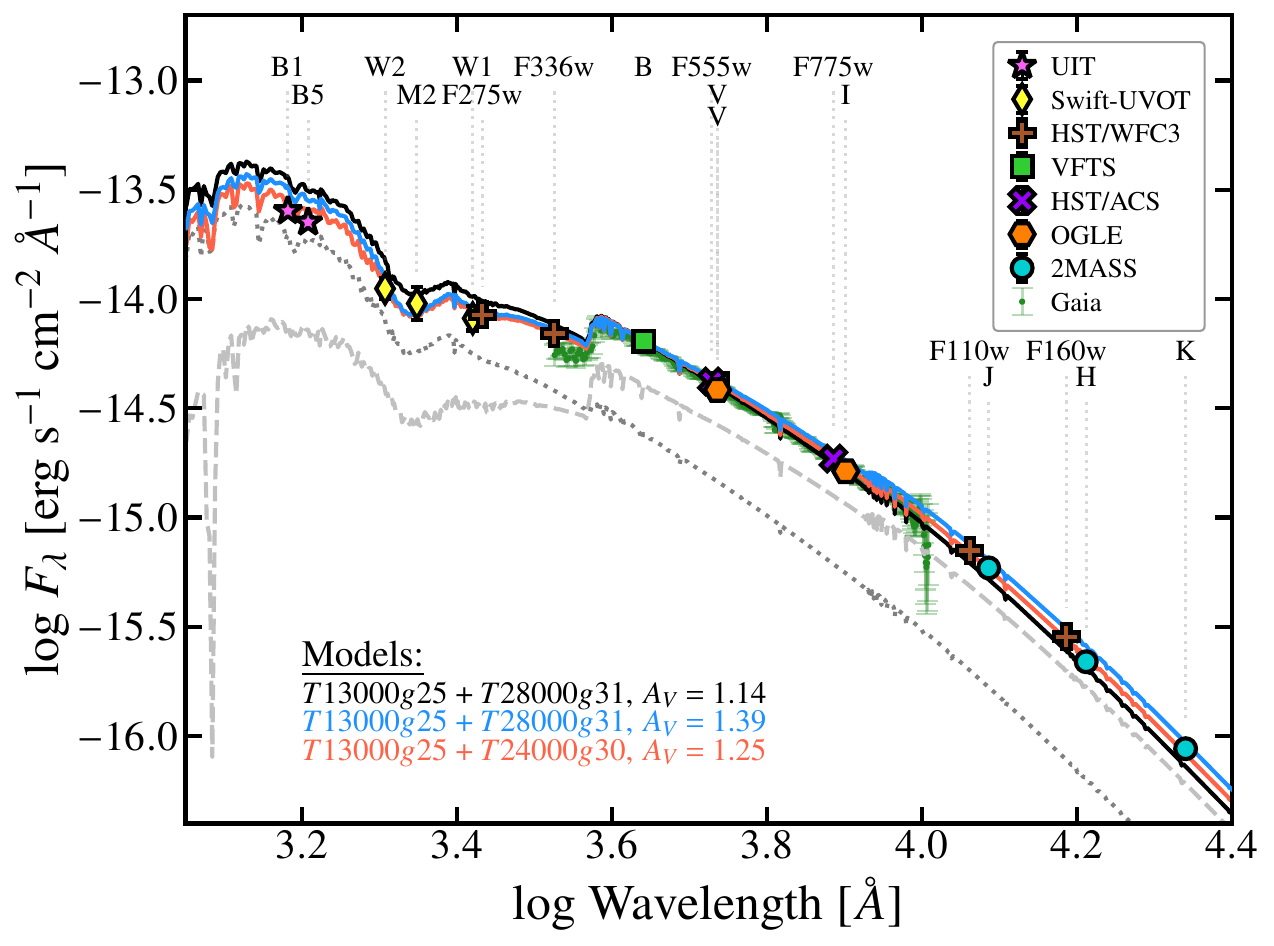}
    \caption{SED fitting to available FUV to NIR photometry, missions/projects are shown with symbols, name of the filters are displayed above each symbol. In black, blue, and red, we show three different composite models, each corresponding to the sum of two ATLAS9 models and a different extinction value, see details at the bottom of the figure and in the main text. The individual contribution of both components to the red model is shown by the dotted-dark grey (24 kK) and dashed-light grey (13 kK) models. Photometric errors are typically the size of the symbols.}\label{fig:sed}
\end{figure}

The available photometry covers a wide range in wavelength, from 1521 to 16300\,\AA, much larger than the range covered by FLAMES. Fitting these data provides a sensitive independent test of our findings from the spectroscopic data. We have used the ATLAS9 flux models provided by \citet{howarth11}, with LMC abundances, to perform the spectral energy distribution (SED) fitting. Figure \ref{fig:sed} shows the results of our fitting procedure. Symbols show the 18 photometric measurements that we have fitted, but also, in green, we show the Gaia XP spectrum for comparison; it covers the wavelength range between 3360 and 10200\,\AA, it is in good agreement with the other optical values, but presents higher uncertanties after the Balmer jump, and falls below the HST F336w filter, so we have not included it in the fitting process. In black we show a composite model corresponding to the closest parameters to our determined values, i.e. a \Teff of 13000\,K and \logg of 2.5 for the narrow-lined star plus a 28000\,K model with \logg of 3.1 for the companion. For the extinction law we have adopted the one computed by \citet{maiz+14} specifically for the 30~Dor region, and assumed a total extinction $A(\lambda)=1.14$ and $R_V=4.5$ as in Sect.~\ref{ssec:physparams}, which gives a reddening of $E(B-V)=0.25$. It is a good fit to the optical data, but overpredicts the UV data while underpredicting the IR photometry. We repeated this experiment by leaving the light ratio as a free parameter, but it resulted in an unconstrained light ratio and returned almost identical results to the previous fit. For the model shown in blue we have used the same composite model as before, but left $R_V$ and $E(B-V)$ as free parameters and employed the same \chisq minimisation procedure as described in Sect.~\ref{ssec:spfit}. This model is a better fit to the NUV data but still slightly overpredicts the FUV UIT measurements and also the NIR values, while the total extinction increased from 1.14 to 1.39. As a final experiment we let the model of the companion star vary in a temperature range of 18000-34000\,K and a range of \logg between 2.5 and 4.5, we note however that the surface gravity has a minimal effect on the SED, and the value obtained from the SED fitting carries little meaning. With the model of the narrow-lined star and the light ratio as fixed parameters we have found a best model shown in red. The temperature of the companion has dropped to 24000\,K and the total extinction lies between the previous values, at 1.25, improving the fit at both ends of our wavelength range. For this model we show the contribution of each component in light and dark grey, scaled by their respective flux contributions. The hotter model clearly dominates the near and far UV, whereas at longer wavelength the cooler model has a larger contribution, which increases towards the IR.

\begin{figure}
\centering
    \includegraphics[width=\columnwidth]{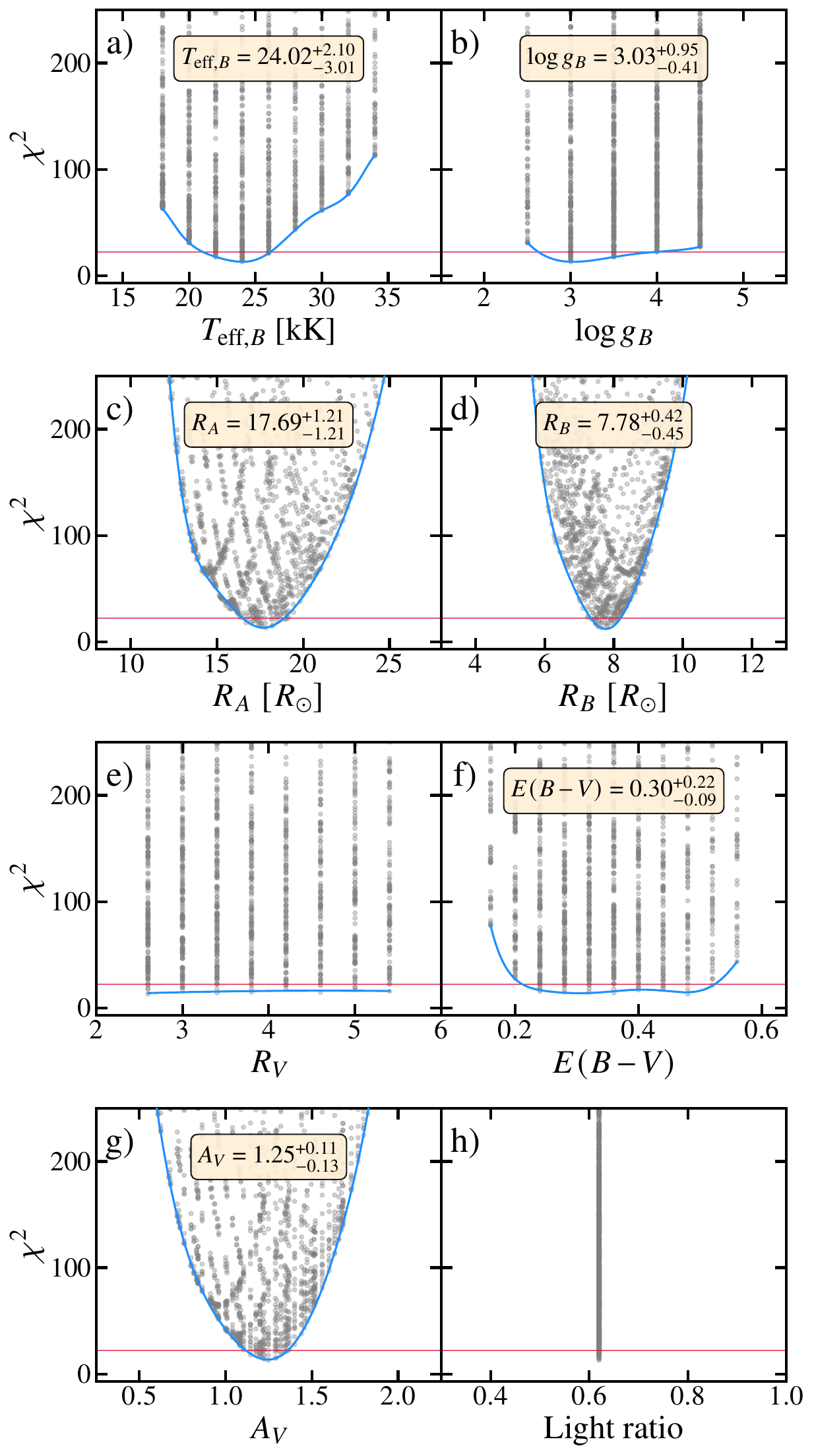}
    \caption{As Fig.~\ref{fig:chi2_er} but for our SED-fitting procedure. The parameters of the narrow-lined star ($T_{{\rm eff}, A}$, $\log g_A$, not shown) and the light ratio (shown here as $F_A/F_{\rm Tot}=0.62$) have been fixed. $R_V$ and $E(B-V)$ are unconstrained but we get a good constraint for $A_V$, in agreement with the value from recent studies.}\label{fig:sederr}
\end{figure}

To properly quantify the difference with our spectroscopic analysis we have computed 2-$\sigma$ errors as in Sect.~\ref{ssec:spfit}, these are shown in Fig.~\ref{fig:sederr}. For the companion star, \Teff and \logg (panels a and b respectively) are in good agreement within errors with our spectroscopic determinations, yielding values of $T_{{\rm eff}, B}=24.0^{+2.1}_{-3.0}$\,kK, and $\log g_B=3.0^{+1.0}_{-0.4}$\,dex. The ratio of total-to-selective extinction (panel e) is completely unconstrained, whereas the reddening (panel f) is also not well constrained, without presenting a clear minimum and values ranging from 0.2 to 0.42. The total extinction (panel g), however, is well constrained with a value of $A_V=1.25^{+0.11}_{-0.13}$, in good agreement with our adopted value to compute the spectroscopic parameters. From the observed and model fluxes (at an adopted $\lambda=5335$\,\AA), we determined angular diameters of $\theta_A=0.0037$\,mas and $\theta_B=0.0013$\,mas, which together with the distance gave us the physical radii shown in panels c and d of Fig.~\ref{fig:sederr}. We found ($R_A/R_\odot)=17.7\pm1.2$ and ($R_B/R_\odot)=7.8\pm0.4$, in excellent agreement with the spectroscopic ones (see Table~\ref{tab:strpstars} and Sec.~\ref{ssec:physparams} for details). 

The results of our SED fitting confirm that, despite the uncertainties in the modelling of the companion's spectrum, our spectroscopic determinations of the physical parameters of the two components are fully consistent with the available photometric measurements for VFTS~291, while the extinction is also consistent with recent findings in Hodge 301 and 30 Dor.  

\begin{figure}
\centering
    \includegraphics[width=\columnwidth]{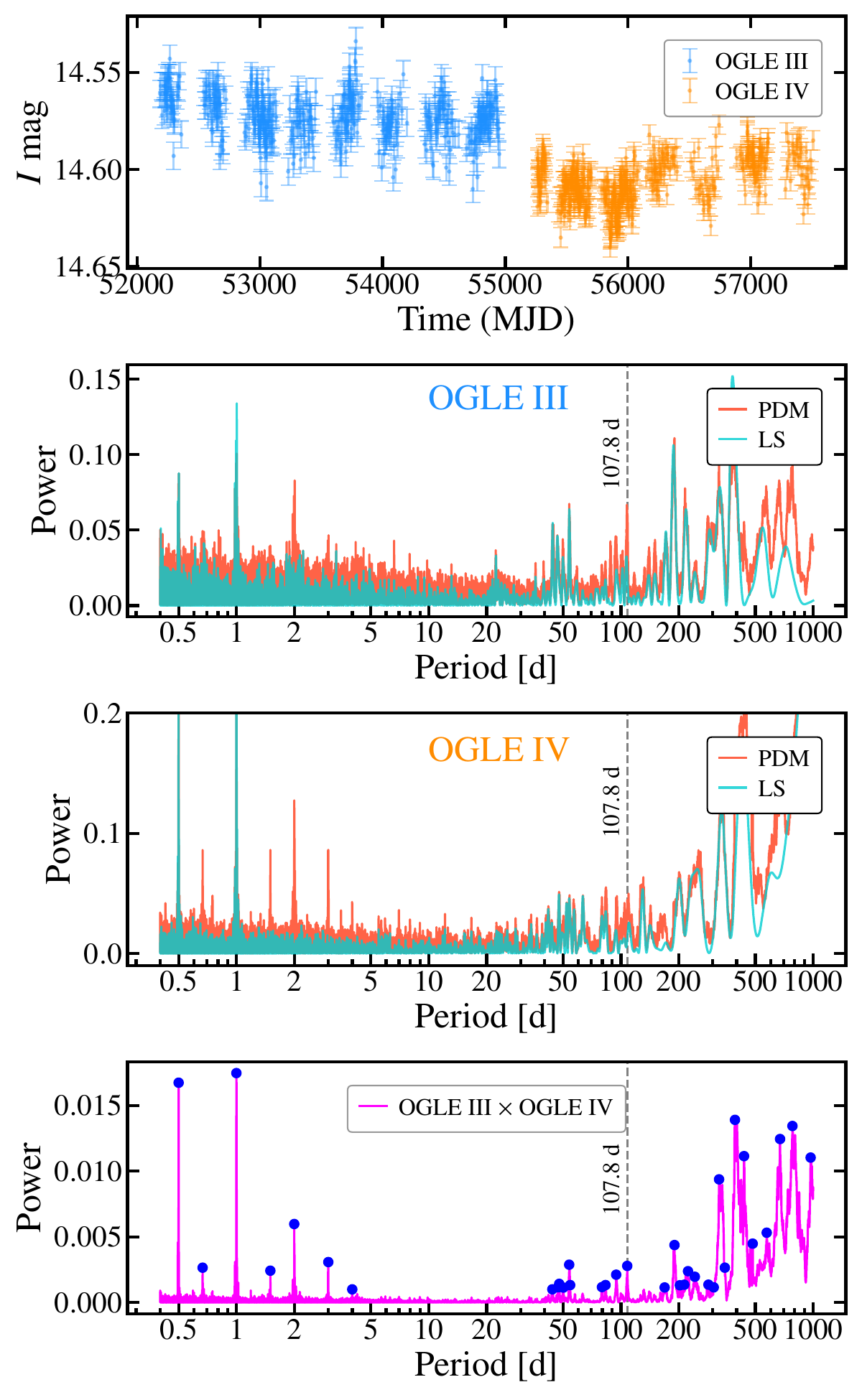}
    \caption{OGLE-III and -IV $I-$band photometry for VFTS~291 (top panel). Magnitude offset between data sets is caused by uneven sky background producing a shift in the zero point of individual stars. We have performed LS and PDM analyses in search for periodicity in each data set separately (second and third panels from the top). The bottom panel shows the multiplication of PDM periodograms for both datasets, and the blue points mark the peaks at the periods that we have used to fold the light curve. The spectroscopic period of 107.8\,d is shown with the vertical dashed line.}\label{fig:ogleLS}
\end{figure}

\subsection{OGLE Photometry}\label{ssec:ogle}

VFTS~291 has also been followed by the OGLE-III and OGLE-IV surveys, in the $V-$ and $I-$bands. We analysed the light curves in both bands but only discuss the $I-$band data, which have 10 and 5 times more observations than the $V-$band in the OGLE-III and -IV data (respectively), as well as producing the more consistent results. In total, the OGLE data cover a time window of about 15\,yr, from 2001 to 2016, as shown in the top panel of Fig.~\ref{fig:ogleLS}.

As in \citetalias{villasenor+21}, we have used the Lomb-Scargle (LS) periodogram \citep{lomb76,scargle82} from \textsc{Astropy} \citep{astropy:2022}, but we have also included a \textit{Phase Dispersion Minimization}\footnote{Available at \url{https://py-pdm.readthedocs.io/en/latest/}} \citep[PDM,][]{stellingwerf78} routine to search for the orbital period. In Fig.~\ref{fig:ogleLS}, second and third panels from the top, we show the resulting periodograms for the OGLE-III and OGLE-IV data respectively. In both cases we can see that PDM did a better job retrieving the spectroscopic period of 108\,d, but there are several stronger peaks that seem to be aliases of the spectroscopic period or of the day/night cycle. To favour the peaks present in both datasets we have multiplied the periodogram obtained with PDM for the OGLE-III and -IV data. The result can be seen in the bottom panel; several strong peaks remained, specially at long periods, that might be related to the baseline of the respective campaigns. 

In order to test the significance of these peaks, we have folded both OGLE-III and -IV data to all the periods at the peaks marked with blue circles. None of these periods resulted in clear signs of periodicity due to orbital motion, except for the OGLE-III data folded at the 107.8\,d period. 
The folded light curve presented a mild close-to-sinusoidal modulation, which could be a signature of ellipsoidal variations from a distorted star. However the filling factor $R_{A}/R_{\rm RL}$ is only 0.3 for our minimum mass ratio, increasing to 0.34 for $q=8$, requiring a high inclination in order to be detectable.

To at least see if this variation is consistent with our findings, we have used PHOEBE \citep[v2.3,][]{conroy+20} to model the light curve. In a first test, we used our improved spectroscopic solution from this work (\porb, $e$, $\omega$, $\gamma$, $T_p$, $a_1\sin i$), and the adopted and determined parameters from our spectral fitting ($d$, $A_V$, $R_A$, $T_{{\rm eff}, A}$) as fixed parameters, only varying the inclination $i$ and the mass ratio $q$, and fitting for $R_B$ and $T_{{\rm eff}, B}$. Once we found a satisfactory solution, we kept $R_B$ and $T_{{\rm eff}, B}$ fixed to fit the orbital period. We found a photometric period of $P_{\rm phot}=108.104$\,d, from which we computed a new orbital solution (available in Table~\ref{tab:phoebe_orbsol}) to try to improve the folded light curve. 

In the second step, we have fixed the orbital inclination to three different values: $83^\circ$, which is the highest inclination the orbit can have before presenting eclipses, $70^\circ$, and $60^\circ$, and tried different values for the mass ratio, while allowing $M$, \logg, $R$, and \Teff for both stars to vary. Finally we let all parameters vary. The details of our procedure can be found in Appendix~\ref{ap:phoebe}, and the individual models are presented in Table~\ref{tab:phoebe}. The selected inclinations explore the full parameter space provided by our previous results, from both spectroscopic analysis and SED fitting. The lower inclination (models \#1 to \#3 in Table~\ref{tab:phoebe}) results in masses for the companion between 17-20\Msun, too high for the lack of strong \ion{He}{ii} lines, from which we conclude that the inclination should be above this value. A mass ratio of 6, favours larger radii for the narrow-lined star ($R_A>20$\Rsun), above the 2-$sigma$ uncertainty of the value obtained from SED fitting, suggesting that the mass ratio is indeed larger than this lower limit. The larger inclination ($i=83^\circ$) result in masses for the narrow-lined star close to 1.5\Msun and overall in lower total masses of the system, which we can not rule out, but models with $i=70^\circ$ are in better agreement with our spectroscopic results. It is also interesting to note that mass ratios close to 8 are in general favoured by the fit, this value matches the mean mass ratio found for 12 Be+sdO binaries by \citet{wangL+23}. The surface gravities present the higher discrepancy with our spectroscopic results. In the case of the narrow-lined star, we obtained in average $\log g_{A} \approx2.1$, which is below the 2-$\sigma$ uncertainty, whereas for the companion, the models result in $\log g_{B} \approx3.8-4.0$, consistent with an evolved MS star.

\begin{figure}
\centering
    \includegraphics[width=\columnwidth]{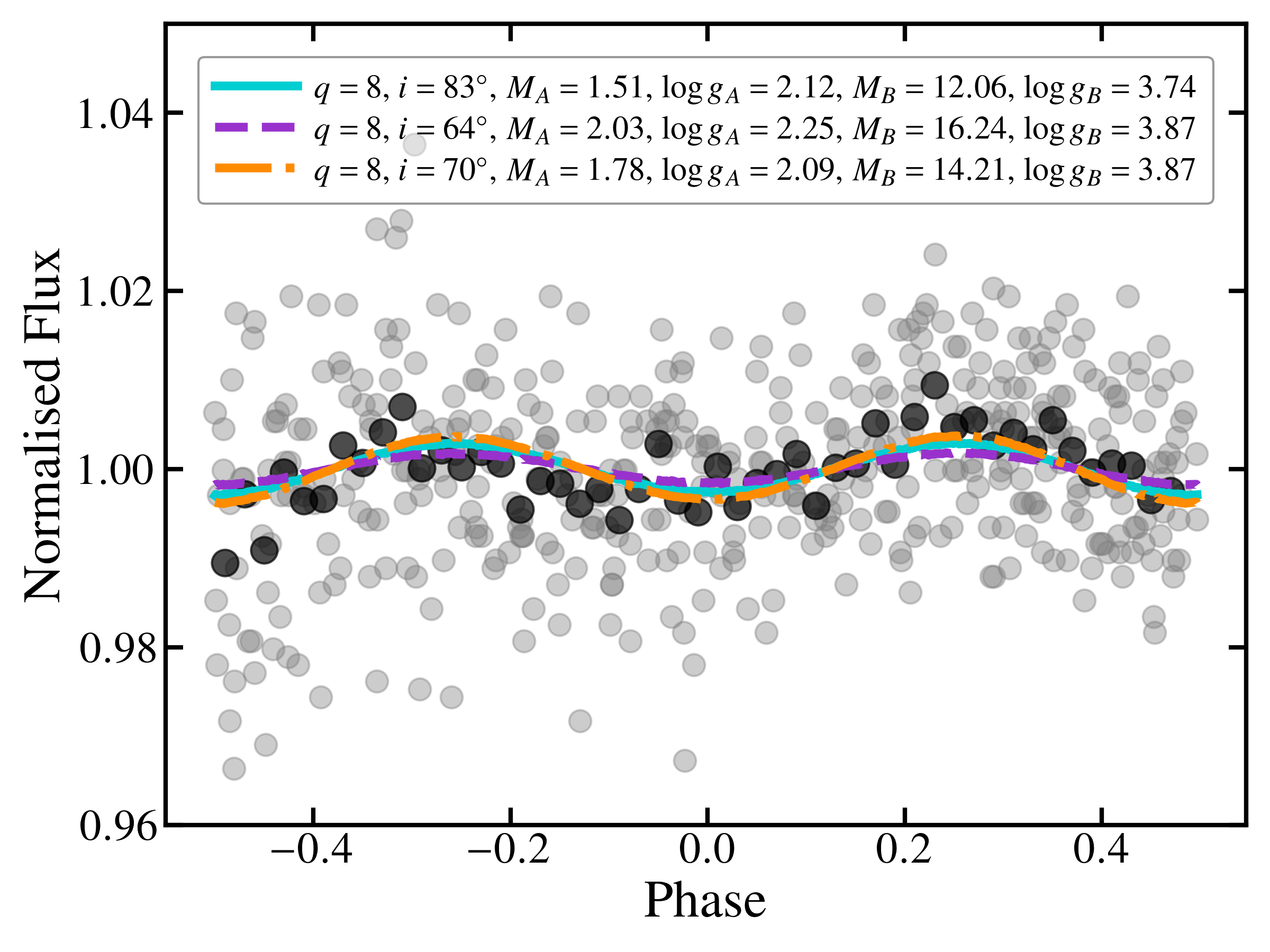}
    \caption{Light curve of VFTS~291 from the OGLE-III data folded to the period of 108.1\,d found with PHOEBE. We have used a binned mean to smooth the light curve (black circles). In orange, we show one of the PHOEBE models that best agrees with our spectroscopic results, see main text for details on the models.}\label{fig:phoebe}
\end{figure}

In Fig.~\ref{fig:phoebe}, we show  the light curve folded to the photometric period in grey circles. We have smoothed the light curve with a binned mean which is plotted with black circles; the sinusoidal modulation is also present here. Two minima are observed at phases 0 and 0.5, if these are produced by ellipsoidal variations, they should be present at inferior and superior conjunction, whereas maxima occur at quadratures, when the projected area of the distorted star is larger. This is consistent with our radial velocity curve (Fig.~\ref{fig:rvcurve}), where our maximum and minimum RVs are also at quadratures (phases 0.25 and 0.75).
One of our PHOEBE models (\#5 in Table~\ref{tab:phoebe}) is also shown in Fig.~\ref{fig:phoebe} (dash-dotted orange curve), none of the other PHOEBE models present significant differences with the shown model, so we have plotted two additional models corresponding to $M_A=1.5$\Msun (turquoise) and $M_A=2.0$\Msun (purple), but with radii and effective temperatures fixed to our spectroscopic values. Given the S/N of the light curve, it is not possible to determine which one is a better fit.

Although these tests are not conclusive, we note that they provide an independent measurement of the temperatures and radii, predominantly constrained by the orbital solution, and that are consistent with our previous determinations of physical properties for the two components of VFTS~291. These results strongly support our proposed scenario of a low-mass stripped star with an early B-type companion.

%%%%%%%%%%%%%%%%%%%%%%%%%%%%%%%%%%%%%%%%%%%%%%
% Sect. 4
\section{Binary evolution} \label{sec:evol}
%%%%%%%%%%%%%%%%%%%%%%%%%%%%%%%%%%%%%%%%%%%%%%

\begin{figure*}
\centering
    \includegraphics[width=0.8\hsize]{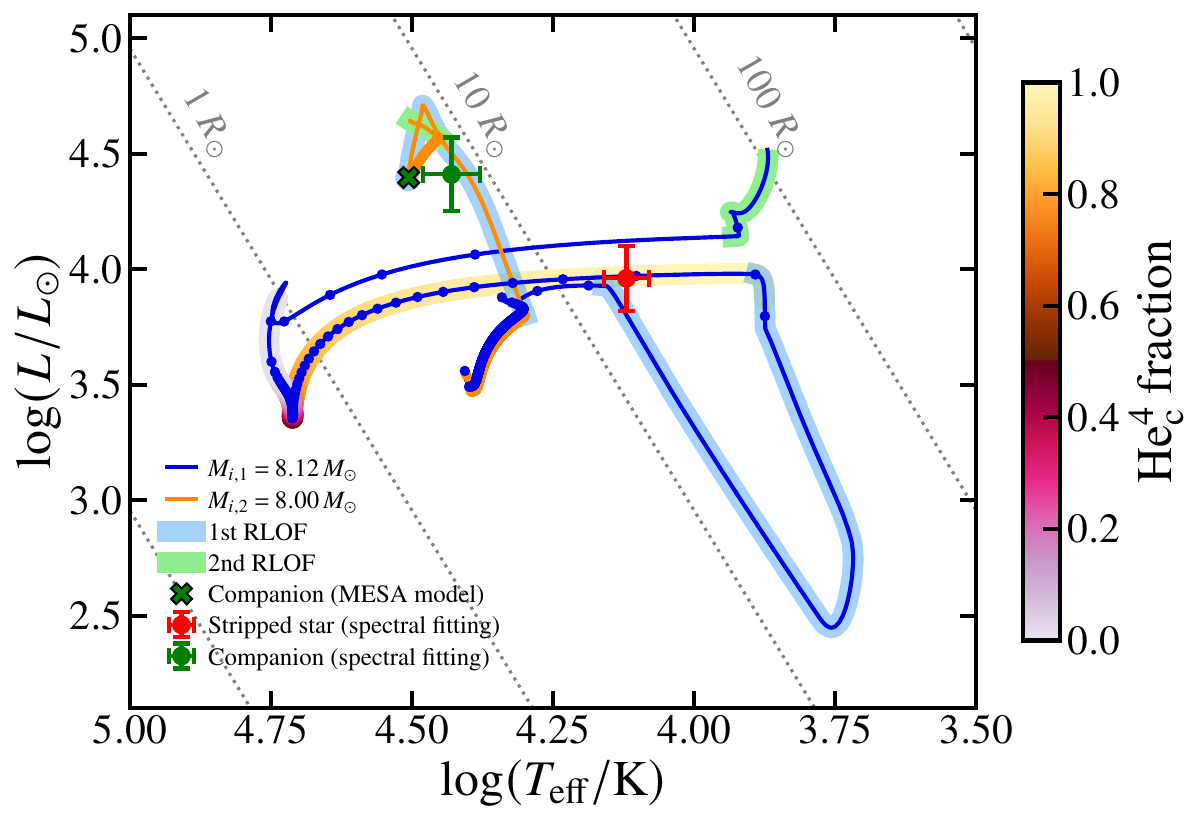}
    \caption{HR diagram presenting the position of the two components of VFTS~291 from our spectroscopic values. The narrow-lined (stripped) star and the companion are shown by the red and green circles respectively, with their corresponding errors. The evolutionary track computed with MESA is shown for the narrow-lined star (in blue) and for the companion (orange), where dots mark intervals of 50\,000\,yr. Thick sky blue (light green) lines mark the first (second) Roche-lobe-filling phase for both stars. The green cross denotes the position of the companion on its evolutionary track at the time in which the narrow-lined star intercepts its evolutionary track, showing that in our computed MESA model, the companion is hotter and smaller, while reaching a similar luminosity in comparison to our spectroscopic measurements. The coloured thick line represents the core-helium fraction of the narrow-lined star, with values given by side colour bar. Core-He burning started before the end of mass transfer and it reaches 0.5 at the minimum luminosity point (darker colours).}\label{fig:HRD}
\end{figure*}

\subsection{Initial properties}

To test our hypothesis on the current state of VFTS~291, we have computed binary evolution models with MESA (version 15140) and followed the evolution of both stars. From the parameters that determine the evolution of a binary, the orbital period is the better constrained for VFTS~291. Using Eq.~(2) from \citet[][see also \citealt{soberman+97}]{bodensteiner+20b} for conservative mass transfer, and the current orbital period and minimum mass ratio of the system (from Sect.~\ref{ssec:disent}) as reference point, we can estimate the initial period of the system by assuming an initial mass ratio. If the initially primary star transferred mass to the secondary---but now more massive star---we can expect an initial mass ratio close to unity. Consequently, using $P_r=108\,$d, $q_r=6$ and $q_i=0.95$ as a first guess, we found an initial orbital period $P_i=12.7\,$d, where subscript $r$ denotes our \textit{reference} point and $i$ the \textit{initial} parameter. Using this initial orbital period, we found current periods longer than the observed one, roughly between 150 and 200\,d, depending on the initial masses. Since post-mass-transfer masses will not be largely affected by small variations of the initial orbital period for case-B mass transfer, due to the current mass being close to the mass of the core at the end of the MS, we have linearly scaled the obtained initial period with the final period, i.e. $P_{i}\times108/P_{f}$. We found $P_{i}=7.025$\,d to produce a post-mass-transfer orbit close to our 108\,d period. For the masses, we have used a small grid of primary masses to find the initial mass that better reproduces the parameters of the stripped star found from the spectroscopic analysis, resulting in $M_{1,i}=8.12$\Msun. In the next step we have fine tuned the initial mass of the secondary with the constraint that at its current value, it must agree with the minimum mass ratio estimate obtained from the disentangling ($q_{\rm min}=6$) while trying to also reproduce its luminosity and temperature, obtaining an initial mass of $M_{2,i}=8$\Msun.

Given the long orbital period of the system, for the following analysis we have assumed that VFTS~291 has reached its current configuration through a, fully conservative, case B mass transfer phase, i.e. the primary has filled its Roche lobe after core hydrogen exhaustion. For other possible configurations see Sect.~\ref{ssec:otherconfigs}. Also, we did not account for stellar rotation in our model due to simplicity, and used a similar setup as the one used by \citet{bodensteiner+20b} but with LMC metallicity ($Z= 0.008$).

\subsection{The age of the system}

H301 is the oldest cluster in 30~Dor, giving enough time for binary evolution and interactions to take place, with estimates that between 40 and 60 type II SN have already exploded \citep{grebel+chu00, cignoni+16}. \citet{grebel+chu00} determined an age for H301 in the range of 15--20 Myr by fitting isochrones to their colour-magnitude diagram (CMD). Later, in the context of the VFTS, \citet{evans+15} obtained effective temperatures from calibrations to their spectral classifications, estimated luminosities from optical photometry (see details in \citealt{evans+15}), and constructed the HRD for the stars in the two smaller clusters in 30~Dor, H301 and SL~639. From the comparison with isochrones for non-rotating models \citep{brott+11a}, they inferred an age of $15\pm5\,$Myr for H301, in agreement with the estimated value from \citet{grebel+chu00}.

The unique capabilities of the HST allowed \citet{cignoni+16} to identify the pre-MS turn-on point close to $V\approx24\text{--}25$. By fitting this feature and the MS turn-off point, they found an age between 26.5 and 31.5\,Myr. They give a detailed discussion regarding the discrepancies with the previous studies, but within their analysis the main source of uncertainty is the metallicity adopted. A lower metallicity ($Z=0.005$) favours the age lower limit, whereas for the usually referenced value of $Z=0.008$, they found $30.5^{+1}_{-2}\,$Myr. 

Finally, \citet{britavskiy+19} determined physical parameters including ages for their sample of candidate RSGs. They determined ages by comparing the observed luminosities to the expected luminosity given by stellar evolution models, then the age of the cluster is estimated by the luminosity of their faintest object. They note however, that red stragglers, i.e. RSGs that are overluminous and overmassive as a product of binary interactions (due to mass transfer or mergers), are expected to represent a large fraction within a cluster, so their estimated age limit might by affected by binarity. From the least luminous RSG they determined an age for H301 of $24^{+5}_{-3}$\,Myr.

In conclusion, if VFTS~291 is a binary interaction product and a member of H301, we do not expect the system to be older than 31.5\,Myr, the highest upper limit from the aforementioned studies. Therefore, the MESA evolutionary model should reach the current configuration of the system within this age.

\subsection{Following the evolution with MESA}

\subsubsection*{From the MS to the end of RLOF}

One of the MESA models that can explain the current state of VFTS~291 is shown in the Hertzsprung-Russell diagram (HRD) of Fig.~\ref{fig:HRD}. The two larger circles show the current positions of the narrow-lined star (red) and its companion (green) in the HRD from our spectroscopic analysis, while the blue and orange lines are the evolutionary tracks for each component respectively, with steps of 50\,000\,yr marked by dots. The originally more massive star (blue track) evolves off the MS after 32\,Myr and starts to move through the Hertzsprung gap while expanding its radius. After just 0.1\,Myr, the primary fills its Roche lobe and case-B mass transfer starts (sky blue thick lines). The luminosity and temperature of the donor star abruptly drop as the star tries to readjust to its rapid mass loss; this is a drop of 1.5 order of magnitudes in luminosity. On the other side, the accretor reacts in the opposite way to the gain of mass, by moving up and to the left of the HRD. As the mass-loss rate stabilises, the donor can more effectively adjust to its decreasing mass, and the luminosity can rise again, however the accretor continues to increase its luminosity until the mass-loss rate starts to decrease. During this period, the central temperature in the core of the donor has steadily increased up to the point it is enough to ignite helium at the core. Mass transfer continues at a decreasing pace until the donor detaches from its Roche lobe and mass transfer ceases. This is an incredibly rapid phase, lasting only 90\,000 years, where the stars go through drastic structural changes. At the end of Roche-lobe overflow (RLOF), the now more massive star is ‘‘rejuvenated'', appearing more luminous and hotter than before and eventually could be seen as a blue straggler, much more luminous and massive than stars at the MS turn-off point of its cluster. On the other side, the donor has lost more than 6\Msun from its initial 8.12\Msun and 98\% of its hydrogen envelope at the start of RLOF, appearing as a core-helium-burning (CHeB) bloated stripped star of 1.8\Msun.

\subsubsection*{Comparing to the current state of VFTS~291}

We can now compare this model to the position in the HRD of the two components of VFTS~291. The green ``x'' marks the current state of the accretor at the moment in time of closest approach between the evolutionary track of the donor and the red circle. We can see that the temperature of the accretor is higher than our determined value from the spectral fitting, but the luminosity is in good agreement. The full set of parameters from our MESA model at the current state of the system is, for the secondary/companion: $\log (L_2/\text{\Lsun}) = 4.40$, $T_{{\rm eff},2}=32\,057$\,K, $\log g_2=4.18$, $R_2 = 5.12$\Rsun, $M_2 = 14.33$\Msun, whereas for the primary/stripped star we found $\log (L_1/\text{\Lsun}) = 3.97$, $T_{{\rm eff},1}=13\,204$\,K, $\log g_1=2.16$, $R_1 = 18.44$\Rsun, $M_1 = 1.80$\Msun. The system reached this configuration in 32.31\,Myr, just above the expected limit, and for an orbital period of 113\,d. 

While we were able to reproduce the luminosity of both components, not all parameters are in good agreement with our spectroscopic analysis. The effective temperature of the stripped star is in good agreement, as expected since it is one of our anchor points, and the radius, since it is computed from $L$ and \Teff. Its surface gravity fell just below our determined lower limit (2.16\,dex), closer to the value found with PHOEBE (\logg=\,2.1). On the other side, increasing the temperature within limits, e.g. to 14\,000\,K, would put \logg within 2-$\sigma$ of our spectroscopic value. The evolutionary mass of the primary is also below our estimated lower limit of 1.85\Msun, but by considering a higher initial mass we can increase the mass of the stripped star while keeping the luminosity matched within errors. The main discrepancy, however, is in the surface helium abundance. In Sect.~\ref{ssec:spfit} we found ${\rm He/H} = 0.105^{+0.014}_{-0.017}$, which would translate to roughly $Y=0.29$, an increase of only 0.06 in the mass fraction of surface helium, when we take ${\rm He/H} = 0.076$ as the initial value \citep[from the ATLAS9 models,][]{howarth11}. Our MESA model predicts an increase from 0.26 to 0.68 in the surface helium mass fraction, which is a significant enrichment compared to our spectroscopic result. We can only speculate, as discussed in Sect.~\ref{ssec:spfit}, that the Balmer lines profile of the stripped star could also be affected by the disentangling, leading to an erroneous determination of the He/H ratio, since we do see significant enrichment in the \spline{N}{ii}{3995} line. Interestingly, \citet{bodensteiner+20b} also did not find evidence of helium enrichment in HR~6819, but \textit{did} find clear signs of nitrogen enrichment. Further testing is needed to investigate if this could be a consequence of the difficulties of disentangling the Balmer lines, or perhaps a real feature, such as the donor retaining a larger fraction of its envelope. A side-by-side comparison of the physical parameters obtained from our spectral analysis and those given by MESA can be seen in Table~\ref{tab:strpstars}.

For the companion, the disagreement between parameters is larger. As mentioned before, \Teff is higher than our upper limit by 2000\,K, which is also reflected in the radius, roughly 1\Rsun below our estimated lower limit. The surface gravity is higher than our maximum value for \logg by 0.7\,dex, but as discussed in Sec.~\ref{ssec:spfit}, this \logg would be more consistent with the signatures in the spectrum of the companion, and supports our hypothesis of \logg being affected by the disentangled process due to nebular contamination, the slow motion of the secondary, and the resolution of our data. Finally, the mass of the companion is significantly higher than our spectroscopic mass (affected by \logg), a factor 2 higher than the upper limit, but in good agreement with the evolutionary mass derived in Sec.~\ref{ssec:physparams} and with our PHOEBE models (see Table~\ref{tab:phoebe}). This high mass results in $q=8$, in agreement with our estimated minimum mass ratio of 6, with the low semi-amplitude velocity found for the companion, and with the mass ratios of the Be+sdO systems found by \citet{wangL+23}.
Overall, our evolutionary model is a good match to the different pieces of evidence resulting from our spectroscopic and photometric analysis, making a strong case for the stripped star scenario as a credible channel able to reproduce the properties of VFTS~291.

\begin{figure}
\centering
    \includegraphics[width=\columnwidth]{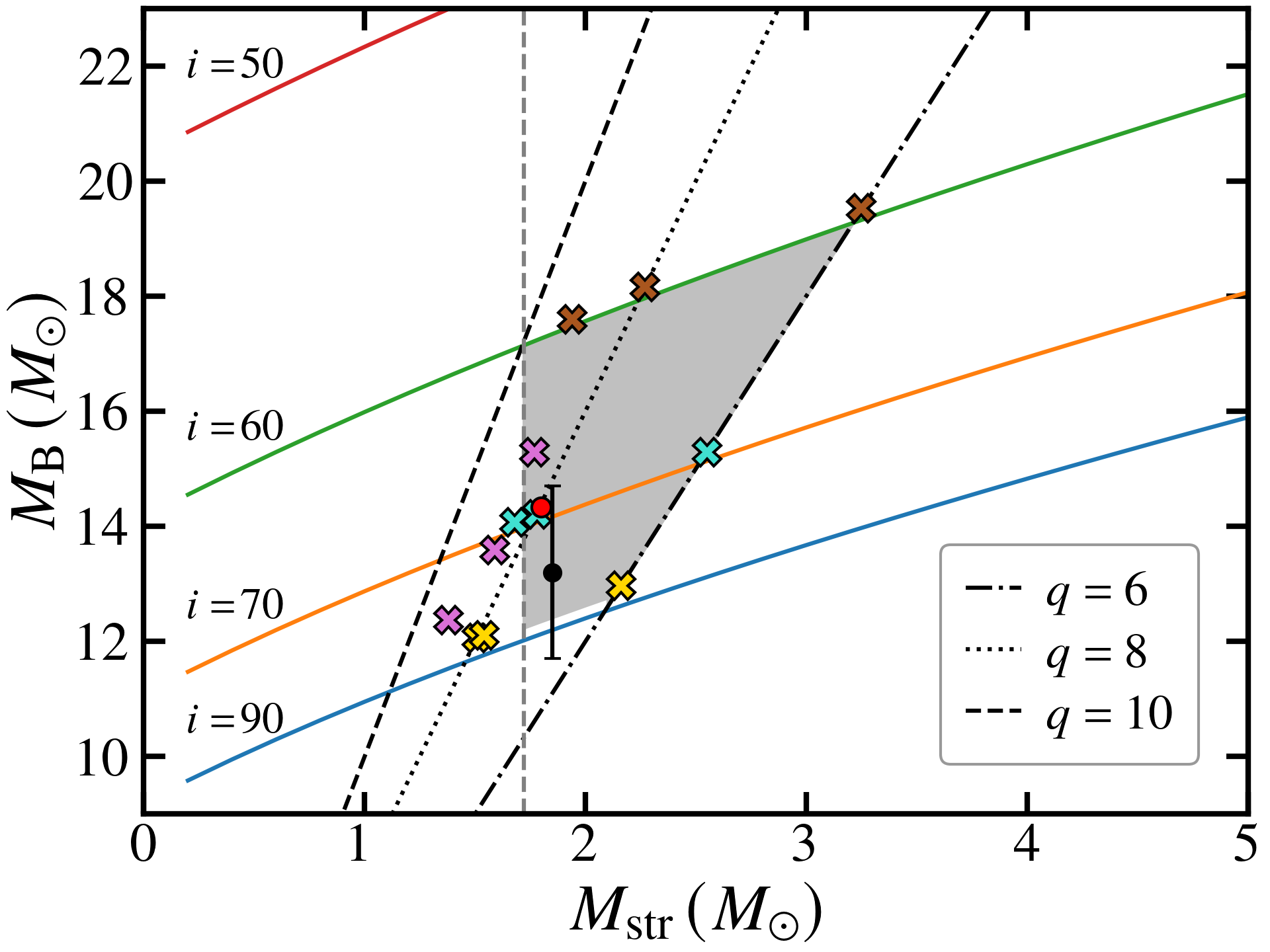}
    \caption{Mass of the early-B companion ($M_{\rm B}$) versus the mass of the stripped star ($M_{\rm str}$). The mass of the companion was computed for a range of masses of the stripped star at different orbital inclinations, which are shown by the labelled solid curves. Our PHOEBE models are represented by the “x'' symbols, the evolutionary mass of the companion with 1-$\sigma$ errors is the black dot, and masses from the MESA model are shown by the red circle.} \label{fig:M1M2plane}
\end{figure}

To visualise the various constraints on the masses, coming from the different analyses, Fig.~\ref{fig:M1M2plane} shows the $M_2$-$M_1$ plane. To construct this diagram we have used Eq.~\ref{eq:massfunc} to compute the mass of the B-type companion ($M_{\rm B}$) for a range of masses of the stripped star ($M_{\rm str}$), assuming different inclinations. An inclination of $90^\circ$ provides an absolute minimum value for $M_B$, but we also have the minimum given by the lack of eclipses in the light curve, absent up to $i=83^\circ$, which is also the maximum inclination that we have used in PHOEBE.
We have defined the upper limit on $M_B$ by $i=60^\circ$, from our test with PHOEBE, resulting in maximum masses roughly between 17 and 19\Msun. It is also a good match with the 3-$\sigma$ uncertainty from the evolutionary mass. The next constrain comes from the minimum mass ratio $q_{\rm min}=6$ (dash-dotted line), which provides a strong limit for both masses, note that this limit greatly reduces the range of possible $M_{\rm str}$ in comparison to our spectroscopic mass or our dynamical mass. We have set the lower limit on $M_{\rm str}$ at 1.72\Msun, which is below the uncertainty on the spectroscopic mass, and is given by the lower limit on \logg from the spectroscopic analysis, and the lower limit on the radius from the SED fitting (which has lower uncertainty than the spectroscopic value). This four limits constraint the parameter space for the masses to the grey area in the figure. 
We have included the results from the light curve modelling with PHOEBE, these are the “x'' symbols; three models at three different inclinations, and three additional models (in magenta) where we let $i$ and $q$ as free parameters. Two of the latter models suggest a lower mass for the stripped star, this could be related to the fixed radius and \Teff used in those models (see Appendix~\ref{ap:phoebe} for details), small variations in these parameters can alleviate the discrepancy. For the third magenta model, we let $R$ and \Teff of the stripped star to vary, resulting in masses within our defined solution space.
The evolutionary mass of the companion is shown by the black circle with its 1-$\sigma$ uncertainty, and is located at $M_{\rm str}=1.85$\Msun, which is our lower limit on the spectroscopic mass. Finally, the red circle marks the results from our evolutionary model computed with MESA. At $q=8$ (dotted line) and $i=69^\circ$, it is within errors from the evolutionary mass of the companion, but it is also a great match with one of our PHOEBE models at $i=70^\circ$, which is shown in Fig.\ref{fig:phoebe}.

We have learned from our evolutionary modelling that the time passed between the end of mass transfer and the current point in the evolution of VFTS~291 is comparable to the total duration of the RLOF phase, i.e. $\sim0.08$\,Myr, making it very unlikely---but not impossible---to observe systems at this evolutionary phase. In fact, a few very similar objects have been found in recent years, and as our understanding of the different outcomes of binary evolution increases, so does the number of observed binary interaction products. We will come back to the difficulties of finding these objects and to a comparison with newly discovered stripped stars in Sect.~\ref{sec:disc}.

%%%%%%%%%%%%%%%%%%%%%%%%%%%%%%%%%%%%%%%%%%%%%%
% Sect. 5
\section{Discussion}\label{sec:disc}
%%%%%%%%%%%%%%%%%%%%%%%%%%%%%%%%%%%%%%%%%%%%%%

\subsection{H301 membership}\label{ssec:membership}

\citet{evans+15} noted that H301 and SL~639 (the other smaller cluster in the 30~Dor region) presented lower mean radial velocities than the two larger clusters (NGC~2060 and NGC~2070) and the field stars. They computed a mean RV of $\overline{\varv_{\rm r}} =271.6\pm12.2\,$\kms excluding H301 and SL~639, whereas for H301 they computed $\overline{\varv_{\rm r}} =261.8\pm5.5\,$\kms, concluding that they seem to be kinematically different, we note however that these values are in agreement within 1-$\sigma$. \citet{patrick+19} also computed a mean RV for H301 from three RSGs finding $\overline{\varv_{\rm r}} =262.1\pm1.4\,$\kms, in agreement with the value from \citet{evans+15} but with a lower dispersion. Due to our much larger number of observations and the constraints provided by the disentangling, we have obtained a precise systemic velocity for VFTS~291 of $270.6\pm0.2$, much closer to the velocity derived by \citet{evans+15} for NGC~2070 ($\overline{\varv_{\rm r}} =271.2\pm12.4\,$\kms) and the field stars ($\overline{\varv_{\rm r}} =271.3\pm11.8\,$\kms). 

\begin{figure}
\centering
    \includegraphics[width=\columnwidth]{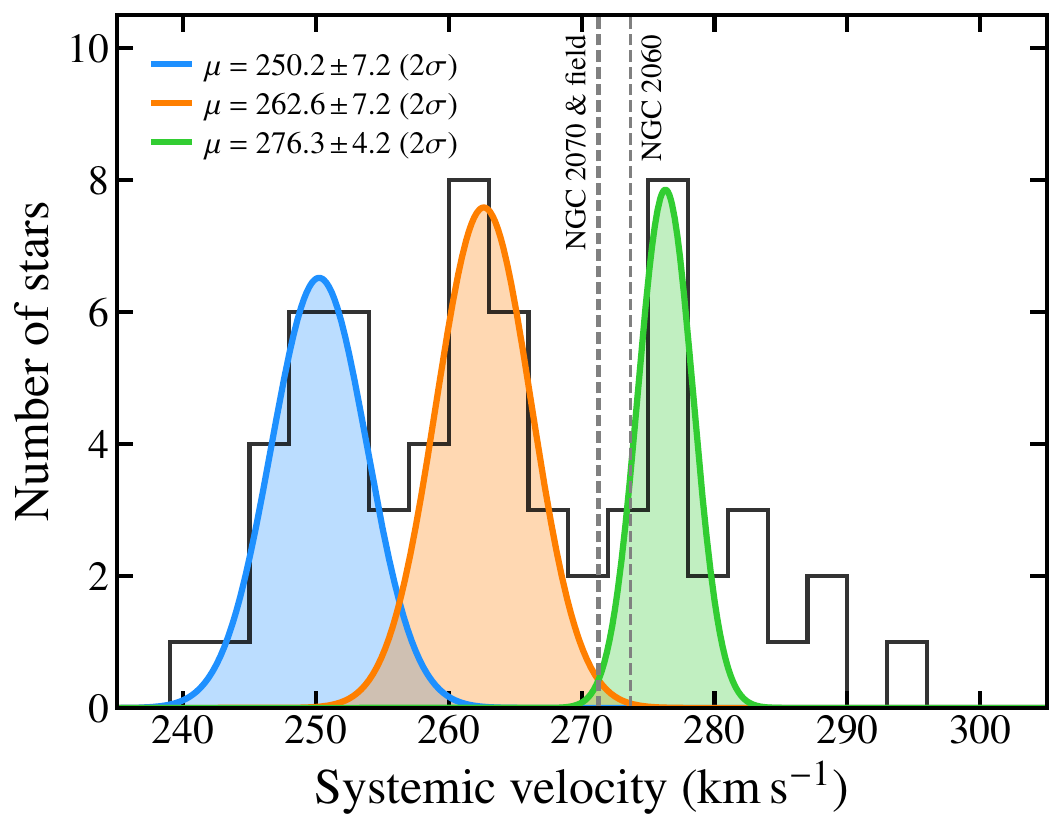}
    \caption{Distribution of systemic velocities of the 50 SB1 and 14 SB2 BBC systems. Mean RVs from \citet{evans+15} for NGC~2060, NGC~2070 and the field are shown with vertical dashed lines. A Gaussian fit to the three main peaks of the distribution is also shown, see legend for the mean and standard deviation.}\label{fig:gammas}
\end{figure}

We have selected all the NGC~2070 members among the BBC systems, those with distances $<33.7\,$pc to the centre of the cluster as defined by \citet{evans+15}, and computed a mean systemic velocity of $\overline{\varv_{\rm r}} =257.9\pm10.7\,$\kms. While it is within 1-$\sigma$ from the results from \citet{evans+15}, the mean value is considerably lower than the mean cluster velocity, suggesting either a different distribution or a biased mean cluster velocity from unidentified binaries. Figure~\ref{fig:gammas} shows the distribution of systemic velocities computed in \citetalias{villasenor+21} excluding the SB1* systems. From this sample of 64 binaries, 95\% of the systems are members of NGC~2060, NGC~2070, or are field stars. We can distinguish three main peaks, which we have fitted with Gaussians, but none of these peaks agree with the mean velocities marked with vertical dashed lines, and the overall distribution is shifted towards lower velocities, with a median of 262.9\kms. While this does not explain the high systemic velocity of VFTS~291 with respect to other H301 stars, it shows that it does not necessarily imply that the system is a member of NGC~2070. 

Using photometric data from the HTTP, \citet{cignoni+16} produced CMDs to reconstruct the star formation history of H301. They defined a radius of 4\,pc as the distance from the centre of the cluster containing 85\% of the member stars. VFTS~291 lies closer to the 4.9\,pc boundary defined by \citet{evans+15}, which is still far from the two annuli that \citeauthor{cignoni+16} defined as containing exclusively field stars (regions between 6.93--8 and 8--8.95\,pc). These two outer annuli present no stars with $\text{F555W} <16\,$mag, whereas VFTS~291 has $\text{F555W} =14.9\,$mag which speaks in favour of its membership. 

Despite the conflict between evidence coming from radial velocities and magnitude, perhaps the most convincing evidence comes from the age distribution of the VFTS stars. \citet{schneider+18} determined the age (and mass) distribution for the single stars in VFTS for the whole 30~Dor region and its main clusters. They found that NGC~2070 is on average the youngest region, with a peak between 2--4 Myr and a median age of 3.6 Myr. The distribution quickly drops after 6 Myr, and for the whole 30~Dor sample it approaches zero at around 40\,Myr. This is in agreement with the study from \citet{sabbi+16} who found that the majority of the stars in NGC~2070 have formed in the last 3 Myr. It is therefore unlikely that VFTS~291 could have formed in NGC~2070 given its evolutionary timescale of $\sim32$\,Myr.

\subsection{Other possible configuration}\label{ssec:otherconfigs}

\subsubsection{Luminosity to initial mass relation}

We have shown that our model with $M_{1,i}=8.12$\Msun is a good match to the observed luminosity of the stripped star. However, this is only an example of one configuration that can explain the system. Using detailed binary evolution models, \citet{schurmann+22} has shown that there is some scatter in the $M_i-L$ relation for stripped stars after Roche-lobe detachment (their Fig. 2). For the case of LB-1, at \Teff$=12500$\,K, the scatter can be of up to 0.6\,dex in $\log L$ for an initial mass of 3\Msun, but it decreases towards higher initial masses. If we extrapolate their results, for our observed luminosity of $\log (L_1/\text{\Lsun}) = 3.96 \pm 0.14$, the relation yields $M_{1,i}=8.2\pm0.5$\Msun, showing that there is a range of possible initial masses that could satisfy the current conditions of the system.

\subsubsection{Case A mass transfer}

Another channel through which we can form stripped stars is through case A mass transfer \citep{wellstein+01}, as in the case of HR~6819 \citep{bodensteiner+20b}. In order to initiate mass transfer during the MS life of the primary star in most cases, the initial orbital period should be $\lesssim3$\,d \citep{wellstein+01, sen+22}. As an example, \citet[][see also \citealt{langer+20b}, their system \#1]{wellstein+01} presented, in their grid of binary evolutionary models, a system with initial properties $M_{1,i}=12$\Msun, $M_{2,i}=8$\Msun, and $P_{{\rm orb},i}=2$\,d, that reached an orbital period of 189\,d after case AB mass transfer ceased, and a luminosity comparable to that of the stripped star in VFTS~291. However, from our tests with MESA models, we were not able to reproduce the long orbital periods and low masses of the stripped star after a case AB mass transfer phase. We believe that the main difference lays in the treatment of overshooting. \citet{wellstein+01} noted that they did not include convective core overshooting in their models, while it is included in MESA. We have used the ‘‘exponential'' scheme (although we also tested the simpler ‘‘step'' one) with parameters $f_0=0.005$ and $f=0.01$ \citep[as in][]{bodensteiner+20b}. By reducing the overshooting, \citeauthor{wellstein+01} reduced the mass of the He core, producing less massive stripped stars. The final orbital period, after the mass transfer episode, is strongly affected by the final mass ratio, and therefore our more massive stripped stars prevent the models from reaching longer orbital periods. We cannot discard a case A mass transfer scenario, but we were not able to find a configuration to reach the large final mass ratios that would lead to the observed orbital period of our system.

\subsubsection{Possibility of the stripped star being at the shell-HeB phase}

It could be argued that the stripped star might be at the shell-HeB (SHeB) phase. Indeed, allowing the stripped star to go further in its evolution would give time to the companion to move through its new MS path, becoming cooler and more luminous. The decreased temperature would now be in agreement with our spectroscopic value within 2-$\sigma$, and \logg would also drop. In order to achieve this, we would have to lower the initial masses, which would have the effect of lowering the luminosity of both tracks, bringing the position of the companion in its evolutionary track back to lower luminosities while placing the stripped star at the SHeB phase. However, there are three problems with this approach: (1) by lowering the initial mass of the primary, the mass of the stripped star will also drop, increasing the conflict with the spectroscopic mass. (2) The mass of the stripped star drops, but also the radius increases, with which \logg of the stripped star would drop even further. (3) By lowering the initial mass we would increase the time it takes the system to reach its current configuration, increasing the difference with the estimated age for H301. In conclusion, while the SHeB phase scenario could reduce the conflict with our determined parameters for the companion from our spectroscopic analysis, the CHeB phase is a better match for the properties of the stripped star, whose physical properties are better constrained than for the companion.

There is also a third possibility. Due to the SHeB phase, the star expands and eventually fills its Roche lobe, starting a case BB mass transfer phase. If mass transfer is stable and the system can avoid merging, the star will once again contract at a higher luminosity \citep[e.g.][]{schurmann+22}. Depending on the mass of the CO core, this second contraction will occur during core-carbon burning or during the pre-white dwarf phase. In order to compare the plausibility of these three scenarios---contraction during CHeB, SHeB, and second contraction---we can use the $L/M$ ratio as defined by \citet[][their Eq. 4]{langer+kudritzki14}, which is proportional to $T_{\rm eff}^4/g$. From our spectroscopic results for the narrow-lined star (i.e., the stripped star), and using the lower limit on the mass, we find $L/M=4930\,L_{\odot} M_{\odot}^{-1}$, and for the three crossings at 13180\,K, in order of increasing luminosity, we find 5185, 7325, and 22590\,$L_{\odot} M_{\odot}^{-1}$, assuming $M_1=1.4$\Msun and $\log (L_1/L_{\odot})=4.5$ for the last crossing. These numbers strongly support the contraction during CHeB as the evolutionary state of the stripped star.

\subsection{Stripped stars disguised as regular B-type stars}

\citet{walborn+blades97} classified VFTS~291 as B5~:p, but from the higher S/N BBC spectra and comparing to 67 Oph, it is very easy to confuse the system with a normal B star in the Hertzsprung gap. Similar examples of such confusion have recently happened in the search for OB+BH systems. \citet{liu+19} reported the discovery of a $\sim$70\Msun BH in the binary LB-1. The primary star (the one dominating the spectrum) in LB-1 was classified as a B3~V star from its \spline{He}{i}{4471}/\spline{Mg}{ii}{4481} ratio and from the weak \spline{N}{ii}{3995} and \spline{Si}{iii}{4553} lines. A hot subdwarf B star \citep[sdB,][]{han+02, han+03, heber16} was ruled out due to the extension of the Balmer series (n$\approx$12) and the weakness of the \spline{He}{i}{4388} line in such stars. HR~6819 was proposed as a triple system containing a BH in the inner binary by \citet{rivinius+20}. The system presents many similarities to LB-1, except that in HR~6819 the H$\alpha$ emission was associated with a Be star in a wider orbit instead of a disk around the BH, as was the case for LB-1. The primary star was also classified as a B3 star, with a luminosity class III but with very similar \logg to LB-1. More recently, \citet{saracino+22} claimed the detection of a 11\Msun BH in the LMC system NGC~1850~BH1, again as a companion to a B-type star. All three systems are remarkably similar---despite the latter being in a short-period orbit and presumably in a semidetached configuration---they all present spectra with narrow spectral lines, near-circular orbits, with large semi-amplitude velocities for their orbital periods, similar effective temperatures between 14 and 18\,kK, and virtually the same surface gravity around \logg$\approx3.5$. However the three of them suffer from the same issue, the misidentification of the main visible star with a regular B-type star and the overestimation of their respective masses. 

In the case of LB-1, \citet{shenar+20} showed that the system could be explained as a 1.5\Msun stripped star with a 7\Msun Be companion \citep[see also][]{irrgang+20} and that the analysis by \citet{liu+19}, based on the anti-phase motion of the H$\alpha$ line, was affected by the wings of the absorption component of the B-type star \citep{abdul-masih+20}. HR~6819 can be explained in the same way without the need of a triple configuration, \citet{bodensteiner+20b} found a mass of 0.46\Msun for the stripped star, and as for LB-1, a fast rotating Be companion. Similar results were found by \citet{gies+wang20} and \citet{el-badry+quataert21}. This scenario was later confirmed by \citet{frost+22} who ruled out the presence of a wide companion with new VLT/MUSE observations while finding a non-degenerate star at $\sim$1.2 mas with interferometric observations using VLTI/GRAVITY. The short-period system NGC~1850~BH1 was also suggested to contain a stripped star by \citet{el-badry+burdge22}, with a mass between 0.65 and 1.5\Msun (which could be larger if the donor is not filling its Roche lobe). This system has been revised by \citet{saracino+23}, but doubts on the true nature of the unseen companion remain. See Table~\ref{tab:strpstars} for a comparison of the properties of the three previously discussed systems. 

\begin{table*}
\caption{Recently discovered stripped stars members of binary systems previous a sub-dwarf phase.}
\label{tab:strpstars}
\centering
\begin{tabular*}{\textwidth}{@{\extracolsep{\fill}\quad}lcccccccccc}
\toprule
\toprule
Parameter & \mc{4}{c}{VFTS~291} & \mc{2}{c}{LB-1} & \mc{2}{c}{HR~6819} & \mc{2}{c}{NGC~1850~BH1}\\
\midrule
Reference & \mc{4}{c}{This work} & \mc{2}{c}{S20} & \mc{2}{c}{B20} & \mc{2}{c}{S23} \\
Configuration  & \mc{4}{c}{strB\,+\,early B~V-IV} & \mc{2}{c}{strB\,+\,B3~Ve} & \mc{2}{c}{strB\,+\,B2-3~Ve} & \mc{2}{c}{strB\,+\,?} \\
\midrule
 & \multicolumn{10}{c}{Orbital parameters}\\
\midrule
$P$ (d)		 & \mc{4}{c}{$108.03\pm0.04$}  & \mc{2}{c}{$78.9\pm0.3$}  & \mc{2}{c}{$40.3\pm0.0$}  & \mc{2}{c}{$5.0\pm0.0$} \\
$e$			 & \mc{4}{c}{$0$ (fixed)} & \mc{2}{c}{$0.03\pm0.01$} & \mc{2}{c}{$0$ (fixed)} & \mc{2}{c}{$0.03\pm0.01$}\\
$K_1$ (km/s) & \mc{4}{c}{$93.6\pm0.2$}  & \mc{2}{c}{$52.9\pm0.1$}  & \mc{2}{c}{$60.4\pm1.0$}  & \mc{2}{c}{$175.6\pm2.6$}\\
$K_2$ (km/s) & \mc{4}{c}{$\lesssim15$}  & \mc{2}{c}{$11.2\pm1.0$}  & \mc{2}{c}{$4.0\pm0.8$}  & \mc{2}{c}{--}\\
$q$ (km/s) & \mc{4}{c}{$\gtrsim6$}  & \mc{2}{c}{$4.7\pm0.4$}  & \mc{2}{c}{$15\pm3$}  & \mc{2}{c}{--}\\
\midrule
 & \multicolumn{10}{c}{Physical parameters -- BH model}\\
\midrule
Original study & \mc{4}{c}{} & \mc{2}{c}{L19} & \mc{2}{c}{R20} & \mc{2}{c}{S22} \\
Spectral type  & \mc{4}{c}{} & \mc{2}{c}{B3~V} & \mc{2}{c}{B3~III} & \mc{2}{c}{B} \\
Configuration  & \mc{4}{c}{} & \mc{2}{c}{B+BH} & \mc{2}{c}{(B+BH)+Be} & \mc{2}{c}{B+BH} \\
$M_{1}$ (\Msun) & \mc{4}{c}{} & \mc{2}{c}{$8.2^{+0.9}_{-1.2}$ (Sp)} & \mc{2}{c}{$6.3\pm0.7$ (SpT)} & \mc{2}{c}{$4.9\pm0.4$ (EM)}\\
$M_{2, \rm min}$ (\Msun) & \mc{4}{c}{} & \mc{2}{c}{$6.3$} & \mc{2}{c}{$5.0\pm0.4$} & \mc{2}{c}{$5.34^{+0.55}_{-0.59}$} \\
$R_1$ (\Rsun) & \mc{4}{c}{} & \mc{2}{c}{$9\pm2$} & \mc{2}{c}{$5.5\pm0.5$} & \mc{2}{c}{6} \\
\Teff (kK)    & \mc{4}{c}{} & \mc{2}{c}{$18.1\pm0.8$} & \mc{2}{c}{16} & \mc{2}{c}{$14.5\pm0.5$}\\
\logg         & \mc{4}{c}{} & \mc{2}{c}{$3.43\pm0.15$}   & \mc{2}{c}{$3.5$} & \mc{2}{c}{3.57} \\
\vsini(\kms) & \mc{4}{c}{} & \mc{2}{c}{10} & \mc{2}{c}{--} & \mc{2}{c}{--} \\
\midrule
 & \multicolumn{10}{c}{Physical parameters -- strB star model}\\
\midrule
Reference & \mc{2}{c}{This work (Sp)} & \mc{2}{c}{This work (EM)} & \mc{2}{c}{S20} & \mc{2}{c}{B20} & \mc{2}{c}{EB22} \\
Component & strB & B & strB & B & strB & Be & strB & Be & \mc{2}{c}{strB} \\
$M_{\rm dyn}$ (\Msun) & $<2.7\pm1.2$ & -- & -- & -- & 1.5 & 7 & 0.46 & 7 & \mc{2}{c}{0.65-1.5} \\
$M_{\rm sp}$ (\Msun)  & $3.05^{+1.98}_{-1.20}$ & $2.51^{+3.51}_{-1.36}$ & -- & -- & 1.1 & 5 & 0.4  & 6 & \mc{2}{c}{--} \\

$M_{\rm ev}$ (\Msun)  & -- & $13.2\pm1.5$ & 1.80 & 14.33 & -- & -- & -- & -- & \mc{2}{c}{--} \\

$R$ (\Rsun) & $17.93^{+2.90}_{-2.49}$ & $7.27^{+1.27}_{-1.08}$ & 18.44 & 5.12 & 5.4 & 3.7 & 4.4 & 3.9 & \mc{2}{c}{4.9-6.5}\\
$R_{\rm SED}$ (\Rsun) & $17.69\pm1.21$ & $7.78^{+0.42}_{-0.45}$ & -- & -- & -- & -- & -- & -- & \mc{2}{c}{--}\\
$\log (L/{\text \Lsun})$ & $3.96\pm0.14$ & $4.41\pm0.16$ & 3.97 & 4.40 & 2.8 & 3.1 & 3.0 & 3.4 & \mc{2}{c}{3.1} \\
\Teff (kK)   & $13.18^{+1.43}_{-1.01}$ & $27.75^{+2.29}_{-3.42}$ & 13.20 & 32.06 & 12.7 & 18 & 16 & 20 & \mc{2}{c}{14.5} \\
\logg      & $2.42^{+0.17}_{-0.18}$ & $3.10^{+0.39}_{-0.29}$ & 2.16 & 4.18 & 3.0 & 4.0 & 2.8 & 4.0 & \mc{2}{c}{2.85-3} \\
\vsini(\kms) & $35.95^{+13.24}_{-14.43}$ & $271.47^{+242.04}_{-85.29}$ & -- & -- & 7 & 300 & $\lesssim$25 & 180 & \mc{2}{c}{--} \\
\bottomrule
\end{tabular*}
\tablefoot{References: \citealt{liu+19} (L19), \citealt{rivinius+20} (R20), \citealt{saracino+22} (S22), \citealt{shenar+20} (S20), \citealt{bodensteiner+20b} (B20), \citealt{el-badry+burdge22} (EB22). Source of determined parameters: from Spectral Analysis (Sp), from spectral types (SpT), from evolutionary models (EM). Other definitions: strB: Stripped star of type B; $M_{\rm dyn}$: dynamical mass; $M_{\rm sp}$: spectroscopic mass; $M_{\rm ev}$: evolutionary mass; $R_{\rm SED}$: radius derived from SED analysis.}
\end{table*}

Perhaps the main conclusion out of these findings is that bloated stripped stars can be easily mistaken with regular B-type stars. However, given their low masses, they should present large radial velocity variations and therefore can be also candidates to harbour BH companions. The evidence also points at the possibility that the predicted large numbers of binary interaction products have been hiding in plain sight among samples of B-type stars, and that the recent race to find dormant BHs has lead to their discovery. The $K$-$P$ diagram is a promising method to select good candidates to both, stripped stars and OB+BH systems, as shown in \citetalias{villasenor+21}, but there are some difficulties that need to be taken into account.

\subsection{Difficulties of finding stripped stars}

Based on the large multiplicity fraction of O-type stars in galactic young clusters \citep{sana+12}, and assuming constant star formation, \citet{demink+14} found that binary interaction products that have not merged, can account for 17\% of all massive stars with masses above 8\Msun, and that they represent 11\% of all systems detectable as binaries. This means that roughly one in every ten systems might contain stripped stars in samples of binary or multiple stars, from which we could expect up to 8 such systems in the BBC sample. 

However, the number of unambiguously identified binary interaction products is still very low despite the great efforts of observing increasingly large and homogeneous samples of massive stars. As discussed by \citet{demink+14}, this might be strongly related to selection effects, e.g. selecting against binaries for stellar atmosphere analysis. Also, most studies targeting binaries focus in young stellar regions to obtain the close-to-birth orbital properties of massive binaries, missing more evolved populations where one would expect to find a larger fractions of binary interaction products. This is the case of BBC and 30~Dor; it is not a coincidence that the first stripped star was found in the oldest cluster of the region, and therefore we can expect a number quite below the prediction of 11\% given the  age distribution of 30~Dor \citep{schneider+18}. Furthermore, monitoring programmes with a large number of epochs necessaries to characterise the orbits and enough S/N and resolving power to constrain the nature of the companions are scarce. 

One more factor influencing the low numbers of known post-mass-transfer systems is the short-duration phases following the end of RLOF; the time spent on the MS for a massive star is about 90\% of their total lifetimes, so a 15\Msun star will only live for about 1-1.5 Myr after core-hydrogen exhaustion. The case of VFTS~291 is indeed uncommon, the large radius determined from our spectral analysis implies that we have caught the system shortly after the end of the mass-transfer phase. Indeed, looking at the HRD of Fig.~\ref{fig:HRD} we can see that it would take just above 50\,000\,yr for the contracting star to reach its current radius, roughly a fourth of the time that it would take to reach a 5\Rsun radius, representative of the other detected bloated stripped stars. 

As a final example, from the study of \citet{cignoni+16} on H301, there are two other stars in the $\sim$14--15\,mag range, above the MS turn-off point, a theoretical region known as the post-MS gap that it is usually found to be populated in clusters of similar ages. If VFTS~291 is a post-interaction system as we propose here, it raises the question of whether those stars are binary interaction products as well, as noted also by \citeauthor{cignoni+16} These two stars, VFTS~270 and VFTS~293, were classified as a B3 supergiant and a B2e giant respectively by \citet{evans+15}, but were not identified as binaries and therefore were not followed up by the BBC programme. They could have lower-mass companions and/or long orbital periods, but their RVs ($\varv_{\rm r} =264.9 \pm 2.3\,$\kms and $\varv_{\rm r} =260.4 \pm 1.9\,$\kms respectively) are in agreement within 1-$\sigma$ with the value from \citet{patrick+19} for H301, and it would be difficult to make a case for individual high-resolution multi-epoch spectroscopy.

\subsection{From sdO to quasi WR stars}\label{ssec:sdO-qWR}

Novel work by \citet{goetberg+17} computed spectral models  using CMFGEN \citep{hillier+miller98} and MESA, to predict the morphology of the spectra of stripped stars produced by binary interactions. In this first paper they used a mock system with a progenitor initial mass of 12\Msun with a 5\Msun companion, motivated by the system HD~45166, a quasi Wolf Rayet (qWR) star plus a B7~V companion \citep{steiner+oliveira05, groh+08}. They evolved the system using MESA and used the stellar structure computed by MESA at the moment the core-helium fraction reached 0.5 to model the spectra with CMFGEN for different metallicities. \citet{goetberg+18} went further and repeated the exercise for a range of initial masses between 2 and 18.2\Msun which produced stripped star masses of 0.35--7.9\Msun, covering the full range of stripped objects, going from sdB to WR stars. They also used an initial mass ratio of $q = M2/M1 = 0.8$ and an orbital period that would guarantee Case B mass transfer. The emerging spectral models from \citet{goetberg+18} were classified into three groups: spectra dominated by emission lines (group E), presenting only absorption features (group A), and a group showing both emission and absorption lines (group A/E). The latter group is particularly interesting for us since it is composed by stripped stars between 1.8 and 5\Msun, this implies that VFTS~291 could present emission lines later in its CHeB phase, including \spline{He}{ii}{4686} and a P-Cygni profile for \spline{N}{v}{4604/20}, as well as strong UV emission lines expected for \spline{He}{ii}{1640}, the Lyman series, and several \ion{N}{v} and \ion{C}{iv} \citep[see][]{goetberg+18}. 

After taking into account the contribution from a wide range of stellar companions between 4 and 18.2\Msun which roughly corresponds to spectral types B5~V to O9~V, \citet{goetberg+18} concluded that searching for a UV excess produced by the stripped component in binary systems is the most effective way to find the majority of these objects (also emission features are a promising diagnostic in the case of the more massive stripped stars, see also \citealt{goetberg+17}). In fact, this has been the most successful technique used to find sdOB stars. \citet{wangL+18} reported the detection of 12 Be+sdO candidates plus the confirmation of four previously known systems using archival FUV spectra from the International Ultraviolet Explorer (IUE). Nine of the candidates were later confirmed by \citet{wangL+21}, adding one new detection previously reported by \citet{chojnowski+18}, increasing the number of known Be+sdO systems to 15, from which 6 have orbital solutions available\footnote{\citet{wangL+23} presented orbital solutions for 5 additional systems.}: FY CMa \citep{peters+08}, $\varphi$~Per \citep{mourard+15}, 59 Cyg \citep{peters+13}, 60 Cyg \citep{wangL+17}, MWC~522 \citep{chojnowski+18}, and V2119 Cyg \citep{klement+22a}.

\begin{figure}
\centering
    \includegraphics[width=\hsize]{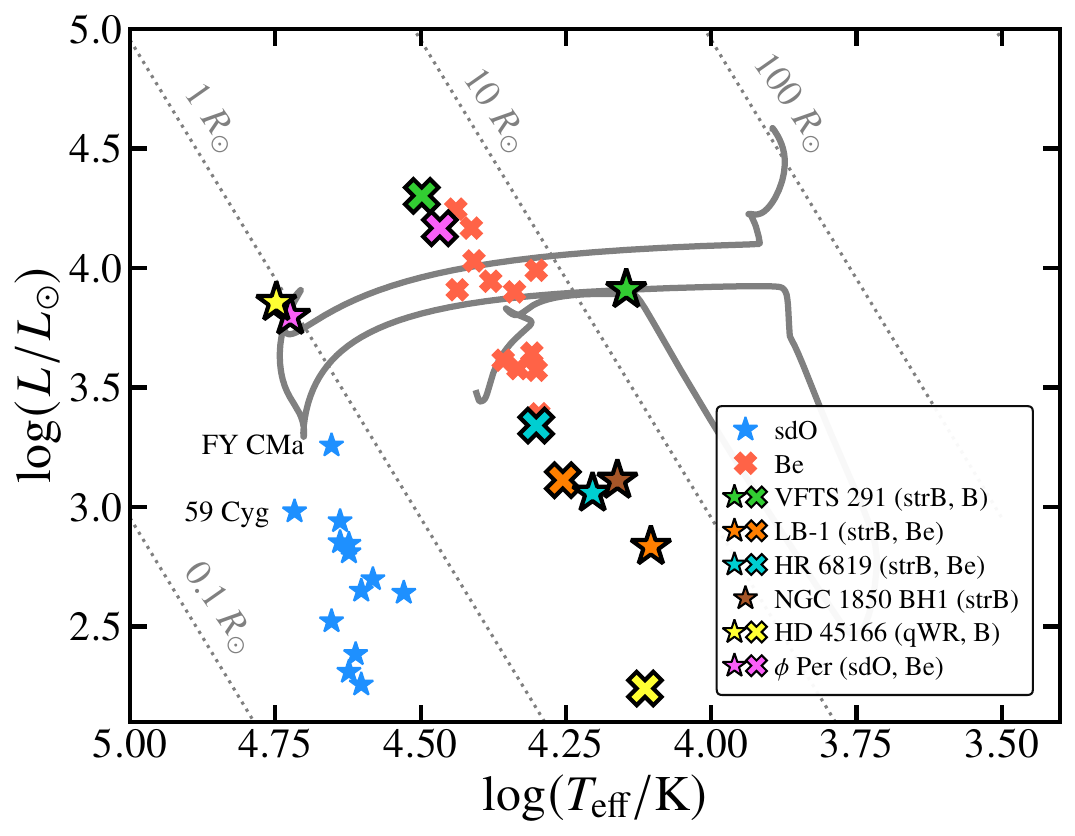}
    \caption{HRD showing the position of both components in known sdO+Be, strB+B/Be, and the only known quasi-WR star, see symbols in legend. The evolutionary track shown in Fig.~\ref{fig:HRD} for the stripped star in VFTS~291 is also shown in grey.}\label{fig:HRDsdO}
\end{figure}

However, these sdO stars---and the ones that are the focus of \citet{goetberg+17,goetberg+18}---are in an advanced state of CHeB and have already contracted and moved towards a hotter part of the HRD. As found by \citet{goetberg+18}, in order for them to strongly contribute to the composite spectrum of their systems they would have to be much more massive than the typical sdO mass and have evolved through non-conservative mass transfer so that the companion does not accrete a large fraction of the transferred mass. The case of VFTS~291 and the other newly discovered stripped stars is different. Since they have recently detached from their RL, they are still bloated and trying to regain thermal equilibrium. For this reason their narrow spectral lines dominate the spectrum; even though the companions are brighter, their fast rotation broadens and dilutes their lines. Their temperatures are also much lower than those of sdO stars, potentially ranging from ${\sim}10\,000$\,K to 30\,000\,K, given the contraction phase in the HRD of our model. These temperatures put these bloated stripped stars in the B-type domain, but they should be differentiated from regular B-type stars. For this reason, we have updated the spectral type of the stripped star in VFTS~291 to BN5: IIp, to highlight its nitrogen enrichment (N) and its peculiarities (p)\footnote{A compilation on qualifiers used in spectral classification can be found in \citet{sota+11}}. For simplicity, in the rest of this work we will refer to this class of objects as strB stars.

In Fig.~\ref{fig:HRDsdO}, a similar HRD to that in Fig.~\ref{fig:HRD} is shown, with the evolutionary track of the strB star in VFTS~291 in grey. The position of both components of VFTS~291 and of the other strB+B/Be systems (LB-1, HR~6819, NGC~1850~BH1) are also shown (star and cross symbols, respectively), and we have also added the known sdO+Be systems (with values from \citet{peters+08, peters+13, wangL+21, klement+22a}), $\varphi$~Per \citep{gies+98}, and the qWR+B system HD~45166 (Shenar et al., in press). All the B/Be companions are in a sequence of roughly equal radius, where we can also find the bloated stripped stars, except for VFTS~291, which has a larger radius and has detached from its Roche lobe more recently. LB-1, HR~6819, NGC~1850~BH1 can be seen as progenitors of the core of the sdO sample; they will reach luminosities roughly in the range 2.25-3 in log scale before the end of the CHeB phase. The stripped star in VFTS~291 however is more massive and it will be more luminous than the typical sdO stars, in many cases an order of magnitude more luminous. The more similar system to a potential descendant might be $\varphi$~Per (yellow star in the HRD), however, it appears to be less massive ($M_{\rm str}=1.2\pm0.2$\Msun \citet{mourard+15}) and overluminous for its mass. The sdO star in $\varphi$~Per and that in 59~Cyg (and possibly the one in FY CMa) were found to be helium shell burning objects by \citet{schootemeijer+18} and therefore more luminous than the CHeB sdO stars, although $\varphi$~Per has roughly the same mass as FY~CMa and it is much more luminous \citep[see discussion by][]{schootemeijer+18}. 

The other possibly related object is HD~45166, the qWR star was found to have a mass of $M_{\rm str}=4.2\pm0.7$\Msun by \citet{steiner+oliveira05} and it is one of the few known binary system together with VFTS~291 to contain a stripped star above the Chandrasekhar limit\footnote{During the revision of this manuscript, a stripped star with a mass $M_{\rm str}\approx3$\Msun was reported by \citet{ramachandran+23}, whereas \citet{drout+23} presented a sample of 25 stars from the Magellanic Clouds that are consistent with predictions for sdO stars with masses between 1 and 9\Msun \citep{goetberg+23}. Also worth noting, V2119~Cyg and LB-1 have determined masses of $1.62\pm0.28$\Msun and $1.5\pm0.4$\Msun respectively, so very close to the Chandrasekhar limit within errors. There is also the case of $\gamma$~Columbae \citep{irrgang+22}, a 5\Msun stripped star, but no companion has been found.}. The qWR star presents a very strong emission spectrum \citep{groh+08}, arguing for a more massive object than the one in VFTS~291 according to the predictions from \citet{goetberg+18}. However, Shenar et al. (in press) has found that HD~45166 is actually in a 22 yr orbit, and that the short periods found by \citet{steiner+oliveira05} might be due to pulsations, therefore this system could have a different origin from binary interactions. Furthermore, given the new orbital solution, they derived a mass of $2.0\pm0.4$\Msun for the qWR star, and explained its emission spectrum by the very strong magnetic properties of the star, making this a very unique object.

\subsection{The uniqueness of VFTS~291}

\subsubsection{The short-lived post mass transfer phase}

In the previous subsections, we have discussed some of the unique characteristics of VFTS~291. One of these is the extremely recent mass transfer episode. We showed in Fig.~\ref{fig:HRDsdO} that the stripped star is in a more bloated state than the other similar stripped stars, having detached from its Roche lobe only about 60\,000\,yr ago. It is possibly that some of the transferred material might have been lost from the system as a circumbinary nebula. If we consider a escape velocity of the system of 150\kms, the ejected material might have expanded through a region of about 12\,pc, with a density too low to produce strong IR excess \citep{deschamps+15}. Accordingly, we do not detect any significant IR excess in the $J$, $H$, and $K$ bands. If the expanding velocity of the ejected material was much lower than the escape velocity, and therefore it would extend to only a few parsecs from the system, the ionised features caused by the hot B-type companion could potentially be visibly in the form of emission lines of CNO, Fe, Si, He, and H elements \citep{deschamps+15}, but we only see the characteristic nebular emission of the Tarantula region in the \ion{He}{i} and Balmer lines, without departure from the systemic velocity.

The recent interaction also suggests that finding stripped stars in such bloated stage should be extremely rare. As a rough estimation, we can use the time span of each evolutionary phase from our MESA model to asses the likeliness of finding a system as VFTS~291. As expected, the highest chance of observing a system like this during its life span, is at the detached phase during the MS with a 86\%. The second longest lasting phase is the sdO phase with a 11\%, which we have defined as the moment from which the stripped star has contracted to a 1\Rsun until it has exhausted helium at its core. The next more likely phase in which it can be observed is at the ``visible'' stripped star phase, which is where VFTS~291, LB-1, and HR~6819 have been observed, with a probability of almost 2\%. We have defined this phase as the period between the end of the case B mass transfer and the sdO phase. Given the uncertainty in the last phases of evolution, after core-He exhaustion, we define this last phase until the second mass-transfer episode, where the stripped star has a probability of 0.8\% of being observed. The two less likely phases to observe the system are at the Hertzsprung gap and during case B mass transfer, with 0.3\% and 0.2\% respectively. 

It should be noted that although sdO stars should be a factor 5 more numerous than bloated stripped stars, their small radii make them only detectable in the UV due to their much higher temperatures, whereas strB stars can be more easily detected with optical spectroscopic surveys, where they can dominate the spectrum as we have learned. In the case of the BBC programme, we could probably expect one or two of these systems at most, but given the complex star-formation history of the region, detailed binary population synthesis studies would be required to make a better prediction.

\subsubsection{Fast rotation and the Be phenomenon}

One final important thing to note is that all companions to the previously mentioned sdO/strB stars, are fast rotating Be stars. However, we do not find evidence of Be-like emission in the spectrum of VFTS~291, nor does the companion seem to be rotating near critical rotation (despite the high uncertainties), challenging the paradigm that mass transfer leads to critical rotation \citep{packet81,pols+91,demink+13}.

In recent years, the scenario where most, if not all, Be stars are formed through binary interactions has gained strength, usually on observational grounds \citep{klement+19,bodensteiner+20c,dallas+22}. From the theoretical point of view, however, the single stellar evolutionary channel could still be relevant \citep{hastings+20,hastings+21}. Despite the general consensus regarding the link between binary interactions and the Be phenomenon, no Roche-lobe-filling companions have been found around classical Be stars \citep{rivinius+13}, although recently a similar system has been reported to be in a case-A mass transfer phase \citep{elbadry+22}, albeit with emission coming from an accretion disk. If mass transfer is responsible for the creation of most Be stars, the lack of mass-transferring binaries with Be stars could be due to the rapid nature of the mass transfer phase (although slow case A mass transfer can lasts for a few Myr), or, it could suggest that the Be behaviour appears after the mass transfer phase. In the case of VFTS~291, that means that a Be-like disk could still be developed. However, we are missing a mechanism that could spin up the companion to velocities closer to critical rotation, if most Be stars are indeed close to critical rotation \citep[>70-80\% of critical velocity,][]{hastings+20}. On the other side, we do not need to look much further for a counter example. \citetalias{villasenor+21} presented six Be binaries that were identified by \citet{evans+15} from their H$\alpha$ emission. Three of them have measured rotational velocities of \vsini$<200$\kms (one is a confirmed SB1 while the others were classified as possible SB1 systems), and all six of them have semi-amplitude velocities of less than 40\kms \citep[see also][]{dufton+22}, suggesting low-to-intermediate-mass companions. More intriguing, all six systems have eccentric orbits ($e>0.2$); if these were binary interaction products with stripped stars, we would expect a circular orbit due to mass transfer, unless the companion is a neutron star that could have caused the eccentricity due to a kick. While most of the sdO+Be systems with orbital solutions have circular orbits, we have found two exceptions in the literature \citep{peters+13, klement+22b}, where the presence of a third companion has been suggested to create perturbations leading to an increase in eccentricity. 

In summary, the B-type companion in VFTS~291 could still produce Be-like features, and there are examples in the literature of Be stars with low projected rotational velocities, but more information on their inclinations would be necessary to determine if VFTS~291 is an isolated case or not. Moreover, given our determined flux contribution from the secondary, $F_B/F_{\rm Tot}=0.38$, we determined an apparent $V$ magnitude for the companion of $m_{V,B} = 15.92 \pm 0.14$\,mag, which will approximately be the magnitude of the system once the strB star contracts and fades in the optical. This magnitude is close to the value determined by \citet{cignoni+16} for the MS turn-off of H301, a region of the CMD strongly populated by Be stars, consistent with the scenario where the companion evolves into a Be star. Independently of how common its rotational velocity is, and of its subsequent evolution, VFTS~291 rises the question of whether mass transfer could result in low rotational velocities of the accretor. High-resolution spectroscopy will certainly help to improve the results from the disentangling to better constrain the rotation rate of the companion. 

One more possibility worth considering, is that of the companion being obscured by residual nebulosity from the recent mass-transfer episode. In fact, the derived physical parameters of VFTS~291 are remarkably similar to those of the well studied $\beta$~Lyrae system. $\beta$~Lyr~A \citep[for a review see][]{harmanec02} is a nearly edge-on, short-period binary (12.9\,d), going through case AB mass transfer. The current accepted picture \citep[see also][]{mennickent+djurasevic13, mourard+18, broz+21} is that the accretor is hidden by a thick accretion disk, with perpendicular jet-like structures and a spherical shell, not allowing the direct observation of the early B-type companion, whereas the donor has already lost a large fraction of its mass ($\sim$7.5\Msun). It is possible that $\beta$~Lyr~A might look similar to VFTS~291 during the late phase of mass transfer, and that not all the mass from the disk was accreted by the companion, leaving material around the star once mass transfer ceased. This scenario could potentially explain why we do not see the companion as a fast rotating star, but also the narrow core of some of the spectral lines, i.e. these could be shell lines originating in the material still surrounding the companion. However, both components of VFTS~291 are well inside their Roche lobes, and any material surrounding the companion would have to survive for more than 50\,000 yr, with the additional downside that the disk would have to be close to edge-on. Speculatively, if this material could escape accretion, for example due to the fast rotation of the underlying companion star, it could become part of the future decretion disk present in the Be companions to the other strB and sdO stars.

In the case of the stripped star, the projected rotational velocity value is in agreement within errors with predictions from detailed binary evolution models for stripped stars \citep{schurmann+22}, that found values between 22 and 25\kms for models with temperatures close to 50\,000\,K, which is the expected temperature for VFTS~291 stripped star at the time of minimum radius. Also in agreement is the surface He mass fraction from \citet{schurmann+22} at 50\,000\,K, which lies at around 0.65-0.70, while in our model we found a surface He mass fraction of 0.68.

\section{Summary and conclusions} \label{sec:concl}

VFTS~291 is a B-type binary in 30~Dor observed by the BBC programme. It was initially labelled as a candidate B+BH system due to its large semi-amplitude velocity and the apparently lacking signal of a massive, non-degenerate, companion in its spectra. We have carried out an extensive analysis of the spectroscopic and photometric data available for the system, and next we present a summary of our work with our main conclusions:

\begin{itemize}
\renewcommand\labelitemi{--}
    \item VFTS~291 can be well explained as a binary interaction product consisting of a stripped star with a B-type companion that has gone through an episode of case-B mass transfer.
    \item To reach this conclusion we have combined the 29 BBC epochs of spectroscopy to disentangle the spectrum of the narrow-lined star and its companion.
    \item The spectral fitting of the disentangled spectrum resulted in a low mass for the narrow lined star of $3.05^{+1.98}_{-1.20}$\Msun and a radius of $17.93^{+2.90}_{-2.49}$\Rsun, suggesting that this star might have lost part of its envelope through an episode of mass transfer given its luminosity of $\log L/L_\odot = 3.96 \pm 0.14$. In the case of the companion, it was not possible to fit all its spectral features, leading to high uncertainties in some of its derived physical parameters, specifically its surface gravity and spectroscopic mass.
    \item Using the temperature and luminosity derived for the companion star, we were able to determine an evolutionary mass of $M_{\rm ev,2}=13.2\pm1.5$\Msun, allowing us to put constrains on the dynamical mass of the strB star, obtaining $M_{\rm dyn,1}<2.7\pm1.2$\Msun. If our deduced minimum mass ratio is correct, the upper limit on the mass of the stripped star can be further reduced to $M_{\rm dyn,1}<2.2\pm0.4$\Msun.
    \item The analysis of the photometry and OGLE light curve supported our proposed scenario of a stripped star with a B-type companion, while also suggesting a lower \Teff for the companion and a higher \logg and mass.
    \item Our modelling of the evolution of the system with MESA revealed that an initial mass of 8.12\Msun is a good match for the current properties of the stripped star, leading to a 1.8\Msun stripped star, close to our lower limit determined from the spectroscopic analysis, and supported the MS phase for the companion, although at a higher temperature. However, it is possible that other model could also reproduce the system adequately. 
    \item The system has finished its interaction only about 60\,000\,yr ago, therefore the stripped star is still bloated, mimicking a regular B-type star in the Hertzsprung gap, and it is now contracting as it moves towards the hotter part of the HRD.
    \item The age of the system of ~32\,Myr found from the evolutionary model is just above the upper limit found for the Hodge~301 cluster in other studies. VFTS~291 is located at the periphery of the cluster and its membership can still be debated due to the proximity of NGC~2070 and its similar systemic velocity.
    \item Although some uncertainties in the mass of the stripped star remain, it is one of the most massive stripped stars (excluding Wolf-Rayet stars) found to date. If the true mass of the stripped star is close to the maximum value ($\sim2.5$\Msun) it would possibly become a HMXB, but if it is closer to the lower limit, then it is more likely than it will end up as a white dwarf, eventually merging with the B-type star during a common-envelope phase.
    \item It belongs to a class of objects that have only recently been identified; it joins LB-1, HR~6819, and possibly NGC~1850 BH1, as systems in an early post-mass-transfer phase, where the stripped star is contracting and increasing its temperature towards the sdO phase, trying to regain thermal equilibrium, while helium is being burned at the core. They are difficult to observe since this is a short-lived phase, lasting less than 1\,Myr before reaching a 1\Rsun radius. After that point, the most long-lived phase after case-B mass transfer takes place, when the fraction of helium at the core is close to 0.5. It would be expected that most of the stripped stars are observed during this longer-lasting subdwarf phase, however, those stripped stars are no longer detectable through optical spectroscopy and have only been found using UV observations.
    \item Our study is inconclusive regarding the rotational velocity of the companion star. The spectral fitting resulted in a wide range of high rotational velocities due to the difficulties encountered when fitting the profile of the Balmer and \ion{He}{i} lines. The inconsistent profiles are likely an outcome of the disentangling process, caused by the slow motion of the companion, nebular contamination, and the quality of our data. High-resolution spectroscopy would help to reduce all these effects, e.g. with UVES we would go from a resolution element of 50\kms in the case of FLAMES to $\sim7$\kms, which would allow the better detection of the companion and of the nebular contribution, substantially improving the results of the disentangling and possibly solving the conflict between \Teff and \logg values obtained from spectroscopy and photometry with those from our evolutionary model.
\end{itemize}

%%%%%%%%%%%%%%%%%%%%%%%%%%%%%%%%%%%%%%%%%%%%%%%%%%
\section*{Acknowledgements}

We are deeply grateful to I. Howarth for kindly providing us the atmosphere models used for the spectral fitting, to S. Simón-Díaz for his useful comments on the rotational velocity analysis, and to M. Abdul-Masih and K. Conroy for their valuable help in the use of Phoebe. We also thank the anonymous referee for the careful review of this manuscript and helpful suggestions. In particular, we are grateful for Fig.~\ref{fig:M1M2plane}, which was suggested by the referee. The authors acknowledge support from the European Research Council (ERC) innovation programme of the Horizon 2020, programme DLV-772225-MULTIPLES. TS acknowledges support from the European Union's Horizon 2020 under the Marie Skłodowska-Curie grant agreement No 101024605. SdM acknowledges funding by the Netherlands Organization for Scientific Research (NWO) as part of the Vidi research program BinWaves with project number 639.042.728. This research was funded in part by the National Science Centre, Poland, grant no. 2022/45/B/ST9/00243.

%%%%%%%%%%%%%%%%%%%%%%%%%%%%%%%%%%%%%%%%%%%%%%%%%%
\section*{Data Availability}

The spectrocopic data from the BBC programme is publicly available through the ESO archive. OGLE light curve is available upon request. The python tools used in this work and in \citetalias{villasenor+21} are available as part of the  Massive bINaries Analysis TOols (MINATO) package at \url{https://github.com/jvillasr/MINATO}.



\bibliographystyle{mnras}
\bibliography{JVbiblio} % if your bibtex file is called example.bib

\begin{thebibliography}{}
\makeatletter
\relax
\def\mn@urlcharsother{\let\do\@makeother \do\$\do\&\do\#\do\^\do\_\do\%\do\~}
\def\mn@doi{\begingroup\mn@urlcharsother \@ifnextchar [ {\mn@doi@}
  {\mn@doi@[]}}
\def\mn@doi@[#1]#2{\def\@tempa{#1}\ifx\@tempa\@empty \href
  {http://dx.doi.org/#2} {doi:#2}\else \href {http://dx.doi.org/#2} {#1}\fi
  \endgroup}
\def\mn@eprint#1#2{\mn@eprint@#1:#2::\@nil}
\def\mn@eprint@arXiv#1{\href {http://arxiv.org/abs/#1} {{\tt arXiv:#1}}}
\def\mn@eprint@dblp#1{\href {http://dblp.uni-trier.de/rec/bibtex/#1.xml}
  {dblp:#1}}
\def\mn@eprint@#1:#2:#3:#4\@nil{\def\@tempa {#1}\def\@tempb {#2}\def\@tempc
  {#3}\ifx \@tempc \@empty \let \@tempc \@tempb \let \@tempb \@tempa \fi \ifx
  \@tempb \@empty \def\@tempb {arXiv}\fi \@ifundefined
  {mn@eprint@\@tempb}{\@tempb:\@tempc}{\expandafter \expandafter \csname
  mn@eprint@\@tempb\endcsname \expandafter{\@tempc}}}

\bibitem[\protect\citeauthoryear{{Abdul-Masih} et~al.,}{{Abdul-Masih}
  et~al.}{2019}]{abdul-masih+19}
{Abdul-Masih} M.,  et~al., 2019, \mn@doi [\apj] {10.3847/1538-4357/ab24d4},
  \href {https://ui.adsabs.harvard.edu/abs/2019ApJ...880..115A} {880, 115}

\bibitem[\protect\citeauthoryear{{Abdul-Masih} et~al.,}{{Abdul-Masih}
  et~al.}{2020}]{abdul-masih+20}
{Abdul-Masih} M.,  et~al., 2020, \mn@doi [\nat] {10.1038/s41586-020-2216-x},
  \href {https://ui.adsabs.harvard.edu/abs/2020Natur.580E..11A} {580, E11}

\bibitem[\protect\citeauthoryear{{Almeida} et~al.,}{{Almeida}
  et~al.}{2017}]{almeida+17}
{Almeida} L.~A.,  et~al., 2017, \mn@doi [\aap] {10.1051/0004-6361/201629844},
  \href {http://adsabs.harvard.edu/abs/2017A%26A...598A..84A} {598, A84}

\bibitem[\protect\citeauthoryear{{Astropy Collaboration} et~al.,}{{Astropy
  Collaboration} et~al.}{2022}]{astropy:2022}
{Astropy Collaboration} et~al., 2022, \mn@doi [apj] {10.3847/1538-4357/ac7c74},
  \href {https://ui.adsabs.harvard.edu/abs/2022ApJ...935..167A} {935, 167}

\bibitem[\protect\citeauthoryear{{Banyard}, {Sana}, {Mahy}, {Bodensteiner},
  {Villase{\~n}or}  \& {Evans}}{{Banyard} et~al.}{2022}]{banyard+22}
{Banyard} G.,  {Sana} H.,  {Mahy} L.,  {Bodensteiner} J.,  {Villase{\~n}or}
  J.~I.,   {Evans} C.~J.,  2022, \mn@doi [\aap] {10.1051/0004-6361/202141037},
  \href {https://ui.adsabs.harvard.edu/abs/2022A&A...658A..69B} {658, A69}

\bibitem[\protect\citeauthoryear{{Barb{\'a}}, {Gamen}, {Arias}  \&
  {Morrell}}{{Barb{\'a}} et~al.}{2017}]{barba+17}
{Barb{\'a}} R.~H.,  {Gamen} R.,  {Arias} J.~I.,   {Morrell} N.~I.,  2017, in
  {Eldridge} J.~J.,  {Bray} J.~C.,  {McClelland} L.~A.~S.,   {Xiao} L.,  eds,
  IAU Symposium Vol. 329, The Lives and Death-Throes of Massive Stars. pp
  89--96, \mn@doi{10.1017/S1743921317003258}

\bibitem[\protect\citeauthoryear{{Bodensteiner}, {Shenar}  \&
  {Sana}}{{Bodensteiner} et~al.}{2020a}]{bodensteiner+20c}
{Bodensteiner} J.,  {Shenar} T.,   {Sana} H.,  2020a, \mn@doi [\aap]
  {10.1051/0004-6361/202037640}, \href
  {https://ui.adsabs.harvard.edu/abs/2020A&A...641A..42B} {641, A42}

\bibitem[\protect\citeauthoryear{{Bodensteiner} et~al.,}{{Bodensteiner}
  et~al.}{2020b}]{bodensteiner+20b}
{Bodensteiner} J.,  et~al., 2020b, \mn@doi [\aap]
  {10.1051/0004-6361/202038682}, \href
  {https://ui.adsabs.harvard.edu/abs/2020A&A...641A..43B} {641, A43}

\bibitem[\protect\citeauthoryear{{Bordier}, {Frost}, {Sana}, {Reggiani},
  {M{\'e}rand}, {Rainot}, {Ram{\'\i}rez-Tannus}  \& {de Wit}}{{Bordier}
  et~al.}{2022}]{bordier+22}
{Bordier} E.,  {Frost} A.~J.,  {Sana} H.,  {Reggiani} M.,  {M{\'e}rand} A.,
  {Rainot} A.,  {Ram{\'\i}rez-Tannus} M.~C.,   {de Wit} W.~J.,  2022, \mn@doi
  [\aap] {10.1051/0004-6361/202141849}, \href
  {https://ui.adsabs.harvard.edu/abs/2022A&A...663A..26B} {663, A26}

\bibitem[\protect\citeauthoryear{{Brands}, {de Koter}, {Bestenlehner},
  {Crowther}, {Kaper}, {Caballero-Nieves}  \& {Gr{\"a}fener}}{{Brands}
  et~al.}{2023}]{brands+23}
{Brands} S.~A.,  {de Koter} A.,  {Bestenlehner} J.~M.,  {Crowther} P.~A.,
  {Kaper} L.,  {Caballero-Nieves} S.~M.,   {Gr{\"a}fener} G.,  2023, \mn@doi
  [arXiv e-prints] {10.48550/arXiv.2303.09374}, \href
  {https://ui.adsabs.harvard.edu/abs/2023arXiv230309374B} {p. arXiv:2303.09374}

\bibitem[\protect\citeauthoryear{{Braun} \& {Langer}}{{Braun} \&
  {Langer}}{1995}]{braun+langer95}
{Braun} H.,  {Langer} N.,  1995, \aap, \href
  {https://ui.adsabs.harvard.edu/abs/1995A&A...297..483B} {297, 483}

\bibitem[\protect\citeauthoryear{{Britavskiy} et~al.,}{{Britavskiy}
  et~al.}{2019}]{britavskiy+19}
{Britavskiy} N.,  et~al., 2019, \mn@doi [\aap] {10.1051/0004-6361/201834564},
  \href {https://ui.adsabs.harvard.edu/abs/2019A&A...624A.128B} {624, A128}

\bibitem[\protect\citeauthoryear{{Brott} et~al.,}{{Brott}
  et~al.}{2011}]{brott+11a}
{Brott} I.,  et~al., 2011, \mn@doi [\aap] {10.1051/0004-6361/201016113}, \href
  {http://adsabs.harvard.edu/abs/2011A%26A...530A.115B} {530, A115}

\bibitem[\protect\citeauthoryear{{Bro{\v{z}}} et~al.,}{{Bro{\v{z}}}
  et~al.}{2021}]{broz+21}
{Bro{\v{z}}} M.,  et~al., 2021, \mn@doi [\aap] {10.1051/0004-6361/202039035},
  \href {https://ui.adsabs.harvard.edu/abs/2021A&A...645A..51B} {645, A51}

\bibitem[\protect\citeauthoryear{{Casares}, {Negueruela}, {Rib{\'o}}, {Ribas},
  {Paredes}, {Herrero}  \& {Sim{\'o}n-D{\'\i}az}}{{Casares}
  et~al.}{2014}]{casares+14}
{Casares} J.,  {Negueruela} I.,  {Rib{\'o}} M.,  {Ribas} I.,  {Paredes} J.~M.,
  {Herrero} A.,   {Sim{\'o}n-D{\'\i}az} S.,  2014, \mn@doi [\nat]
  {10.1038/nature12916}, \href
  {https://ui.adsabs.harvard.edu/abs/2014Natur.505..378C} {505, 378}

\bibitem[\protect\citeauthoryear{{Chojnowski} et~al.,}{{Chojnowski}
  et~al.}{2018}]{chojnowski+18}
{Chojnowski} S.~D.,  et~al., 2018, \mn@doi [\apj] {10.3847/1538-4357/aad964},
  \href {https://ui.adsabs.harvard.edu/abs/2018ApJ...865...76C} {865, 76}

\bibitem[\protect\citeauthoryear{{Cignoni} et~al.,}{{Cignoni}
  et~al.}{2016}]{cignoni+16}
{Cignoni} M.,  et~al., 2016, \mn@doi [\apj] {10.3847/1538-4357/833/2/154},
  \href {https://ui.adsabs.harvard.edu/abs/2016ApJ...833..154C} {833, 154}

\bibitem[\protect\citeauthoryear{{Conroy} et~al.,}{{Conroy}
  et~al.}{2020}]{conroy+20}
{Conroy} K.~E.,  et~al., 2020, \mn@doi [\apjs] {10.3847/1538-4365/abb4e2},
  \href {https://ui.adsabs.harvard.edu/abs/2020ApJS..250...34C} {250, 34}

\bibitem[\protect\citeauthoryear{{Cutri} et~al.,}{{Cutri}
  et~al.}{2003}]{cutri+03}
{Cutri} R.~M.,  et~al., 2003, {2MASS All-Sky Catalog of Point Sources.
  Available at: http://irsa.ipac.caltech.edu/applications/Gator/}

\bibitem[\protect\citeauthoryear{{Dallas}, {Oey}  \& {Castro}}{{Dallas}
  et~al.}{2022}]{dallas+22}
{Dallas} M.~M.,  {Oey} M.~S.,   {Castro} N.,  2022, \mn@doi [\apj]
  {10.3847/1538-4357/ac8988}, \href
  {https://ui.adsabs.harvard.edu/abs/2022ApJ...936..112D} {936, 112}

\bibitem[\protect\citeauthoryear{{De Marchi} et~al.,}{{De Marchi}
  et~al.}{2016}]{DeMarchi+16}
{De Marchi} G.,  et~al., 2016, \mn@doi [\mnras] {10.1093/mnras/stv2528}, \href
  {https://ui.adsabs.harvard.edu/abs/2016MNRAS.455.4373D} {455, 4373}

\bibitem[\protect\citeauthoryear{{Deschamps}, {Braun}, {Jorissen}, {Siess},
  {Baes}  \& {Camps}}{{Deschamps} et~al.}{2015}]{deschamps+15}
{Deschamps} R.,  {Braun} K.,  {Jorissen} A.,  {Siess} L.,  {Baes} M.,   {Camps}
  P.,  2015, \mn@doi [\aap] {10.1051/0004-6361/201424772}, \href
  {https://ui.adsabs.harvard.edu/abs/2015A&A...577A..55D} {577, A55}

\bibitem[\protect\citeauthoryear{{Didelon}}{{Didelon}}{1982}]{didelon82}
{Didelon} P.,  1982, \aaps, \href
  {https://ui.adsabs.harvard.edu/abs/1982A&AS...50..199D} {50, 199}

\bibitem[\protect\citeauthoryear{{Drout}, {G{\"o}tberg}, {Ludwig}, {Groh}, {de
  Mink}, {O'Grady}  \& {Smith}}{{Drout} et~al.}{2023}]{drout+23}
{Drout} M.~R.,  {G{\"o}tberg} Y.,  {Ludwig} B.~A.,  {Groh} J.~H.,  {de Mink}
  S.~E.,  {O'Grady} A.~J.~G.,   {Smith} N.,  2023, \mn@doi [arXiv e-prints]
  {10.48550/arXiv.2307.00061}, \href
  {https://ui.adsabs.harvard.edu/abs/2023arXiv230700061D} {p. arXiv:2307.00061}

\bibitem[\protect\citeauthoryear{{Dufton}, {Lennon}, {Villase{\~n}or},
  {Howarth}, {Evans}, {de Mink}, {Sana}  \& {Taylor}}{{Dufton}
  et~al.}{2022}]{dufton+22}
{Dufton} P.~L.,  {Lennon} D.~J.,  {Villase{\~n}or} J.~I.,  {Howarth} I.~D.,
  {Evans} C.~J.,  {de Mink} S.~E.,  {Sana} H.,   {Taylor} W.~D.,  2022, \mn@doi
  [\mnras] {10.1093/mnras/stac630}, \href
  {https://ui.adsabs.harvard.edu/abs/2022MNRAS.512.3331D} {512, 3331}

\bibitem[\protect\citeauthoryear{{Dunstall} et~al.,}{{Dunstall}
  et~al.}{2015}]{dunstall+15}
{Dunstall} P.~R.,  et~al., 2015, \mn@doi [\aap] {10.1051/0004-6361/201526192},
  \href {http://adsabs.harvard.edu/abs/2015A%26A...580A..93D} {580, A93}

\bibitem[\protect\citeauthoryear{{El-Badry} \& {Burdge}}{{El-Badry} \&
  {Burdge}}{2022}]{el-badry+burdge22}
{El-Badry} K.,  {Burdge} K.~B.,  2022, \mn@doi [\mnras]
  {10.1093/mnrasl/slab135}, \href
  {https://ui.adsabs.harvard.edu/abs/2022MNRAS.511L..24E} {511, 24}

\bibitem[\protect\citeauthoryear{{El-Badry} \& {Quataert}}{{El-Badry} \&
  {Quataert}}{2021}]{el-badry+quataert21}
{El-Badry} K.,  {Quataert} E.,  2021, \mn@doi [\mnras] {10.1093/mnras/stab285},
  \href {https://ui.adsabs.harvard.edu/abs/2021MNRAS.502.3436E} {502, 3436}

\bibitem[\protect\citeauthoryear{{El-Badry} et~al.,}{{El-Badry}
  et~al.}{2022}]{elbadry+22}
{El-Badry} K.,  et~al., 2022, \mn@doi [\mnras] {10.1093/mnras/stac2422}, \href
  {https://ui.adsabs.harvard.edu/abs/2022MNRAS.516.3602E} {516, 3602}

\bibitem[\protect\citeauthoryear{{Eldridge}, {Fraser}, {Smartt}, {Maund}  \&
  {Crockett}}{{Eldridge} et~al.}{2013}]{eldridge+13}
{Eldridge} J.~J.,  {Fraser} M.,  {Smartt} S.~J.,  {Maund} J.~R.,   {Crockett}
  R.~M.,  2013, \mn@doi [\mnras] {10.1093/mnras/stt1612}, \href
  {https://ui.adsabs.harvard.edu/abs/2013MNRAS.436..774E} {436, 774}

\bibitem[\protect\citeauthoryear{{Evans} et~al.,}{{Evans}
  et~al.}{2005}]{evans+05}
{Evans} C.~J.,  et~al., 2005, \mn@doi [\aap] {10.1051/0004-6361:20042446},
  \href {http://adsabs.harvard.edu/abs/2005A%26A...437..467E} {437, 467}

\bibitem[\protect\citeauthoryear{{Evans}, {Lennon}, {Smartt}  \&
  {Trundle}}{{Evans} et~al.}{2006}]{evans+06}
{Evans} C.~J.,  {Lennon} D.~J.,  {Smartt} S.~J.,   {Trundle} C.,  2006, \mn@doi
  [\aap] {10.1051/0004-6361:20064988}, \href
  {http://adsabs.harvard.edu/abs/2006A%26A...456..623E} {456, 623}

\bibitem[\protect\citeauthoryear{{Evans} et~al.,}{{Evans}
  et~al.}{2011}]{evans+11}
{Evans} C.~J.,  et~al., 2011, \mn@doi [\aap] {10.1051/0004-6361/201116782},
  \href {http://adsabs.harvard.edu/abs/2011A%26A...530A.108E} {530, A108}

\bibitem[\protect\citeauthoryear{{Evans} et~al.,}{{Evans}
  et~al.}{2015}]{evans+15}
{Evans} C.~J.,  et~al., 2015, \mn@doi [A$\&$A] {10.1051/0004-6361/201424414},
  \href {http://adsabs.harvard.edu/abs/2015A%26A...574A..13E} {574, A13}

\bibitem[\protect\citeauthoryear{{Fahrion} \& {De Marchi}}{{Fahrion} \& {De
  Marchi}}{2023}]{fahrion+demarchi23}
{Fahrion} K.,  {De Marchi} G.,  2023, \mn@doi [\aap]
  {10.1051/0004-6361/202346240}, \href
  {https://ui.adsabs.harvard.edu/abs/2023A&A...671L..14F} {671, L14}

\bibitem[\protect\citeauthoryear{{Flower}}{{Flower}}{1996}]{flowers96}
{Flower} P.~J.,  1996, \mn@doi [\apj] {10.1086/177785}, \href
  {https://ui.adsabs.harvard.edu/abs/1996ApJ...469..355F} {469, 355}

\bibitem[\protect\citeauthoryear{{Frost} et~al.,}{{Frost}
  et~al.}{2022}]{frost+22}
{Frost} A.~J.,  et~al., 2022, \mn@doi [\aap] {10.1051/0004-6361/202143004},
  \href {https://ui.adsabs.harvard.edu/abs/2022A&A...659L...3F} {659, L3}

\bibitem[\protect\citeauthoryear{{Gao}, {Jiang}, {Li}  \& {Xue}}{{Gao}
  et~al.}{2013}]{gao+13}
{Gao} J.,  {Jiang} B.~W.,  {Li} A.,   {Xue} M.~Y.,  2013, \mn@doi [\apj]
  {10.1088/0004-637X/776/1/7}, \href
  {https://ui.adsabs.harvard.edu/abs/2013ApJ...776....7G} {776, 7}

\bibitem[\protect\citeauthoryear{{Garland} et~al.,}{{Garland}
  et~al.}{2017}]{garland+17}
{Garland} R.,  et~al., 2017, \mn@doi [\aap] {10.1051/0004-6361/201629982},
  \href {https://ui.adsabs.harvard.edu/abs/2017A&A...603A..91G} {603, A91}

\bibitem[\protect\citeauthoryear{{Gies} \& {Wang}}{{Gies} \&
  {Wang}}{2020}]{gies+wang20}
{Gies} D.~R.,  {Wang} L.,  2020, \mn@doi [\apjl] {10.3847/2041-8213/aba51c},
  \href {https://ui.adsabs.harvard.edu/abs/2020ApJ...898L..44G} {898, L44}

\bibitem[\protect\citeauthoryear{{Gies}, {Bagnuolo}, {Ferrara}, {Kaye},
  {Thaller}, {Penny}  \& {Peters}}{{Gies} et~al.}{1998}]{gies+98}
{Gies} D.~R.,  {Bagnuolo} William~G. J.,  {Ferrara} E.~C.,  {Kaye} A.~B.,
  {Thaller} M.~L.,  {Penny} L.~R.,   {Peters} G.~J.,  1998, \mn@doi [\apj]
  {10.1086/305113}, \href
  {https://ui.adsabs.harvard.edu/abs/1998ApJ...493..440G} {493, 440}

\bibitem[\protect\citeauthoryear{{Gonz{\'a}lez} \& {Levato}}{{Gonz{\'a}lez} \&
  {Levato}}{2006}]{gonzalez+levato06}
{Gonz{\'a}lez} J.~F.,  {Levato} H.,  2006, \mn@doi [\aap]
  {10.1051/0004-6361:20053177}, \href
  {https://ui.adsabs.harvard.edu/abs/2006A&A...448..283G} {448, 283}

\bibitem[\protect\citeauthoryear{{Gordon}, {Clayton}, {Misselt}, {Landolt}  \&
  {Wolff}}{{Gordon} et~al.}{2003}]{gordon+03}
{Gordon} K.~D.,  {Clayton} G.~C.,  {Misselt} K.~A.,  {Landolt} A.~U.,   {Wolff}
  M.~J.,  2003, \mn@doi [\apj] {10.1086/376774}, \href
  {https://ui.adsabs.harvard.edu/abs/2003ApJ...594..279G} {594, 279}

\bibitem[\protect\citeauthoryear{{G{\"o}tberg}, {de Mink}  \&
  {Groh}}{{G{\"o}tberg} et~al.}{2017}]{goetberg+17}
{G{\"o}tberg} Y.,  {de Mink} S.~E.,   {Groh} J.~H.,  2017, \mn@doi [\aap]
  {10.1051/0004-6361/201730472}, \href
  {https://ui.adsabs.harvard.edu/abs/2017A&A...608A..11G} {608, A11}

\bibitem[\protect\citeauthoryear{{G{\"o}tberg}, {de Mink}, {Groh}, {Kupfer},
  {Crowther}, {Zapartas}  \& {Renzo}}{{G{\"o}tberg} et~al.}{2018}]{goetberg+18}
{G{\"o}tberg} Y.,  {de Mink} S.~E.,  {Groh} J.~H.,  {Kupfer} T.,  {Crowther}
  P.~A.,  {Zapartas} E.,   {Renzo} M.,  2018, \mn@doi [\aap]
  {10.1051/0004-6361/201732274}, \href
  {https://ui.adsabs.harvard.edu/abs/2018A&A...615A..78G} {615, A78}

\bibitem[\protect\citeauthoryear{{Gotberg} et~al.,}{{Gotberg}
  et~al.}{2023}]{goetberg+23}
{Gotberg} Y.,  et~al., 2023, \mn@doi [arXiv e-prints]
  {10.48550/arXiv.2307.00074}, \href
  {https://ui.adsabs.harvard.edu/abs/2023arXiv230700074G} {p. arXiv:2307.00074}

\bibitem[\protect\citeauthoryear{{Grebel} \& {Chu}}{{Grebel} \&
  {Chu}}{2000}]{grebel+chu00}
{Grebel} E.~K.,  {Chu} Y.-H.,  2000, \mn@doi [\aj] {10.1086/301218}, \href
  {https://ui.adsabs.harvard.edu/abs/2000AJ....119..787G} {119, 787}

\bibitem[\protect\citeauthoryear{{Groh}, {Oliveira}  \& {Steiner}}{{Groh}
  et~al.}{2008}]{groh+08}
{Groh} J.~H.,  {Oliveira} A.~S.,   {Steiner} J.~E.,  2008, \mn@doi [\aap]
  {10.1051/0004-6361:200809511}, \href
  {https://ui.adsabs.harvard.edu/abs/2008A&A...485..245G} {485, 245}

\bibitem[\protect\citeauthoryear{{Han}, {Podsiadlowski}, {Maxted}, {Marsh}  \&
  {Ivanova}}{{Han} et~al.}{2002}]{han+02}
{Han} Z.,  {Podsiadlowski} P.,  {Maxted} P.~F.~L.,  {Marsh} T.~R.,   {Ivanova}
  N.,  2002, \mn@doi [\mnras] {10.1046/j.1365-8711.2002.05752.x}, \href
  {https://ui.adsabs.harvard.edu/abs/2002MNRAS.336..449H} {336, 449}

\bibitem[\protect\citeauthoryear{{Han}, {Podsiadlowski}, {Maxted}  \&
  {Marsh}}{{Han} et~al.}{2003}]{han+03}
{Han} Z.,  {Podsiadlowski} P.,  {Maxted} P.~F.~L.,   {Marsh} T.~R.,  2003,
  \mn@doi [\mnras] {10.1046/j.1365-8711.2003.06451.x}, \href
  {https://ui.adsabs.harvard.edu/abs/2003MNRAS.341..669H} {341, 669}

\bibitem[\protect\citeauthoryear{{Harmanec}}{{Harmanec}}{2002}]{harmanec02}
{Harmanec} P.,  2002, \mn@doi [Astronomische Nachrichten]
  {10.1002/1521-3994(200207)323:2<87::AID-ASNA87>3.0.CO;2-P}, \href
  {https://ui.adsabs.harvard.edu/abs/2002AN....323...87H} {323, 87}

\bibitem[\protect\citeauthoryear{{Hastings}, {Wang}  \& {Langer}}{{Hastings}
  et~al.}{2020}]{hastings+20}
{Hastings} B.,  {Wang} C.,   {Langer} N.,  2020, \mn@doi [\aap]
  {10.1051/0004-6361/201937018}, \href
  {https://ui.adsabs.harvard.edu/abs/2020A&A...633A.165H} {633, A165}

\bibitem[\protect\citeauthoryear{{Hastings}, {Langer}, {Wang}, {Schootemeijer}
  \& {Milone}}{{Hastings} et~al.}{2021}]{hastings+21}
{Hastings} B.,  {Langer} N.,  {Wang} C.,  {Schootemeijer} A.,   {Milone} A.~P.,
   2021, \mn@doi [\aap] {10.1051/0004-6361/202141269}, \href
  {https://ui.adsabs.harvard.edu/abs/2021A&A...653A.144H} {653, A144}

\bibitem[\protect\citeauthoryear{{Heber}}{{Heber}}{2016}]{heber16}
{Heber} U.,  2016, \mn@doi [\pasp] {10.1088/1538-3873/128/966/082001}, \href
  {https://ui.adsabs.harvard.edu/abs/2016PASP..128h2001H} {128, 082001}

\bibitem[\protect\citeauthoryear{{Hilditch}}{{Hilditch}}{2001}]{hilditch01}
{Hilditch} R.~W.,  2001, {An Introduction to Close Binary Stars. Cambridge
  University Press, Cambridge, UK}

\bibitem[\protect\citeauthoryear{{Hillier} \& {Miller}}{{Hillier} \&
  {Miller}}{1998}]{hillier+miller98}
{Hillier} D.~J.,  {Miller} D.~L.,  1998, \mn@doi [\apj] {10.1086/305350}, \href
  {https://ui.adsabs.harvard.edu/abs/1998ApJ...496..407H} {496, 407}

\bibitem[\protect\citeauthoryear{{Holgado}, {Sim{\'o}n-D{\'\i}az}, {Herrero}
  \& {Barb{\'a}}}{{Holgado} et~al.}{2022}]{holgado+22}
{Holgado} G.,  {Sim{\'o}n-D{\'\i}az} S.,  {Herrero} A.,   {Barb{\'a}} R.~H.,
  2022, \mn@doi [\aap] {10.1051/0004-6361/202243851}, \href
  {https://ui.adsabs.harvard.edu/abs/2022A&A...665A.150H} {665, A150}

\bibitem[\protect\citeauthoryear{{Howarth}}{{Howarth}}{2011}]{howarth11}
{Howarth} I.~D.,  2011, \mn@doi [\mnras] {10.1111/j.1365-2966.2011.18122.x},
  \href {https://ui.adsabs.harvard.edu/abs/2011MNRAS.413.1515H} {413, 1515}

\bibitem[\protect\citeauthoryear{{Ikonnikova}, {Parthasarathy}, {Dodin},
  {Hubrig}  \& {Sarkar}}{{Ikonnikova} et~al.}{2020}]{ikonnikova+20}
{Ikonnikova} N.~P.,  {Parthasarathy} M.,  {Dodin} A.~V.,  {Hubrig} S.,
  {Sarkar} G.,  2020, \mn@doi [\mnras] {10.1093/mnras/stz3355}, \href
  {https://ui.adsabs.harvard.edu/abs/2020MNRAS.491.4829I} {491, 4829}

\bibitem[\protect\citeauthoryear{{Irrgang}, {Geier}, {Kreuzer}, {Pelisoli}  \&
  {Heber}}{{Irrgang} et~al.}{2020}]{irrgang+20}
{Irrgang} A.,  {Geier} S.,  {Kreuzer} S.,  {Pelisoli} I.,   {Heber} U.,  2020,
  \mn@doi [\aap] {10.1051/0004-6361/201937343}, \href
  {https://ui.adsabs.harvard.edu/abs/2020A&A...633L...5I} {633, L5}

\bibitem[\protect\citeauthoryear{{Irrgang}, {Przybilla}  \& {Meynet}}{{Irrgang}
  et~al.}{2022}]{irrgang+22}
{Irrgang} A.,  {Przybilla} N.,   {Meynet} G.,  2022, \mn@doi [Nature Astronomy]
  {10.1038/s41550-022-01809-6}, \href
  {https://ui.adsabs.harvard.edu/abs/2022NatAs...6.1414I} {6, 1414}

\bibitem[\protect\citeauthoryear{{Kiminki} \& {Kobulnicky}}{{Kiminki} \&
  {Kobulnicky}}{2012}]{kiminki+kobulnicky12}
{Kiminki} D.~C.,  {Kobulnicky} H.~A.,  2012, \mn@doi [\apj]
  {10.1088/0004-637X/751/1/4}, \href
  {https://ui.adsabs.harvard.edu/abs/2012ApJ...751....4K} {751, 4}

\bibitem[\protect\citeauthoryear{{Klement} et~al.,}{{Klement}
  et~al.}{2019}]{klement+19}
{Klement} R.,  et~al., 2019, \mn@doi [\apj] {10.3847/1538-4357/ab48e7}, \href
  {https://ui.adsabs.harvard.edu/abs/2019ApJ...885..147K} {885, 147}

\bibitem[\protect\citeauthoryear{{Klement} et~al.,}{{Klement}
  et~al.}{2022a}]{klement+22a}
{Klement} R.,  et~al., 2022a, \mn@doi [\apj] {10.3847/1538-4357/ac4266}, \href
  {https://ui.adsabs.harvard.edu/abs/2022ApJ...926..213K} {926, 213}

\bibitem[\protect\citeauthoryear{{Klement} et~al.,}{{Klement}
  et~al.}{2022b}]{klement+22b}
{Klement} R.,  et~al., 2022b, \mn@doi [\apj] {10.3847/1538-4357/ac98b8}, \href
  {https://ui.adsabs.harvard.edu/abs/2022ApJ...940...86K} {940, 86}

\bibitem[\protect\citeauthoryear{{Kobulnicky} \& {Fryer}}{{Kobulnicky} \&
  {Fryer}}{2007}]{kobulnicky+fryer07}
{Kobulnicky} H.~A.,  {Fryer} C.~L.,  2007, \mn@doi [\apj] {10.1086/522073},
  \href {https://ui.adsabs.harvard.edu/abs/2007ApJ...670..747K} {670, 747}

\bibitem[\protect\citeauthoryear{{Kobulnicky} et~al.,}{{Kobulnicky}
  et~al.}{2014}]{kobulnicky+14}
{Kobulnicky} H.~A.,  et~al., 2014, \mn@doi [\apjs]
  {10.1088/0067-0049/213/2/34}, \href
  {http://adsabs.harvard.edu/abs/2014ApJS..213...34K} {213, 34}

\bibitem[\protect\citeauthoryear{{Kurucz}}{{Kurucz}}{2005}]{kurucz05}
{Kurucz} R.~L.,  2005, Memorie della Societa Astronomica Italiana Supplementi,
  \href {https://ui.adsabs.harvard.edu/abs/2005MSAIS...8...14K} {8, 14}

\bibitem[\protect\citeauthoryear{{Langer} \& {Kudritzki}}{{Langer} \&
  {Kudritzki}}{2014}]{langer+kudritzki14}
{Langer} N.,  {Kudritzki} R.~P.,  2014, \mn@doi [\aap]
  {10.1051/0004-6361/201423374}, \href
  {https://ui.adsabs.harvard.edu/abs/2014A&A...564A..52L} {564, A52}

\bibitem[\protect\citeauthoryear{{Langer}, {Baade}, {Bodensteiner}, {Greiner},
  {Rivinius}, {Martayan}  \& {Borre}}{{Langer} et~al.}{2020a}]{langer+20b}
{Langer} N.,  {Baade} D.,  {Bodensteiner} J.,  {Greiner} J.,  {Rivinius} T.,
  {Martayan} C.,   {Borre} C.~C.,  2020a, \mn@doi [\aap]
  {10.1051/0004-6361/201936736}, \href
  {https://ui.adsabs.harvard.edu/abs/2020A&A...633A..40L} {633, A40}

\bibitem[\protect\citeauthoryear{{Langer} et~al.,}{{Langer}
  et~al.}{2020b}]{langer+20a}
{Langer} N.,  et~al., 2020b, \mn@doi [\aap] {10.1051/0004-6361/201937375},
  \href {https://ui.adsabs.harvard.edu/abs/2020A&A...638A..39L} {638, A39}

\bibitem[\protect\citeauthoryear{{Lanz} \& {Hubeny}}{{Lanz} \&
  {Hubeny}}{2003}]{lanz+hubeny03}
{Lanz} T.,  {Hubeny} I.,  2003, \mn@doi [\apjs] {10.1086/374373}, \href
  {https://ui.adsabs.harvard.edu/abs/2003ApJS..146..417L} {146, 417}

\bibitem[\protect\citeauthoryear{{Lanz} \& {Hubeny}}{{Lanz} \&
  {Hubeny}}{2007}]{lanz+hubeny07}
{Lanz} T.,  {Hubeny} I.,  2007, \mn@doi [\apjs] {10.1086/511270}, \href
  {http://adsabs.harvard.edu/abs/2007ApJS..169...83L} {169, 83}

\bibitem[\protect\citeauthoryear{{Lennon}, {Dufton}  \& {Fitzsimmons}}{{Lennon}
  et~al.}{1992}]{lennon+92}
{Lennon} D.~J.,  {Dufton} P.~L.,   {Fitzsimmons} A.,  1992, A$\&$AS, \href
  {http://adsabs.harvard.edu/abs/1992A%26AS...94..569L} {94, 569}

\bibitem[\protect\citeauthoryear{{Liu} et~al.,}{{Liu} et~al.}{2019}]{liu+19}
{Liu} J.,  et~al., 2019, \mn@doi [\nat] {10.1038/s41586-019-1766-2}, \href
  {https://ui.adsabs.harvard.edu/abs/2019Natur.575..618L} {575, 618}

\bibitem[\protect\citeauthoryear{{Lomb}}{{Lomb}}{1976}]{lomb76}
{Lomb} N.~R.,  1976, \mn@doi [\apss] {10.1007/BF00648343}, \href
  {https://ui.adsabs.harvard.edu/abs/1976Ap%26SS..39..447L} {39, 447}

\bibitem[\protect\citeauthoryear{{Mahy} et~al.,}{{Mahy} et~al.}{2022}]{mahy+22}
{Mahy} L.,  et~al., 2022, \mn@doi [\aap] {10.1051/0004-6361/202243147}, \href
  {https://ui.adsabs.harvard.edu/abs/2022A&A...664A.159M} {664, A159}

\bibitem[\protect\citeauthoryear{{Ma{\'\i}z Apell{\'a}niz} et~al.,}{{Ma{\'\i}z
  Apell{\'a}niz} et~al.}{2014}]{maiz+14}
{Ma{\'\i}z Apell{\'a}niz} J.,  et~al., 2014, \mn@doi [\aap]
  {10.1051/0004-6361/201423439}, \href
  {https://ui.adsabs.harvard.edu/abs/2014A&A...564A..63M} {564, A63}

\bibitem[\protect\citeauthoryear{{Marchenko}, {Moffat}  \&
  {Eenens}}{{Marchenko} et~al.}{1998}]{marchenko+98}
{Marchenko} S.~V.,  {Moffat} A. F.~J.,   {Eenens} P. R.~J.,  1998, \mn@doi
  [\pasp] {10.1086/316280}, \href
  {https://ui.adsabs.harvard.edu/abs/1998PASP..110.1416M} {110, 1416}

\bibitem[\protect\citeauthoryear{{Markova} \& {Puls}}{{Markova} \&
  {Puls}}{2008}]{markova+puls}
{Markova} N.,  {Puls} J.,  2008, \mn@doi [\aap] {10.1051/0004-6361:20077919},
  \href {https://ui.adsabs.harvard.edu/abs/2008A&A...478..823M} {478, 823}

\bibitem[\protect\citeauthoryear{{Mason}, {Hartkopf}, {Gies}, {Henry}  \&
  {Helsel}}{{Mason} et~al.}{2009}]{mason+09}
{Mason} B.~D.,  {Hartkopf} W.~I.,  {Gies} D.~R.,  {Henry} T.~J.,   {Helsel}
  J.~W.,  2009, \mn@doi [\aj] {10.1088/0004-6256/137/2/3358}, \href
  {https://ui.adsabs.harvard.edu/abs/2009AJ....137.3358M} {137, 3358}

\bibitem[\protect\citeauthoryear{{McEvoy} et~al.,}{{McEvoy}
  et~al.}{2015}]{mcevoy+15}
{McEvoy} C.~M.,  et~al., 2015, \mn@doi [\aap] {10.1051/0004-6361/201425202},
  \href {http://adsabs.harvard.edu/abs/2015A%26A...575A..70M} {575, A70}

\bibitem[\protect\citeauthoryear{{Mennickent} \&
  {Djura{\v{s}}evi{\'c}}}{{Mennickent} \&
  {Djura{\v{s}}evi{\'c}}}{2013}]{mennickent+djurasevic13}
{Mennickent} R.~E.,  {Djura{\v{s}}evi{\'c}} G.,  2013, \mn@doi [\mnras]
  {10.1093/mnras/stt515}, \href
  {https://ui.adsabs.harvard.edu/abs/2013MNRAS.432..799M} {432, 799}

\bibitem[\protect\citeauthoryear{{Moe} \& {Di Stefano}}{{Moe} \& {Di
  Stefano}}{2017}]{moe+distefano17}
{Moe} M.,  {Di Stefano} R.,  2017, \mn@doi [\apjs] {10.3847/1538-4365/aa6fb6},
  \href {https://ui.adsabs.harvard.edu/abs/2017ApJS..230...15M} {230, 15}

\bibitem[\protect\citeauthoryear{{Mourard} et~al.,}{{Mourard}
  et~al.}{2015}]{mourard+15}
{Mourard} D.,  et~al., 2015, \mn@doi [\aap] {10.1051/0004-6361/201425141},
  \href {https://ui.adsabs.harvard.edu/abs/2015A&A...577A..51M} {577, A51}

\bibitem[\protect\citeauthoryear{{Mourard} et~al.,}{{Mourard}
  et~al.}{2018}]{mourard+18}
{Mourard} D.,  et~al., 2018, \mn@doi [\aap] {10.1051/0004-6361/201832952},
  \href {https://ui.adsabs.harvard.edu/abs/2018A&A...618A.112M} {618, A112}

\bibitem[\protect\citeauthoryear{{Offner}, {Moe}, {Kratter}, {Sadavoy},
  {Jensen}  \& {Tobin}}{{Offner} et~al.}{2023}]{offner+22}
{Offner} S.~S.~R.,  {Moe} M.,  {Kratter} K.~M.,  {Sadavoy} S.~I.,  {Jensen}
  E.~L.~N.,   {Tobin} J.~J.,  2023, in {Inutsuka} S.,  {Aikawa} Y.,  {Muto} T.,
   {Tomida} K.,   {Tamura} M.,  eds,  Astronomical Society of the Pacific
  Conference Series Vol. 534, Astronomical Society of the Pacific Conference
  Series. p.~275

\bibitem[\protect\citeauthoryear{{Packet}}{{Packet}}{1981}]{packet81}
{Packet} W.,  1981, \aap, \href
  {https://ui.adsabs.harvard.edu/abs/1981A&A...102...17P} {102, 17}

\bibitem[\protect\citeauthoryear{{Paczy{\'n}ski}}{{Paczy{\'n}ski}}{1967}]{paczynski67}
{Paczy{\'n}ski} B.,  1967, \actaa, \href
  {https://ui.adsabs.harvard.edu/abs/1967AcA....17..355P} {17, 355}

\bibitem[\protect\citeauthoryear{{Parker} et~al.,}{{Parker}
  et~al.}{1998}]{parker+98}
{Parker} J.~W.,  et~al., 1998, \mn@doi [\aj] {10.1086/300419}, \href
  {https://ui.adsabs.harvard.edu/abs/1998AJ....116..180P} {116, 180}

\bibitem[\protect\citeauthoryear{{Patrick} et~al.,}{{Patrick}
  et~al.}{2019}]{patrick+19}
{Patrick} L.~R.,  et~al., 2019, \mn@doi [\aap] {10.1051/0004-6361/201834951},
  \href {https://ui.adsabs.harvard.edu/abs/2019A&A...624A.129P} {624, A129}

\bibitem[\protect\citeauthoryear{{Peters}, {Gies}, {Grundstrom}  \&
  {McSwain}}{{Peters} et~al.}{2008}]{peters+08}
{Peters} G.~J.,  {Gies} D.~R.,  {Grundstrom} E.~D.,   {McSwain} M.~V.,  2008,
  \mn@doi [\apj] {10.1086/591145}, \href
  {https://ui.adsabs.harvard.edu/abs/2008ApJ...686.1280P} {686, 1280}

\bibitem[\protect\citeauthoryear{{Peters}, {Pewett}, {Gies}, {Touhami}  \&
  {Grundstrom}}{{Peters} et~al.}{2013}]{peters+13}
{Peters} G.~J.,  {Pewett} T.~D.,  {Gies} D.~R.,  {Touhami} Y.~N.,
  {Grundstrom} E.~D.,  2013, \mn@doi [\apj] {10.1088/0004-637X/765/1/2}, \href
  {https://ui.adsabs.harvard.edu/abs/2013ApJ...765....2P} {765, 2}

\bibitem[\protect\citeauthoryear{{Pietrzy{\'n}ski} et~al.,}{{Pietrzy{\'n}ski}
  et~al.}{2019}]{pietrzynski+19}
{Pietrzy{\'n}ski} G.,  et~al., 2019, \mn@doi [\nat]
  {10.1038/s41586-019-0999-4}, \href
  {https://ui.adsabs.harvard.edu/abs/2019Natur.567..200P} {567, 200}

\bibitem[\protect\citeauthoryear{{Podsiadlowski}, {Joss}  \&
  {Hsu}}{{Podsiadlowski} et~al.}{1992}]{podsiadlowski+92}
{Podsiadlowski} P.,  {Joss} P.~C.,   {Hsu} J.~J.~L.,  1992, \mn@doi [\apj]
  {10.1086/171341}, \href {http://adsabs.harvard.edu/abs/1992ApJ...391..246P}
  {391, 246}

\bibitem[\protect\citeauthoryear{{Pols}, {Cote}, {Waters}  \& {Heise}}{{Pols}
  et~al.}{1991}]{pols+91}
{Pols} O.~R.,  {Cote} J.,  {Waters} L.~B.~F.~M.,   {Heise} J.,  1991, \aap,
  \href {https://ui.adsabs.harvard.edu/abs/1991A&A...241..419P} {241, 419}

\bibitem[\protect\citeauthoryear{{Quintero}, {Eenens}  \& {Rauw}}{{Quintero}
  et~al.}{2020}]{Quintero+2020}
{Quintero} E.~A.,  {Eenens} P.,   {Rauw} G.,  2020, \mn@doi [Astronomische
  Nachrichten] {10.1002/asna.202013696}, \href
  {https://ui.adsabs.harvard.edu/abs/2020AN....341..628Q} {341, 628}

\bibitem[\protect\citeauthoryear{{Ramachandran}, {Klencki}, {Sander}, {Pauli},
  {Shenar}, {Oskinova}  \& {Hamann}}{{Ramachandran}
  et~al.}{2023}]{ramachandran+23}
{Ramachandran} V.,  {Klencki} J.,  {Sander} A.~A.~C.,  {Pauli} D.,  {Shenar}
  T.,  {Oskinova} L.~M.,   {Hamann} W.~R.,  2023, \mn@doi [\aap]
  {10.1051/0004-6361/202346818}, \href
  {https://ui.adsabs.harvard.edu/abs/2023A&A...674L..12R} {674, L12}

\bibitem[\protect\citeauthoryear{{Reed}}{{Reed}}{2003}]{reed03}
{Reed} B.~C.,  2003, \mn@doi [\aj] {10.1086/374771}, \href
  {https://ui.adsabs.harvard.edu/abs/2003AJ....125.2531R} {125, 2531}

\bibitem[\protect\citeauthoryear{{Rivinius}, {Carciofi}  \&
  {Martayan}}{{Rivinius} et~al.}{2013}]{rivinius+13}
{Rivinius} T.,  {Carciofi} A.~C.,   {Martayan} C.,  2013, \mn@doi [\aapr]
  {10.1007/s00159-013-0069-0}, \href
  {https://ui.adsabs.harvard.edu/abs/2013A&ARv..21...69R} {21, 69}

\bibitem[\protect\citeauthoryear{{Rivinius}, {Baade}, {Hadrava}, {Heida}  \&
  {Klement}}{{Rivinius} et~al.}{2020}]{rivinius+20}
{Rivinius} T.,  {Baade} D.,  {Hadrava} P.,  {Heida} M.,   {Klement} R.,  2020,
  \mn@doi [\aap] {10.1051/0004-6361/202038020}, \href
  {https://ui.adsabs.harvard.edu/abs/2020A&A...637L...3R} {637, L3}

\bibitem[\protect\citeauthoryear{{Rivinius}, {Klement}, {Chojnowski}, {Baade},
  {Shepard}  \& {Hadrava}}{{Rivinius} et~al.}{2022}]{rivinius+22}
{Rivinius} T.,  {Klement} R.,  {Chojnowski} S.~D.,  {Baade} D.,  {Shepard} K.,
   {Hadrava} P.,  2022, arXiv e-prints, \href
  {https://ui.adsabs.harvard.edu/abs/2022arXiv220812315R} {p. arXiv:2208.12315}

\bibitem[\protect\citeauthoryear{{Roming} et~al.,}{{Roming}
  et~al.}{2005}]{roming+05}
{Roming} P. W.~A.,  et~al., 2005, \mn@doi [\ssr] {10.1007/s11214-005-5095-4},
  \href {https://ui.adsabs.harvard.edu/abs/2005SSRv..120...95R} {120, 95}

\bibitem[\protect\citeauthoryear{{Sabbi} et~al.,}{{Sabbi}
  et~al.}{2013}]{sabbi+13}
{Sabbi} E.,  et~al., 2013, \mn@doi [\aj] {10.1088/0004-6256/146/3/53}, \href
  {https://ui.adsabs.harvard.edu/abs/2013AJ....146...53S} {146, 53}

\bibitem[\protect\citeauthoryear{{Sabbi} et~al.,}{{Sabbi}
  et~al.}{2016}]{sabbi+16}
{Sabbi} E.,  et~al., 2016, \mn@doi [\apjs] {10.3847/0067-0049/222/1/11}, \href
  {https://ui.adsabs.harvard.edu/abs/2016ApJS..222...11S} {222, 11}

\bibitem[\protect\citeauthoryear{{Sana} et~al.,}{{Sana} et~al.}{2012}]{sana+12}
{Sana} H.,  et~al., 2012, \mn@doi [Science] {10.1126/science.1223344}, \href
  {http://adsabs.harvard.edu/abs/2012Sci...337..444S} {337, 444}

\bibitem[\protect\citeauthoryear{{Sana} et~al.,}{{Sana} et~al.}{2013}]{sana+13}
{Sana} H.,  et~al., 2013, \mn@doi [\aap] {10.1051/0004-6361/201219621}, \href
  {http://adsabs.harvard.edu/abs/2013A%26A...550A.107S} {550, A107}

\bibitem[\protect\citeauthoryear{{Sana} et~al.,}{{Sana} et~al.}{2014}]{sana+14}
{Sana} H.,  et~al., 2014, \mn@doi [\apjs] {10.1088/0067-0049/215/1/15}, \href
  {https://ui.adsabs.harvard.edu/abs/2014ApJS..215...15S} {215, 15}

\bibitem[\protect\citeauthoryear{{Saracino} et~al.,}{{Saracino}
  et~al.}{2022}]{saracino+22}
{Saracino} S.,  et~al., 2022, \mn@doi [\mnras] {10.1093/mnras/stab3159}, \href
  {https://ui.adsabs.harvard.edu/abs/2022MNRAS.511.2914S} {511, 2914}

\bibitem[\protect\citeauthoryear{{Saracino} et~al.,}{{Saracino}
  et~al.}{2023}]{saracino+23}
{Saracino} S.,  et~al., 2023, \mn@doi [\mnras] {10.1093/mnras/stad764}, \href
  {https://ui.adsabs.harvard.edu/abs/2023MNRAS.521.3162S} {521, 3162}

\bibitem[\protect\citeauthoryear{{Scargle}}{{Scargle}}{1982}]{scargle82}
{Scargle} J.~D.,  1982, \mn@doi [\apj] {10.1086/160554}, \href
  {https://ui.adsabs.harvard.edu/abs/1982ApJ...263..835S} {263, 835}

\bibitem[\protect\citeauthoryear{{Schneider}, {Langer}, {de Koter}, {Brott},
  {Izzard}  \& {Lau}}{{Schneider} et~al.}{2014}]{schneider+14}
{Schneider} F.~R.~N.,  {Langer} N.,  {de Koter} A.,  {Brott} I.,  {Izzard}
  R.~G.,   {Lau} H.~H.~B.,  2014, \mn@doi [\aap] {10.1051/0004-6361/201424286},
  \href {https://ui.adsabs.harvard.edu/abs/2014A&A...570A..66S} {570, A66}

\bibitem[\protect\citeauthoryear{{Schneider} et~al.,}{{Schneider}
  et~al.}{2018}]{schneider+18}
{Schneider} F.~R.~N.,  et~al., 2018, \mn@doi [\aap]
  {10.1051/0004-6361/201833433}, \href
  {https://ui.adsabs.harvard.edu/abs/2018A&A...618A..73S} {618, A73}

\bibitem[\protect\citeauthoryear{{Schootemeijer}, {G{\"o}tberg}, {de Mink},
  {Gies}  \& {Zapartas}}{{Schootemeijer} et~al.}{2018}]{schootemeijer+18}
{Schootemeijer} A.,  {G{\"o}tberg} Y.,  {de Mink} S.~E.,  {Gies} D.,
  {Zapartas} E.,  2018, \mn@doi [\aap] {10.1051/0004-6361/201731194}, \href
  {https://ui.adsabs.harvard.edu/abs/2018A&A...615A..30S} {615, A30}

\bibitem[\protect\citeauthoryear{{Sch{\"u}rmann}, {Langer}, {Xu}  \&
  {Wang}}{{Sch{\"u}rmann} et~al.}{2022}]{schurmann+22}
{Sch{\"u}rmann} C.,  {Langer} N.,  {Xu} X.,   {Wang} C.,  2022, \mn@doi [\aap]
  {10.1051/0004-6361/202244153}, \href
  {https://ui.adsabs.harvard.edu/abs/2022A&A...667A.122S} {667, A122}

\bibitem[\protect\citeauthoryear{{Sen}, {Xu}, {Langer}, {El Mellah},
  {Sch{\"u}rmann}  \& {Quast}}{{Sen} et~al.}{2021}]{sen+21}
{Sen} K.,  {Xu} X.~T.,  {Langer} N.,  {El Mellah} I.,  {Sch{\"u}rmann} C.,
  {Quast} M.,  2021, \mn@doi [\aap] {10.1051/0004-6361/202141214}, \href
  {https://ui.adsabs.harvard.edu/abs/2021A&A...652A.138S} {652, A138}

\bibitem[\protect\citeauthoryear{{Sen} et~al.,}{{Sen} et~al.}{2022}]{sen+22}
{Sen} K.,  et~al., 2022, \mn@doi [\aap] {10.1051/0004-6361/202142574}, \href
  {https://ui.adsabs.harvard.edu/abs/2022A&A...659A..98S} {659, A98}

\bibitem[\protect\citeauthoryear{{Shenar} et~al.,}{{Shenar}
  et~al.}{2019}]{Shenar+2019}
{Shenar} T.,  et~al., 2019, \mn@doi [\aap] {10.1051/0004-6361/201935684}, \href
  {https://ui.adsabs.harvard.edu/abs/2019A&A...627A.151S} {627, A151}

\bibitem[\protect\citeauthoryear{{Shenar} et~al.,}{{Shenar}
  et~al.}{2020}]{shenar+20}
{Shenar} T.,  et~al., 2020, \mn@doi [\aap] {10.1051/0004-6361/202038275}, \href
  {https://ui.adsabs.harvard.edu/abs/2020A&A...639L...6S} {639, L6}

\bibitem[\protect\citeauthoryear{{Shenar} et~al.,}{{Shenar}
  et~al.}{2022a}]{shenar+22a}
{Shenar} T.,  et~al., 2022a, \mn@doi [Nature Astronomy]
  {10.1038/s41550-022-01730-y}, \href
  {https://ui.adsabs.harvard.edu/abs/2022NatAs...6.1085S} {6, 1085}

\bibitem[\protect\citeauthoryear{{Shenar} et~al.,}{{Shenar}
  et~al.}{2022b}]{shenar+22b}
{Shenar} T.,  et~al., 2022b, \mn@doi [\aap] {10.1051/0004-6361/202244245},
  \href {https://ui.adsabs.harvard.edu/abs/2022A&A...665A.148S} {665, A148}

\bibitem[\protect\citeauthoryear{{Sim{\'o}n-D{\'\i}az} \&
  {Herrero}}{{Sim{\'o}n-D{\'\i}az} \& {Herrero}}{2014}]{simon-diaz+herrero14}
{Sim{\'o}n-D{\'\i}az} S.,  {Herrero} A.,  2014, \mn@doi [\aap]
  {10.1051/0004-6361/201322758}, \href
  {https://ui.adsabs.harvard.edu/#abs/2014A&A...562A.135S} {562, A135}

\bibitem[\protect\citeauthoryear{{Soberman}, {Phinney}  \& {van den
  Heuvel}}{{Soberman} et~al.}{1997}]{soberman+97}
{Soberman} G.~E.,  {Phinney} E.~S.,   {van den Heuvel} E.~P.~J.,  1997, \aap,
  \href {https://ui.adsabs.harvard.edu/abs/1997A&A...327..620S} {327, 620}

\bibitem[\protect\citeauthoryear{{Sota}, {Ma{\'{\i}}z Apell{\'a}niz},
  {Walborn}, {Alfaro}, {Barb{\'a}}, {Morrell}, {Gamen}  \& {Arias}}{{Sota}
  et~al.}{2011}]{sota+11}
{Sota} A.,  {Ma{\'{\i}}z Apell{\'a}niz} J.,  {Walborn} N.~R.,  {Alfaro} E.~J.,
  {Barb{\'a}} R.~H.,  {Morrell} N.~I.,  {Gamen} R.~C.,   {Arias} J.~I.,  2011,
  \mn@doi [\apjs] {10.1088/0067-0049/193/2/24}, \href
  {http://adsabs.harvard.edu/abs/2011ApJS..193...24S} {193, 24}

\bibitem[\protect\citeauthoryear{{Steiner} \& {Oliveira}}{{Steiner} \&
  {Oliveira}}{2005}]{steiner+oliveira05}
{Steiner} J.~E.,  {Oliveira} A.~S.,  2005, \mn@doi [\aap]
  {10.1051/0004-6361:20052782}, \href
  {https://ui.adsabs.harvard.edu/abs/2005A&A...444..895S} {444, 895}

\bibitem[\protect\citeauthoryear{{Stellingwerf}}{{Stellingwerf}}{1978}]{stellingwerf78}
{Stellingwerf} R.~F.,  1978, \mn@doi [\apj] {10.1086/156444}, \href
  {https://ui.adsabs.harvard.edu/abs/1978ApJ...224..953S} {224, 953}

\bibitem[\protect\citeauthoryear{{The LIGO Scientific Collaboration}
  et~al.,}{{The LIGO Scientific Collaboration} et~al.}{2021}]{abbott+21}
{The LIGO Scientific Collaboration} et~al., 2021, \mn@doi [arXiv e-prints]
  {10.48550/arXiv.2111.03606}, \href
  {https://ui.adsabs.harvard.edu/abs/2021arXiv211103606T} {p. arXiv:2111.03606}

\bibitem[\protect\citeauthoryear{{Torres}}{{Torres}}{2010}]{torres10}
{Torres} G.,  2010, \mn@doi [\aj] {10.1088/0004-6256/140/5/1158}, \href
  {https://ui.adsabs.harvard.edu/abs/2010AJ....140.1158T} {140, 1158}

\bibitem[\protect\citeauthoryear{{Udalski}, {Szymanski}, {Soszynski}  \&
  {Poleski}}{{Udalski} et~al.}{2008}]{udalski+08}
{Udalski} A.,  {Szymanski} M.~K.,  {Soszynski} I.,   {Poleski} R.,  2008,
  \mn@doi [\actaa] {10.48550/arXiv.0807.3884}, \href
  {https://ui.adsabs.harvard.edu/abs/2008AcA....58...69U} {58, 69}

\bibitem[\protect\citeauthoryear{{Udalski}, {Szyma{\'n}ski}  \&
  {Szyma{\'n}ski}}{{Udalski} et~al.}{2015}]{udalski+15}
{Udalski} A.,  {Szyma{\'n}ski} M.~K.,   {Szyma{\'n}ski} G.,  2015, \mn@doi
  [\actaa] {10.48550/arXiv.1504.05966}, \href
  {https://ui.adsabs.harvard.edu/abs/2015AcA....65....1U} {65, 1}

\bibitem[\protect\citeauthoryear{{Villase{\~n}or} et~al.,}{{Villase{\~n}or}
  et~al.}{2021}]{villasenor+21}
{Villase{\~n}or} J.~I.,  et~al., 2021, \mn@doi [\mnras]
  {10.1093/mnras/stab2197}, \href
  {https://ui.adsabs.harvard.edu/abs/2021MNRAS.507.5348V} {507, 5348}

\bibitem[\protect\citeauthoryear{{Vogt}, {Barrera}  \& {Navarro}}{{Vogt}
  et~al.}{1990}]{vogt+90}
{Vogt} N.,  {Barrera} L.~H.,   {Navarro} M.,  1990, \mn@doi [\apss]
  {10.1007/BF00642569}, \href
  {https://ui.adsabs.harvard.edu/abs/1990Ap&SS.173..145V} {173, 145}

\bibitem[\protect\citeauthoryear{{Walborn} \& {Blades}}{{Walborn} \&
  {Blades}}{1997}]{walborn+blades97}
{Walborn} N.~R.,  {Blades} J.~C.,  1997, \mn@doi [\apjs] {10.1086/313043},
  \href {https://ui.adsabs.harvard.edu/abs/1997ApJS..112..457W} {112, 457}

\bibitem[\protect\citeauthoryear{{Walborn} \& {Bohlin}}{{Walborn} \&
  {Bohlin}}{1996}]{walborn+bohlin96}
{Walborn} N.~R.,  {Bohlin} R.~C.,  1996, \mn@doi [\pasp] {10.1086/133753},
  \href {https://ui.adsabs.harvard.edu/abs/1996PASP..108..477W} {108, 477}

\bibitem[\protect\citeauthoryear{{Wang}, {Gies}  \& {Peters}}{{Wang}
  et~al.}{2017}]{wangL+17}
{Wang} L.,  {Gies} D.~R.,   {Peters} G.~J.,  2017, \mn@doi [\apj]
  {10.3847/1538-4357/aa740a}, \href
  {https://ui.adsabs.harvard.edu/abs/2017ApJ...843...60W} {843, 60}

\bibitem[\protect\citeauthoryear{{Wang}, {Gies}  \& {Peters}}{{Wang}
  et~al.}{2018}]{wangL+18}
{Wang} L.,  {Gies} D.~R.,   {Peters} G.~J.,  2018, \mn@doi [\apj]
  {10.3847/1538-4357/aaa4b8}, \href
  {https://ui.adsabs.harvard.edu/abs/2018ApJ...853..156W} {853, 156}

\bibitem[\protect\citeauthoryear{{Wang}, {Gies}, {Peters}, {G{\"o}tberg},
  {Chojnowski}, {Lester}  \& {Howell}}{{Wang} et~al.}{2021}]{wangL+21}
{Wang} L.,  {Gies} D.~R.,  {Peters} G.~J.,  {G{\"o}tberg} Y.,  {Chojnowski}
  S.~D.,  {Lester} K.~V.,   {Howell} S.~B.,  2021, \mn@doi [\aj]
  {10.3847/1538-3881/abf144}, \href
  {https://ui.adsabs.harvard.edu/abs/2021AJ....161..248W} {161, 248}

\bibitem[\protect\citeauthoryear{{Wang}, {Gies}, {Peters}  \& {Han}}{{Wang}
  et~al.}{2023}]{wangL+23}
{Wang} L.,  {Gies} D.~R.,  {Peters} G.~J.,   {Han} Z.,  2023, \mn@doi [\aj]
  {10.3847/1538-3881/acc6ca}, \href
  {https://ui.adsabs.harvard.edu/abs/2023AJ....165..203W} {165, 203}

\bibitem[\protect\citeauthoryear{{Wellstein}, {Langer}  \& {Braun}}{{Wellstein}
  et~al.}{2001}]{wellstein+01}
{Wellstein} S.,  {Langer} N.,   {Braun} H.,  2001, \mn@doi [\aap]
  {10.1051/0004-6361:20010151}, \href
  {https://ui.adsabs.harvard.edu/abs/2001A&A...369..939W} {369, 939}

\bibitem[\protect\citeauthoryear{{de Mink}, {Langer}, {Izzard}, {Sana}  \& {de
  Koter}}{{de Mink} et~al.}{2013}]{demink+13}
{de Mink} S.~E.,  {Langer} N.,  {Izzard} R.~G.,  {Sana} H.,   {de Koter} A.,
  2013, \mn@doi [\apj] {10.1088/0004-637X/764/2/166}, \href
  {http://adsabs.harvard.edu/abs/2013ApJ...764..166D} {764, 166}

\bibitem[\protect\citeauthoryear{{de Mink}, {Sana}, {Langer}, {Izzard}  \&
  {Schneider}}{{de Mink} et~al.}{2014}]{demink+14}
{de Mink} S.~E.,  {Sana} H.,  {Langer} N.,  {Izzard} R.~G.,   {Schneider}
  F.~R.~N.,  2014, \mn@doi [\apj] {10.1088/0004-637X/782/1/7}, \href
  {http://adsabs.harvard.edu/abs/2014ApJ...782....7D} {782, 7}

\bibitem[\protect\citeauthoryear{{van den Heuvel}}{{van den
  Heuvel}}{2019}]{vandenheuvel19}
{van den Heuvel} E. P.~J.,  2019, \mn@doi [IAU Symposium]
  {10.1017/S1743921319001315}, \href
  {https://ui.adsabs.harvard.edu/abs/2019IAUS..346....1V} {346, 1}

\makeatother
\end{thebibliography}

%\begin{thebibliography}{99}
%\end{thebibliography}

%%%%%%%%%%%%%%%%%%%%%%%%%%%%%%%%%%%%%%%%%%%%%%%%%%

%%%%%%%%%%%%%%%%% APPENDICES %%%%%%%%%%%%%%%%%%%%%

 \appendix

\section{Disentangling results}\label{ap:dist_chi2}

In Sect.~\ref{ssec:disent}, we presented the disentangling results for $K_2$ based on the \spline{He}{i}{4388} line. Here we show the reduced \chisq as a function of $K_2$ for other three strong \ion{He}{i} lines. It is clear from the different minima found for each line that $K_2$ is not very well constrained, with values as low as 0\kms, but suggesting a slowly-moving companion. 

\begin{figure}
    \centering
    \begin{subfigure}[t]{\columnwidth}
        \centering
        \caption{\spline{He}{i}{4009}}
        \includegraphics[width=\columnwidth]{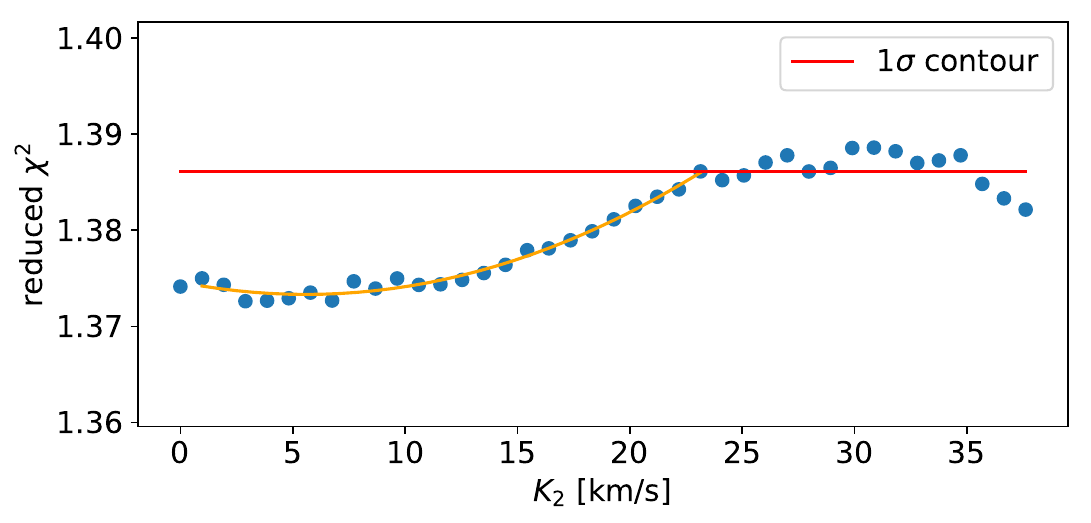} 
    \end{subfigure}

    \begin{subfigure}[t]{\columnwidth}
        \centering
        \caption{\spline{He}{i}{4144}}
        \includegraphics[width=\columnwidth]{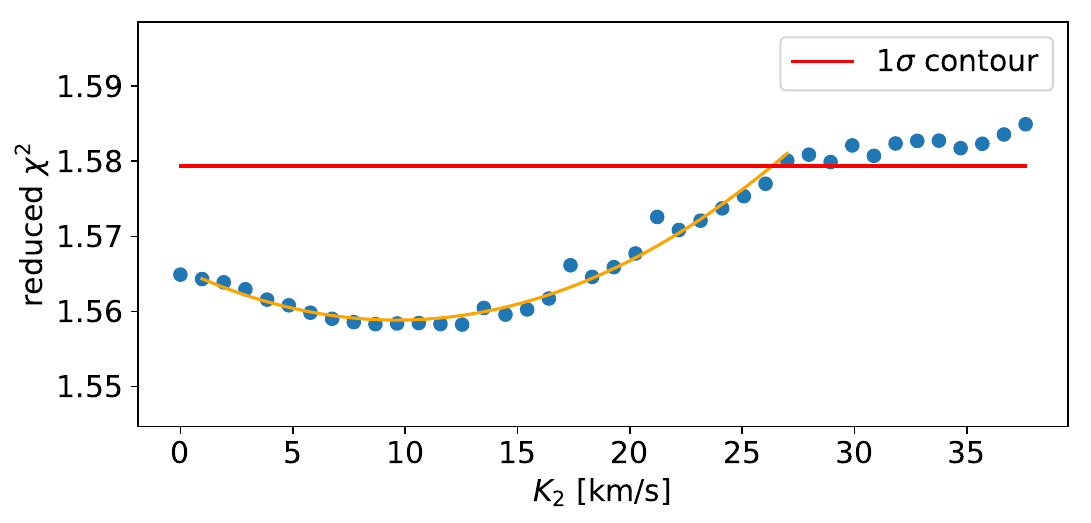} 
    \end{subfigure}

    \begin{subfigure}[t]{\columnwidth}
        \centering
        \caption{\spline{He}{i}{4471}}
        \includegraphics[width=\columnwidth]{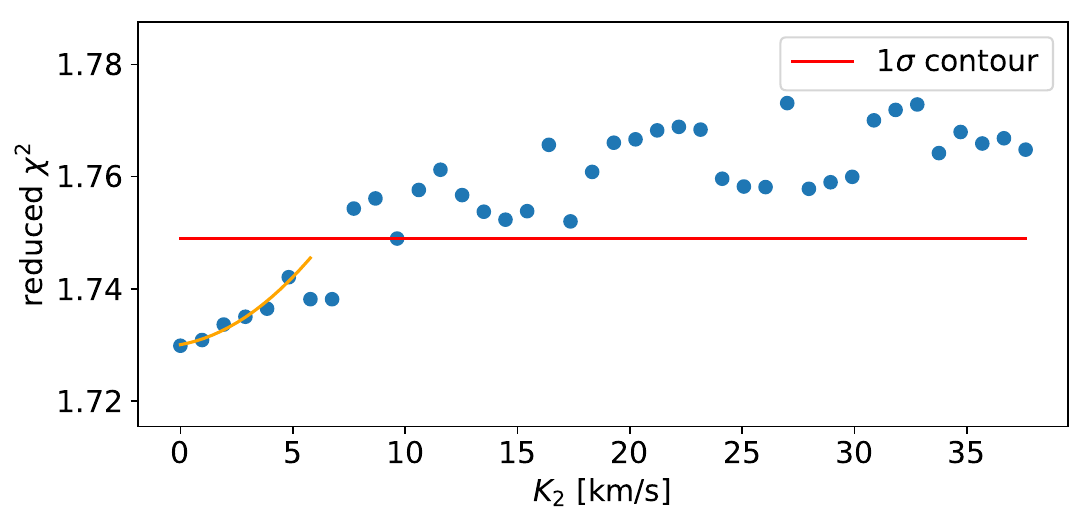} 
    \end{subfigure}
    \caption{As Fig.~\ref{fig:chi2-K2} but for three different \ion{He}{i} lines.}
    \label{fig:AppDisent}
\end{figure}
 
\section{Phoebe Models}\label{ap:phoebe}
To improve the folded light curve of the OGLE-III data in the $I$ band, we have fitted the light curve with PHOEBE allowing the orbital period to vary. As explained in Sec.~\ref{ssec:ogle}, with the orbital period found by PHOEBE we computed a new orbital solution which is shown in Table~\ref{tab:phoebe_orbsol}. We have used the values of this new solution to constrain the physical properties of both stars by fitting the light curve, leaving all orbital parameters, distance, and extinction, fixed. 

For the physical parameters, we tested different combinations of fixed and fitted parameters. The results of these tests can be seen in Table~\ref{tab:phoebe}. First, we fixed the orbital inclination to three values ($60^\circ$, $70^\circ$, $83^\circ$), these correspond to models \#1 to \#9. We let the mass ratio fixed to values of 6 and 8, but also allowed $q$ to vary, the latter in models \#3, \#6, and \#9. All other parameters in these first nine models were allowed to vary. In the next two models, we let $i$ and $q$ to vary, and now fixed $T_{{\rm eff},A}$ (model \#10) and $R_A$ (model \#11). Finally, for model \#12, we let all parameters to vary. For the computation of all these models, we provided the spectroscopic values as initial guesses.
\begin{table}
\caption{Orbital solution obtained by fixing the period found with PHOEBE from the light curve fitting.}
\label{tab:phoebe_orbsol}
\center
\begin{tabular}{lr}
\toprule
\toprule
Parameter			&Value\\
\midrule
$P_{\rm orb}$ (d)  & 108.104 (fixed)\\
$T_p$ (HJD)		   & 2457374.49 $\pm$ 1.82\\
$e$			       & 0.020 $\pm$ 0.005 \\
$\omega$ (deg)	   & 253.64 $\pm$ 6.03 \\
$\gamma$ (km/s)	   & 270.94 $\pm$ 0.35 \\
$K_1$ (km/s)	   & 94.10 $\pm$ 0.47 \\
$a_1\sin i$ (\Rsun) & 200.96 $\pm$ 1.01\\
$f(m_1,m_2)$ ($M_\odot$) & 9.33 $\pm$ 0.19\\
\bottomrule
\end{tabular}
\end{table}
\begin{table}
\caption{Phoebe models computed for different inclinations ($i$) and mass ratios ($q$). Model number is given in column $N$.}
\label{tab:phoebe}
\center
\resizebox{\columnwidth}{!}{%
\begin{tabular}{ccccccccccc}
\toprule
\toprule
 $N$ & $i$ & $q$ & $M_A$ & $\log g_A$ & $R_A$ & $T_{{\rm eff},A}$ & $M_B$ & $\log g_B$ & $R_B$ & $T_{{\rm eff},B}$  \\
 & [deg] & & [\Msun] & [dex] & [\Rsun] & [K] & [\Msun] & [dex] & [\Rsun] & [K] \\ 
\midrule
1 & 60 & 6 & 3.25 & 2.16 & 24.87 & 14689 & 19.53 & 3.97 & 7.60 & 27402 \\
2 & 60 & 8 & 2.27 & 2.11 & 21.87 & 14513 & 18.16 & 4.07 & 6.53 & 22859 \\
3 & 60 & 9.13 & 1.94 & 2.09 & 20.80 & 13171 & 17.66 & 3.99 & 7.06 & 21630 \\
[0.1cm]
4 & 70 & 6 & 2.55 & 2.15 & 22.29 & 14692 & 15.29 & 3.88 & 7.43 & 25713 \\
5 & 70 & 8 & 1.78 & 2.09 & 19.82 & 14316 & 14.21 & 3.87 & 7.22 & 25276 \\
6 & 70 & 8.37 & 1.68 & 2.09 & 19.44 & 13501 & 14.07 & 3.84 & 7.43 & 24851 \\
[0.1cm]
7 & 83 & 6 & 2.16 & 2.14 & 20.63 & 13923 & 12.97 & 3.87 & 6.93 & 25307 \\
8 & 83 & 8 & 1.51 & 2.08 & 18.54 & 13996 & 12.06 & 3.79 & 7.33 & 28179 \\
9 & 83 & 7.87 & 1.54 & 2.10 & 18.33 & 13582 & 12.11 & 3.77 & 7.55 & 28393 \\
[0.1cm]
10 & 77.47 & 8.98 & 1.38 & 2.08 & 17.69 & 14045 & 12.37 & 3.93 & 6.33 & 24662 \\
11 & 71.70 & 8.53 & 1.59 & 2.07 & 19.28 & 13180 & 13.59 & 3.86 & 7.17 & 27123 \\
12 & 65.77 & 8.65 & 1.77 & 2.09 & 19.88 & 12646 & 15.29 & 3.90 & 7.30 & 23405 \\
\bottomrule
\end{tabular}}
\tablefoot{Subscripts $A$ and $B$ refer to the narrow-lined star and companion respectively. The radius and effective temperature of component $A$ have been fixed to the values obtained in our SED fitting and atmosphere analysis in models \#10 and \#11 respectively, i.e., $R_A=17.69$\Rsun and $T_{{\rm eff},A}=\,13180$\,K.}
\end{table}

% Don't change these lines
\bsp	% typesetting comment
\label{lastpage}
\end{document}